\documentclass[a4paper,11pt]{article}
\usepackage[english]{babel}
\usepackage{enumerate}
\usepackage{hyperref}
\hypersetup{
    colorlinks=true,
    citecolor=blue,
    linkcolor=blue,
    filecolor=red,      
    urlcolor=blue,
}
 
\usepackage{doi}
\usepackage{ragged2e}
\usepackage{setspace}
\usepackage{graphics}
\usepackage{paralist}
\usepackage{epstopdf}
\usepackage{natbib}
\usepackage{mathrsfs}
\usepackage{widetext}
\usepackage{eqnarray}
\usepackage{amsmath}
\usepackage{graphicx,amssymb,amsfonts,epsfig}
\usepackage[usenames,dvipsnames]{color}
\usepackage{cleveref}
\usepackage{subfigure}
\title{Inertial Frame Dragging as a Probe to Differentiate Kerr–Newman Naked Singularities from Black Holes}
\author{
	{Arindam Kumar Chatterjee$^{a}$\thanks{72arindam@gmail.com}},
	{Parthapratim Pradhan$^{b,}$\thanks{pppradhan77@gmail.com}}, 
	\\[0.3cm]
	{\small $^a$ Department of Physics, Gurukul Kangri (Deemed to be University), Haridwar - 249407, India.}\\[0.1cm]
	{\small $^b$ Department of Physics, Hiralal Mazumdar Memorial College For Women, Kolkata-700035, India.}\\[0.1cm]
}
\begin{document}
\baselineskip=0.5 cm
\date{}
\maketitle
\begin{abstract}
	\baselineskip=0.5 cm
	\noindent
	We study here the precession of the spin of a test gyroscope attached to a stationary observer in the Kerr-Newman spacetime, specifically, to distinguish a naked singularity from a black hole. Extending previous work on Kerr, we thoroughly investigate the role of the electric charge $Q$ and its effects on the precession of the spin of a test gyroscope for black hole and naked singularity. For gyros attached to stationary observers with nonzero angular velocity $\Omega$, we derive closed-form expressions for the general spin-precession, Lense-Thirring, and geodetic precession frequencies. For Kerr-Newman black holes, the spin-precession frequency generically diverges as the event horizon is approached from any direction, remaining finite only for the zero-angular-momentum observer (ZAMO) family. In contrast, for Kerr-Newman naked singularities, the precession remains finite throughout the spacetime, with divergences confined solely to the ring singularity on the equatorial plane. We show that $Q$ systematically modifies these behaviours, particularly in rapidly rotating regimes, and that the acceleration scalar of stationary observers further sharpens the black hole/naked singularity distinction. We then investigate the Lense-Thirring (nodal) precession frequency of equatorial circular orbits in accretion disks around Kerr-Newman black holes and naked singularities. For black holes, the nodal frequency decreases monotonically with increasing radius, whereas for naked singularities it increases, attains a finite maximum, and then decreases; for sufficiently large spin and charge it can even change sign, signalling a reversal of the precession direction. We also compute the fundamental orbital frequencies (Keplerian, radial and vertical epicyclic) and the periastron precession frequency, highlighting the role of the electric charge $Q$ on these frequencies, and showing how the charge parameter modifies the innermost stable circular orbit (ISCO) and the characteristic frequency hierarchy. Since these precession frequencies are intimately related to observed quasiperiodic oscillations (QPOs), such features provide a robust strong-field probe to determine whether a given rotating compact object is a black hole or a naked singularity, with implications for testing the cosmic censorship conjecture using future high-precision timing measurements.
\end{abstract}
\newpage
\tableofcontents
	
\section{Introduction}

The observational signatures distinguishing a naked singularity from a black hole remain an unresolved and highly debated issue in gravitational physics. Nonetheless, several theoretical studies have proposed scenarios where a visible singularity may arise. In particular, the results \citep{Joshi_2011} demonstrate that a spacetime containing a central naked singularity can emerge as a final equilibrium configuration from the gravitational collapse of a general matter distribution. The Kerr spacetime, an axially symmetric solution to the Einstein field equations in general relativity, describes a Kerr black hole when the singularity is hidden behind an event horizon. However, if the event horizon is absent, such as when the spin parameter exceeds a critical value, the same metric represents a Kerr naked singularity. The transition between these regimes is of fundamental interest, as it challenges the validity of the cosmic censorship conjecture~\citep{1969NCimR,Wald:1997wa} and opens up the possibility of observable effects that could differentiate a naked singularity from conventional black holes. The properties of geodesic motion are fundamentally connected to the underlying geometry of the spacetime. As such, the study of particle trajectories around various types of compact objects has become a central topic of investigation in gravitational physics. In the context of black hole shadows, where the geodesic motion of photons plays a critical role, it has been shown that the shadow cast by a naked singularity spacetime can closely resemble that of a Schwarzschild black hole with the same mass \citep{Shaikh_2018}. Further investigations into the shadow features of naked singularities have been carried out in \citep{PhysRevD.102.024022,Dey_2020,Dey_2021}. Beyond photon trajectories, another promising observational signature arises from the orbital dynamics of massive particles or stars around horizonless ultra-compact objects. These trajectories may, in principle, differ from those observed in the vicinity of a black hole. In this regard, the timelike geodesics for different compact spacetimes and the motion of massive test particles or stars have been extensively analysed in \citep{Pugliese_2011,Mart_nez_2019,Hackmann_2014,Potashov_2019,bhattacharya2020newclassnakedsingularities,Bambhaniya_2019,Deng:2020hxw,Lin:2021noq,bambhaniya2021lensethirringeffectprecessiontimelike,Ota_2022} and references therein. Particularly relevant are the relativistic precession effects of bound orbits, which are expected to reveal subtle characteristics of the central object. The ongoing observations by the GRAVITY \citep{GRAVITY:2020lpa} and SINFONI \citep{Eisenhauer_2005} collaborations, which are tracking the orbital motion of the so-called “S-stars” around the Galactic Center (Sgr~A*), are expected to provide critical data to distinguish between black holes and other exotic compact objects.

\vspace{0.2cm}
\noindent
In the context of timelike orbits around rotating compact objects, two of the most striking relativistic phenomena predicted by general relativity are the geodetic precession~\citep{deSitter:1916zz} and the Lense-Thirring (LT) precession~\citep{Lense:1918zz}. The geodetic precession, also known as de Sitter precession, arises due to the spacetime curvature induced by a massive central object. In contrast, LT precession originates from the rotation of the central mass, which leads to the dragging of inertial frames, a phenomenon often referred to as frame dragging. These dragging effects can be examined using a test gyroscope, which tends to maintain its spin axis pointing in a fixed direction relative to distant stars. The deviations caused by the spacetime geometry then serve as a direct measure of frame dragging. Several satellite-based experiments have been conducted to test and measure these relativistic effects in Earth's gravitational field. Among them, the Gravity Probe B (GP-B) mission employed a satellite carrying four gyroscopes and a telescope in a 650 km polar orbit to measure geodetic and LT precessions \citep{Everitt_2011}. The Laser Relativity Satellite (LARES) was launched to improve LT precession measurements with an accuracy of about \(10^{-2}\) \citep{Capozziello_2015}. The LAser GEOdynamic Satellite (LAGEOS) \citep{Ciufolini:2004rq} and the Earth-based Gyroscopes IN General Relativity (GINGER) project \citep{Di_Virgilio_2021} have also contributed significantly, with GINGER aiming for measurement uncertainties of one part in \(10^4\) and \(10^3\) for geodetic and LT effects, respectively. Future missions, such as the Gravity Probe Spin (GPS) Satellite, are designed to explore spin-related relativistic effects, including the intrinsic spin of the electron \citep{Fadeev_2021}. The geodetic precession in Schwarzschild and Kerr black hole spacetimes has been previously studied in \citep{Chakraborty:2016mhx}. LT precession, however, is more intricate, often requiring approximations for analytic treatment. In the weak-field regime, the LT precession frequency is directly proportional to the spin parameter of the central body and exhibits a characteristic decay proportional to \(r^{-3}\), where \(r\) is the distance between the test gyroscope and the central source \citep{hartle2003gravity}. In the strong-field regime, LT precession has been examined in various compact rotating spacetimes, including the Kerr black hole~\citep{Chakraborty:2013naa,Bini:2016iym,Chakraborty_2014,Chakraborty_2015,Ghosh:2024arw,Zahra_2025,Rizwan_2019,PhysRevD.98.024015}, its generalisations \citep{Chakraborty:2012wv}, rotating traversable wormholes \citep{Chakraborty:2016oja}, and rotating neutron stars \citep{Chakraborty:2014qba}. These studies reveal that the behaviour of LT precession frequencies in strong gravity regions is highly sensitive to the geometry and nature of the central rotating object. Remarkably, it was proposed in \citep{Chakraborty:2016mhx} that the spin precession of a test gyroscope could serve as a diagnostic tool to distinguish Kerr naked singularity from black hole.

\vspace{0.2cm}
\noindent
In a recent study, Rizwan et al. \citep{Rizwan_2019,PhysRevD.98.024015} investigated the possibility of distinguishing a rotating Kiselev black hole and a Kerr-like black hole in perfect fluid dark matter from a naked singularity by analysing the spin precession of a test gyroscope. This approach highlights the sensitivity of gyroscopic precession to the geometry of the central object; see~\citep{PhysRevD.95.044006,Zahra_2025,Pradhan_2024,Iyer_2025,wang2025characteristicprecessionssphericalorbit,Iyer:2025uyb,Zhen:2025nah,Holme:2025eey,Ashraf:2025aen,Ashraf:2024xwm,Stepanian:2020vwk,Pradhan2024,Wu:2025xtn} for the latest advancements. Further efforts to distinguish black holes from naked singularities have explored alternative observational techniques. For instance, iron-line spectroscopy has been proposed as a means to probe the spacetime geometry near compact objects, offering a potential method to distinguish between black holes and naked singularities~\citep{Liu_2018}. Additionally, rotating naked singularities have been examined in the context of gravitational lensing \citep{PhysRevD.78.083004}, where their lensing signatures could be observationally distinct from those of black holes. In another study, Ranea-Sandoval and collaborators~\citep{Ranea_2015} investigated how external magnetic fields modify the thermal continuum and iron-line profiles emitted by thin accretion discs around Kerr black holes and naked singularities. They argued that such spectroscopic signatures can serve as a diagnostic of magnetic fields in the immediate vicinity of compact objects and, simultaneously, as a potential observational probe of the cosmic censorship conjecture in these astrophysical settings. Furthermore, they developed a ray-tracing algorithm~\citep{Garcia_2016} to compute the light curves and power spectra of disc hot spots as observed at infinity, considering both uniform and dipolar magnetic-field configurations and assuming a weak coupling between the magnetic field and the disc plasma. In another recent work \citep{Jusufi_2019}, the deflection angles of massive particles were used to compare rotating naked singularities with Kerr-like wormholes. Recent horizon-scale observations have been used to test both the nature of the central compact object and possible deviations from general relativity. In particular, Bambi~\citep{Bambi_2019} proposed using the size and near-circularity of the first EHT image of M$87^{*}$ to constrain its rotational properties and quantify departures from the Kerr expectation. Subsequently, Vagnozzi~\citep{Vagnozzi_2023} performed a broad set of horizon-scale tests with the EHT image of Sgr~A*, placing constraints on a wide class of alternative gravity models and black-hole mimickers, including wormholes and naked-singularity spacetimes. More recently, Khodadi~\citep{Khodadi_2024} showed that the shadow predictions of baseline mimetic gravity are pathological (the naked singularity casts no shadow, while the matched black-hole shadow is too small), arguing that EHT observations of M$87^{*}$ and Sgr~$A^{*}$ exclude this baseline scenario. These studies collectively suggest that a combination of relativistic observables such as spin precession, Shadow profile, spectral features, and lensing deflection may offer viable pathways to test the cosmic censorship conjecture and reveal the true nature of compact objects.

\vspace{0.2cm}
\noindent
Interestingly, accretion of matter onto rotating neutron stars and black holes is accompanied by intense electromagnetic emission, predominantly in the X-ray and gamma-ray bands~\citep{nowak1998stableoscillationsblackhole}. Quasi-periodic oscillations (QPOs) are observed in several X-ray binary systems and are commonly classified into high-frequency (HF) QPOs and three distinct categories of low-frequency (LF) QPOs~\citep{1998ApJ...492L..59S,Stella_1999}. A particularly intriguing possibility is that these timing features are connected to the LT frame-dragging effect. It is found that if the inner accretion disk is slightly tilted with respect to the equatorial plane of a rotating black hole, the resulting nodal (orbital-plane) precession can modulate the emitted radiation and thereby contribute to the observed QPO phenomenology~\citep{Motta_2017}. Zhen~\citep{Zhen:2025nah} computed the three fundamental frequencies of particle motion in the Horndeski rotating spacetime, which are closely connected to the QPOs observed from the surrounding accretion disk, and discussed the implications of their results. Recently, Zahra~\citep{Zahra_2025} investigated the fundamental frequencies associated with particle motion around a rotating naked singularity and reported several noteworthy results. In particular, they showed that an ISCO may be absent for null geodesics in the naked singularity spacetime and that the orbital-plane (nodal) precession frequency exhibits behaviour analogous to the LT precession arising from frame dragging. Moreover, by fitting the relativistic precession model to the observed QPOs of five sources, they employed Markov Chain Monte Carlo (MCMC) simulations to place constraints on the parameters of the rotating NS. Hence, in recent years, the study of QPOs in the vicinity of rotating black holes and other exotic compact objects has attracted considerable attention~\citep{Mustafa:2025mkc,Wang:2025zis,Guo_2025,Jiang_2021,Banerjee_2022,Belloni_2012,kolovs2017possible}.

\vspace{0.2cm}
\noindent
In this work, we investigate the spin precession phenomena of test gyroscopes in the spacetime of a rotating, charged black hole. In particular, we examine the precession of spin and the associated epicyclic oscillations that arise in such a gravitational background. To explore these effects, we consider a test gyroscope attached to a stationary observer orbiting the compact object. Within this framework, we derive and analyse the Lense-Thirring (frame-dragging) precession, geodetic precession, and general spin precession frequencies. Our results show that these frequencies are significantly influenced by the intrinsic parameters of the black hole, namely its mass $M$, angular momentum $a$, and electric charge $Q$. To assess the robustness of our findings, we construct a theoretical model corresponding to a black hole of mass $10M_{\odot}$. This model allows for a comprehensive study of QPOs, enabling us to clearly distinguish LF QPOs from HF QPOs within the proposed scenario. Finally, we constrain the influence of the charge parameter $Q$ through its effect on the nodal precession frequency. The tabulated values derived from the theoretical model illustrate how variations in charge modify the characteristic frequencies, providing deeper insight into the role of electromagnetic contributions in rotating charged black hole spacetimes.

\vspace{0.2cm}
\noindent
The paper is structured as follows. In Section~\ref{sec_2}, we summarise the spacetime structure of the Kerr–Newman geometry, emphasizing the horizon/ergoregion properties and the parameter ranges relevant for black hole versus naked singularity configurations. In Section~\ref{sec_3}, we formulate the generalised spin precession of a test gyroscope around Kerr–Newman and obtain the corresponding Lense–Thirring, geodetic, and total spin-precession frequencies for stationary observers. In Section~\ref{sec_4}, we develop and apply the spin-precession frequency criterion to differentiate a black hole from a naked singularity by analysing the divergence/regularity of the precession frequency along different approach directions and for different observer angular velocities. In Section~\ref{sec_5}, we discuss the role of frame dragging in shaping accretion flows in Kerr–Newman, connecting the qualitative behavior of the precession characteristics with disk-related phenomenology, particularly the QPOs. Finally, Section~\ref{sec_6} presents our discussion and summary, highlighting the main outcomes and their implications. Throughout this paper, we adopt geometrised units with $(8\pi G = c = M = 1)$, unless stated otherwise.

\section{The Spacetime Structure}\label{sec_2}

The Kerr-Newman solution in general involves the mass of the BH $M$, the rotation parameter $a$, the electric charge $Q$ and the spacetime metric~\citep{adamo2016kerrnewmanmetricreview,Newman:1965my} is given by   
\begin{align}
	ds^{2}=-\frac{\Delta}{\rho ^{2}}\left(dt-a\sin ^{2}\theta d\phi\right)^{2}
	+\frac{\sin ^{2}\theta}{\rho ^{2}}\left\{\left(r^{2}+a^{2}\right)d\phi -adt\right\}^{2}
	+\frac{\rho ^{2}}{\Delta}dr^{2}+\rho ^{2}d\theta ^{2},
	\label{kn_1}
\end{align}
where the the metric functions $\Delta$ and $\rho ^{2}$ have the following explicit expressions:
\begin{align}
	\Delta = r^{2}-2Mr+a^{2}+Q^{2},\qquad \rho ^{2}=r^{2}+a^{2}\cos^2 \theta. \notag
\end{align}

\begin{figure}[h!]
	\centering
	\subfigure[]{\includegraphics[width=7.3cm,height=7.7cm]{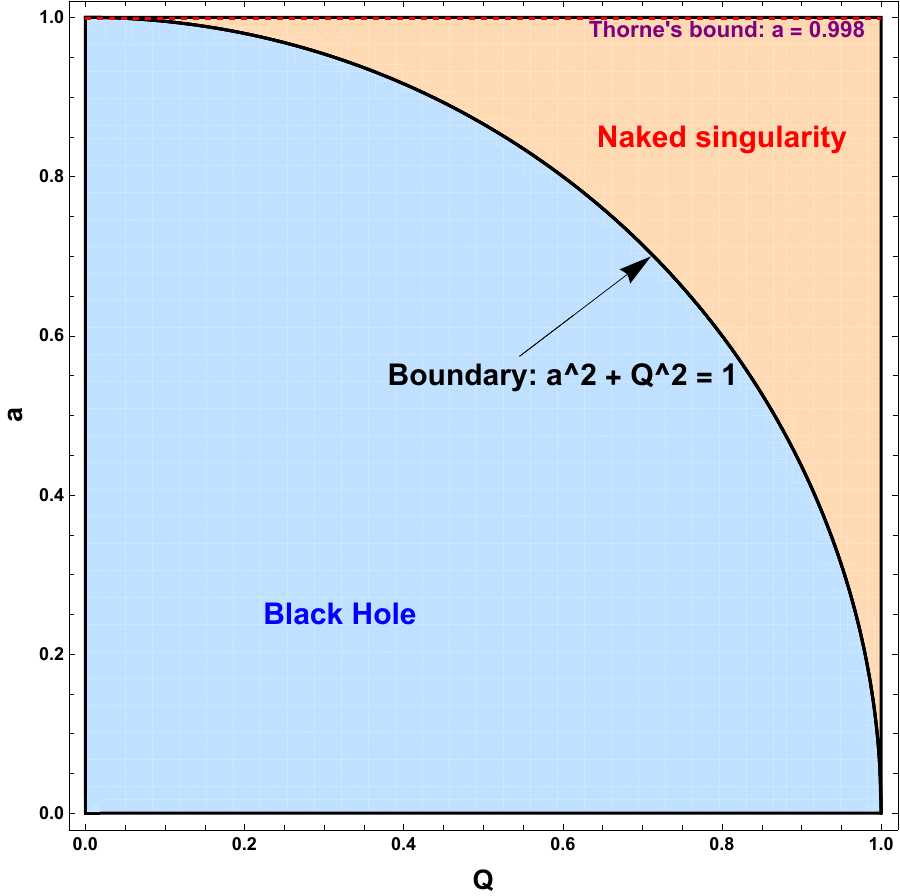}\label{bhr_1}} \hfill
	\subfigure[]{\includegraphics[width=7.5cm,height=8.5cm]{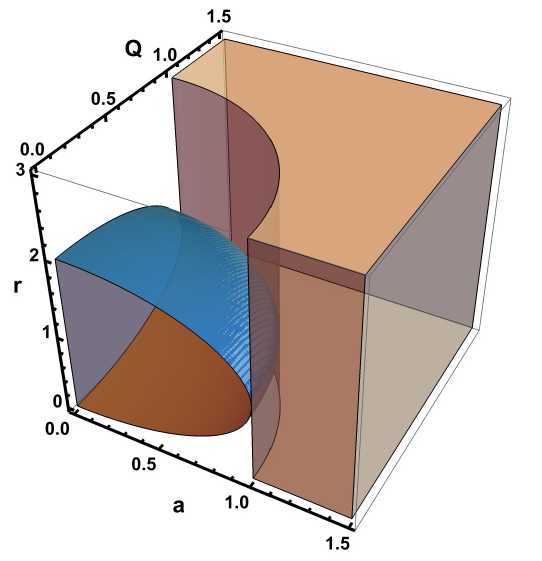}\label{bhr_2}}
	\caption{Illustration showing the spin-charge parameter plane of a rotating Kerr-Newman spacetime with $M=1$. In Fig.~(i), the curve separates the black hole region from the naked singularity region, corresponding to configurations without an event horizon. Fig.~(ii) represents the three-dimensional visualisation of the black hole with inner and outer event horizons.} \label{bhr_regions}
\end{figure}

The metric~(\ref{kn_1}), which extends the Kerr solution by incorporating electric charge, exhibits distinct behaviours based on the values of its parameters, i.e. mass $M$, spin parameter $a$ and electric charge $Q$. The structure of the Kerr-Newman spacetime depends critically on the roots of the equation $\Delta=0$, which determine the locations of the horizons. Solving this equation yields

\begin{equation}
	r_{\pm}=M\pm\sqrt{M^2-a^2-Q^2},
\end{equation}

where $r_{+}$ denotes the outer (event) horizon and $r_{-}$ corresponds to the inner (Cauchy) horizon. The event horizon represents the boundary beyond which causal signals cannot escape to an external observer. In contrast, the inner horizon marks a secondary causal boundary within the BH interior. The region $r<r_+$, therefore, constitutes the black hole interior, containing both the Cauchy horizon and the central curvature singularity.

\vspace{0.2cm}
\noindent
The nature of the spacetime is determined by the quantity $M^2-a^2-Q^2$. When $M^2>a^2+Q^2$, the spacetime possesses two distinct horizons and corresponds to a Kerr-Newman black hole. In the extremal case, defined by the condition $M^2=a^2+Q^2$~\citep{adamo2016kerrnewmanmetricreview}, the inner and outer horizons coincide at $r=M$. However, when $M^2<a^2+Q^2$, the equation $\Delta=0$ admits no real solutions and therefore no event horizon exists. In this situation, the curvature singularity is not hidden behind a horizon and becomes visible to distant observers, giving rise to a naked singularity. This classification of the spacetime structure can be clearly illustrated in the spin-charge parameter space shown in Figure~(\ref{bhr_regions}). In panel (i), the curve defined by the extremality condition $a^2+Q^2=1$ (with $M=1$) separates the parameter space into two distinct regions: the black hole region, where the inequality $a^2+Q^2\leq 1$ is satisfied, and the naked singularity region, where $a^2+Q^2>1$. Thus, configurations lying below the curve correspond to Kerr-Newman black holes with well-defined event horizons, whereas those above the curve represent horizonless spacetimes with exposed singularities. Panel (ii) provides a three-dimensional visualisation of the Kerr-Newman geometry, illustrating the presence of both the inner and outer horizons that characterise the black hole configuration. The parameter-space representation shown in Figure~(\ref{bhr_regions}) therefore plays a crucial role in distinguishing between physically admissible black hole solutions and naked singularity configurations. This distinction is particularly relevant for the analysis presented in this chapter, since the behaviour of frame dragging and gyroscopic precession differs qualitatively depending on whether the spacetime contains an event horizon or corresponds to a naked singularity. 

\vspace{0.2cm}
\noindent
Having discussed the horizon structure, we next consider the hypersurface determined by the condition \(g_{tt}=0\) in Boyer--Lindquist coordinates, which plays a key role in the causal and energetic properties of the Kerr--Newman spacetime. Consider a static worldline, i.e., an observer whose spatial coordinates remain fixed,

\[
(r,\theta,\phi)=\text{const.}
\]

\vspace{0.2cm}
Along such a trajectory the line element reduces to

\begin{equation}
	ds^{2}\Big|_{r,\theta,\phi=\text{const.}} = g_{tt}\,dt^{2}
	= -\,\frac{r^{2}+a^{2}\cos^{2}\theta-2Mr+Q^{2}}{\rho^{2}}\,dt^{2}.\label{eq:gtt_static}
\end{equation}

\vspace{0.2cm}
The boundary beyond which static worldlines cease to be timelike is determined by the condition \(g_{tt}=0\). Solving this equation gives the static limit surface,

\begin{equation}
	r=r_{ergo}^{\pm}(\theta)=M\pm\sqrt{M^{2}-a^{2}\cos^{2}\theta-Q^{2}}.
	\label{eq:static_limit_pm}
\end{equation}

For a Kerr-Newman black hole, the physically relevant solution is the outer root,

\begin{equation}
	r=r_{ergo}^{+}(\theta)=M+\sqrt{M^{2}-a^{2}\cos^{2}\theta-Q^{2}},
	\label{eq:ergosurface}
\end{equation}

\vspace{0.2cm}
\noindent
since the inner root $r_{ergo}^{-}(\theta)$ lies at or within the event horizon and is therefore not of direct physical interest. Eq.~\eqref{eq:gtt_static} makes the causal structure transparent. For $r>r_{ergo}^{+}(\theta)$, the stationary Killing vector field $k=\partial_{t}$, tangent to the static worldline, is timelike, and static observers can exist. In contrast, for $r<r_{ergo}^{+}(\theta)$ it becomes spacelike, implying that no observer can remain at fixed $(r,\theta,\phi)$. The hypersurface \eqref{eq:ergosurface} is thus identified as the stationary limit surface (or ergosurface). The region bounded externally by the ergosurface and internally by the outer event horizon,

\[
r_{+}<r<r_{ergo}^{+}(\theta),
\]

\vspace{0.2cm}
\noindent
is known as the ergosphere. Within the Kerr–Newman ergoregion, the geometric hallmark is that the time-translation Killing vector field ($k=\partial_t$) ceases to be timelike and instead becomes spacelike. This change in causal character is accompanied by a distinctive deformation of the local light cones: frame dragging tilts them in the azimuthal direction, ensuring that every future-directed timelike vector necessarily acquires a nonzero rotational ($\phi$) component. As a result, no physical observer can remain stationary with respect to infinity inside the ergosphere; equivalently, timelike worldlines with fixed spatial coordinates do not exist there in any time-independent coordinate system.

\section{Generalised Spin Precession of a Test Gyroscope in Kerr-Newman Spacetime}\label{sec_3}

In this section, we examine the spin precession frequency of a test gyro attached to a stationary observer with respect to a distant fixed star, arising due to the frame-dragging effects of the Kerr-Newman black hole. The gyro is assumed to move along a timelike Killing trajectory, and its spin vector is subjected to Fermi-Walker transport. The corresponding four-velocity of the observer is expressed as $u = (-K^{2})^{-1/2} K$, where $K$ denotes the timelike Killing vector field of the Kerr-Newman geometry. Therefore, the general spin precession frequency of a test gyro, $\Omega_{s}$, can be recast as \citep{straumann_2025,Straumann:2013spu}.

\begin{equation}
	\begin{aligned}
		\tilde{\Omega}_{s} & =\frac{1}{2 K^2} *(\tilde{K} \wedge d \tilde{K}) \\[10pt]
        \text{or},
        \left(\Omega_{s}\right)_d & =\frac{1}{2 K^2} \eta_d^{a b c} K_a \partial_b K_c, \label{sp_1}
	\end{aligned}
\end{equation}

\vspace{0.2cm}
\noindent
where $\Omega_{s}$ is the spin precession frequency in the coordinate basis, $\ast$ denoted by the Hodge star operator or Hodge dual, $\eta^{abc}$ denoting the components of the volume-form in spacetime and $\tilde{K},\tilde{\Omega}_{s}$ the one-form of $K$ and $\Omega_{s}$ respectively. It has already been derived in \citep{Chakraborty:2014qba,Chatterjee_2017} the LT precession frequency of a test gyro due to the rotation of any stationary and axisymmetric spacetime as

\begin{equation}
\vec{\Omega}_{s}|_{\Omega=0}=\frac{1}{2\sqrt{-g}}\left[-\sqrt{g_{rr}}\left(g_{0\phi,\theta}-\frac{g_{0\phi}}{g_{00}}g_{00,\theta}\right)\hat{r}+\sqrt{g_{\theta\theta}}\left(g_{0\phi,r}-\frac{g_{0\phi}}{g_{00}}g_{00,r}\right)\hat{\theta}\right], \label{sp_2}
\end{equation}

\vspace{0.2cm}
This result is valid for outside the ergoregion and $\Omega = 0$. For an arbitrary value of $\Omega$, the formalism is derived in the following according to~\citep{Straumann:2013spu,Chakraborty:2016mhx}. The timelike Killing vector field in generalised spacetime can be written as

\begin{equation}
	\begin{matrix}K&=&\partial_0+\Omega\partial_s,\end{matrix} \label{sp_3}
\end{equation}

where $\partial_{s}$ is the spacelike Killing vector of the said stationary spacetime and $\Omega$ is the angular velocity of an observer moving along integral curves of $K$. The metric for this stationary spacetime is independent of $x^0$ and $x^s$ coordinates. Thus, the corresponding co-vector of $K$ is

\begin{equation}
	\begin{matrix}
    \tilde{K}&=&g_{0\alpha}dx^\alpha+\Omega g_{\beta s}dx^\beta, \label{sp_4}
    \end{matrix}
\end{equation}

where $\alpha,\beta=0,1,2,3$ in 4D spacetime. Now one can write $\tilde{K}$ by separating into space and time components as

\begin{equation}
	\tilde{K}=(g_{00}dx^0+g_{0s}dx^s+g_{0i}dx^i)+\Omega\left(g_{0s}dx^0+g_{ss}dx^s+g_{is}dx^i\right), \label{sp_5}
\end{equation}

\vspace{0.2cm}
where $i = 2, 3$. Since we are interested in the ergoregion of the said stationary, axisymmetric spacetime thus, we are neglecting the terms $g_{0i}$ and $g_{is}$, and we get

\begin{equation}
	\tilde{K}=(g_{00}dx^0+g_{0s}dx^s)+\Omega(g_{0s}dx^0+g_{ss}dx^s). \label{sp_6}
\end{equation}

and

\begin{equation}
	d\tilde{K}=(g_{00,k}dx^{k}\wedge dx^{0}+g_{0s,k}dx^{k}\wedge dx^{s})+\Omega(g_{0s,k}dx^{k}\wedge dx^{0}+g_{ss,k}dx^{k}\wedge dx^{s}). \label{sp_7}
\end{equation}

\vspace{0.3cm}
Now, Eq.~\eqref{sp_1} can rewritten as

\begin{equation}
	\tilde{\Omega}_{s}=\frac{1}{2K^{2}}*(\tilde{K}\wedge d\tilde{K}). \label{sp_8}
\end{equation}

\vspace{0.2cm}
Substituting the values of $\tilde{K}$ and $d\tilde{K}$ in Eq.~\eqref{sp_8}, one finds the one-form of the precession frequency

\begin{equation}
\begin{aligned}
\tilde{\Omega}_{s}=&\frac{\varepsilon_{skl}g_{l\mu}dx^{\mu}}{2\sqrt{-g}\left(1+2\Omega\frac{g_{0s}}{g_{00}}+\Omega^{2}\frac{g_{ss}}{g_{00}}\right)} \\[10pt] \qquad
&\left[\left(g_{0s,k}-\frac{g_{0s}}{g_{00}}g_{00,k}\right)+\Omega\left(g_{ss,k}-\frac{g_{ss}}{g_{00}}g_{00,k}\right)+\Omega^{2}\left(\frac{g_{0s}}{g_{00}}g_{ss,k}-\frac{g_{ss}}{g_{00}}g_{0s,k}\right)\right], \label{sp_9}
\end{aligned}
\end{equation}

\vspace{0.2cm}
\noindent where we have used $\ast(dx^{0}\wedge dx^{k}\wedge dx^{s})=\eta^{0ksl}g_{l\mu}dx^{\mu}=-\frac{1}{\sqrt{-g}}\varepsilon_{ksl}g_{l\mu}dx^{\mu}$ and $K^{2}=g_{00}+2\Omega g_{0s}+\Omega^{2}g_{ss}$. 

\vspace{0.3cm}
\noindent
The corresponding precession frequency vector $(\Omega_{s})$ of a test gyro of the co-vector $\tilde{\Omega}_s$, as measured by a stationary observer with four-velocity $u = (-K^{2})^{-1/2}K$ in a stationary spacetime, is defined in terms of the timelike Killing vector field $K$, as given in~\citep{Chakraborty:2016mhx}.

\begin{equation}
	\begin{aligned}
    \Omega_{s}&=\frac{\varepsilon_{skl}}{2\sqrt{-g}\left(1+2\Omega\frac{g_{0s}}{g_{00}}+\Omega^{2}\frac{g_{ss}}{g_{00}}\right)} \\[10pt] \qquad
    &\left[\left(g_{0s,k}-\frac{g_{0s}}{g_{00}}g_{00,k}\right)+\Omega\left(g_{ss,k}-\frac{g_{ss}}{g_{00}}g_{00,k}\right)+\Omega^{2}\left(\frac{g_{0s}}{g_{00}}g_{ss,k}-\frac{g_{ss}}{g_{00}}g_{0s,k}\right)\right]\partial_{l}. \label{sp_10}
    \end{aligned}
\end{equation}

\vspace{0.2cm}
For a stationary and axisymmetric spacetime with coordinates $(0,r,\theta,\phi)$ the Eq.~\eqref{sp_10} becomes

\begin{equation}
	\begin{aligned}
		\vec{\Omega}_s= & \frac{1}{2 \sqrt{-g}\left(1+2 \Omega \frac{g_{0 \phi}}{g_{00}}+\Omega^2 \frac{g_{\phi \phi}}{g_{00}}\right)}  \\[10pt]
		& {\left[-\sqrt{g_{r r}}\left[\left(g_{0 \phi, \theta}-\frac{g_{0 \phi}}{g_{00}} g_{00, \theta}\right)+\Omega\left(g_{\phi \phi, \theta}-\frac{g_{\phi \phi}}{g_{00}} g_{00, \theta}\right)+\Omega^2\left(\frac{g_{0 \phi}}{g_{00}} g_{\phi \phi, \theta}-\frac{g_{\phi \phi}}{g_{00}} g_{0 \phi, \theta}\right)\right] \hat{r}\right.} \\[10pt]
		& \left.+\sqrt{g_{\theta \theta}}\left[\left(g_{0 \phi, r}-\frac{g_{0 \phi}}{g_{00}} g_{00, r}\right)+\Omega\left(g_{\phi \phi, r}-\frac{g_{\phi \phi}}{g_{00}} g_{00, r}\right)+\Omega^2\left(\frac{g_{0 \phi}}{g_{00}} g_{\phi \phi, r}-\frac{g_{\phi \phi}}{g_{00}} g_{0 \phi, r}\right)\right] \hat{\theta}\right]. \label{sp_11}
	\end{aligned}
\end{equation}

\vspace{0.2cm}
Having established the necessary theoretical framework, the next step is to evaluate the generalised spin precession frequency for a test gyro in a stationary and axisymmetric spacetime. This expression remains applicable both outside and inside the ergoregion. The metric components of the Kerr-Newman black hole~(\ref{kn_1}) in Boyer–Lindquist coordinates are therefore written as

\begin{equation}
\begin{aligned}
g_{tt} &= -\frac{r^2 - 2 M r + a^2 \cos^2\theta + Q^2}{\rho^2}, \quad
g_{rr} = \frac{\rho^2}{\Delta}, \quad
g_{\theta\theta} = \rho^2, \\[10pt]
g_{\phi\phi} &= \left[r^2 + a^2 + \frac{a(2 M r - Q^2)\sin^2\theta}{\rho^2}\right]\sin^2\theta, \quad
g_{t\phi} = -\frac{a(2 M r - Q^2)\sin^2\theta}{\rho^2}, \label{sp_12}
\end{aligned}
\end{equation}

\vspace{0.2cm}
where $a$ is the specific angular momentum, defined as $a=J/M$ and,

\begin{equation}
	\rho^2=r^2+a^2\cos^2\theta,\quad\Delta=r^2-2Mr+a^2+Q^2, \quad \sqrt{-g}=\rho^2\sin\theta. \notag
\end{equation}

\vspace{0.2cm}
Putting these metric components in Eq.~\eqref{sp_11}, one obtains the spin precession frequency of a test gyro for a Kerr-Newman black hole as

\begin{equation}    \vec{\Omega}_s=\frac{\xi\sqrt{\Delta}\cos\theta\hat{r}+\eta\sin\theta\hat{\theta}}{\zeta}, \label{sp_13}
\end{equation}

where
\begin{equation}
\begin{aligned}
\xi =\;& a\left(2 M r - Q^2\right)
 - \frac{\Omega}{8}\Big[8 r^4 + 8 a^2 r^2 + 16 a^2 M r - 8 a^2 Q^2 + 3 a^4 
 + 4 a^2\left(2 \Delta - a^2\right)\cos 2\theta \\[10pt]
 &+ a^4 \cos 4\theta\Big] 
 + \Omega^2 a^3\left(2 M r - Q^2\right)\sin^4\theta, \\[10pt]
\eta =\;& a M\left(r^2 - a^2 \cos^2\theta\right) - a Q^2 r 
 + \Omega\Big[r^5 - 3 M r^4 + 2 a^2 r^3 \cos^2\theta - 2 M a^2 r^2 
 + a^4 r \cos^4\theta \\
 &+ M a^4 \cos^2\theta\left(1 + \sin^2\theta\right) 
 + 2 Q^2 r\left(r^2 + a^2\right)\Big] \\[10pt]
 &+ \Omega^2 a \sin^2\theta\Big[
 M\!\left(3 r^4 + a^2 r^2 + a^2 \cos^2\theta (r^2 - a^2)\right) - Q^2 r\!\left(2 r^2 + a^2(1 + \cos^2\theta)\right)
 \Big], \\[10pt]
\zeta =\;& \rho^3\Big[
\left(\rho^2 - 2 M r + Q^2\right)
 + 2 a \Omega \left(2 M r - Q^2\right)\sin^2\theta 
 - \Omega^2 \sin^2\theta\Big\{
 \rho^2\left(r^2 + a^2\right) \\[10pt]
 &+ a^2\left(2 M r - Q^2\right)\sin^2\theta
 \Big\}
 \Big].
\end{aligned} \notag
\end{equation}

\vspace{0.2cm}
\noindent
Where $\hat{r}$ and $\hat{\theta}$ denote the unit vectors along the radial and polar directions, respectively. The above expression represents the generalised spin precession frequency for a test gyro in the background of a Kerr-Newman black hole, which remains valid both outside and inside the ergosphere. It is noteworthy that, in the limiting case $Q = 0$, the derived expression for the spin precession coincides with that of the Kerr black hole, thereby successfully reproducing the well-established result reported in \citep{Chakraborty:2016mhx}. Therefore, the magnitude of the generalised spin precession frequency of a test gyro for a Kerr-Newman black hole  (Eq.~\ref{sp_13}) is given by

\begin{equation}
    \left|\vec{\Omega}_s\right|
    =
    \frac{
        \sqrt{
            \Delta\cos^2\theta\,\xi^2
            +
            \sin^2\theta\,\eta^2
        }
    }{
        \left|\zeta\right|
    }.
    \label{sp_13_mag}
\end{equation}

%
%

\vspace{0.3cm}
It is important to emphasise that the expression for the precession frequency given in Eq.~(\ref{sp_10}) is valid only for a timelike observer situated at fixed $r$ and $\theta$. Hence, this condition consequently imposes a constraint on the allowable range of the observer's angular velocity $\Omega$, i.e. $K^{2}=g_{\phi\phi}\Omega^{2}+2g_{t\phi}\Omega+g_{tt}<0$. At any fixed position $(r,\theta)$, the angular velocity $\Omega$ of a timelike observer must lie within the bounds  

\begin{equation}
    \Omega_{-}(r,\theta) < \Omega(r,\theta) < \Omega_{+}(r,\theta), \label{sp_14}
\end{equation}

where the limiting values $\Omega_{\pm}$ are expressed as  

\begin{equation}
\Omega_{\pm} = \frac{-g_{t\phi} \pm \sqrt{g_{t\phi}^{2} - g_{\phi\phi} g_{tt}}}{g_{\phi\phi}}. \label{sp_15}
\end{equation}

For the Kerr-Newman black hole, the above equation reads as 

\begin{equation}
	\Omega_{\pm}=\frac{a(2Mr-Q^2)\sin^2\theta\pm\rho^2\sqrt{\Delta}\sin\theta}{\sin^2\theta[\rho^2(r^2+a^2)+a^2(2Mr-Q^2)\sin^2\theta]}, \label{sp_16}
\end{equation}

\begin{figure}[tbp!]
	\centering
	\subfigure[]{\includegraphics[width=7.3cm,height=7.7cm]{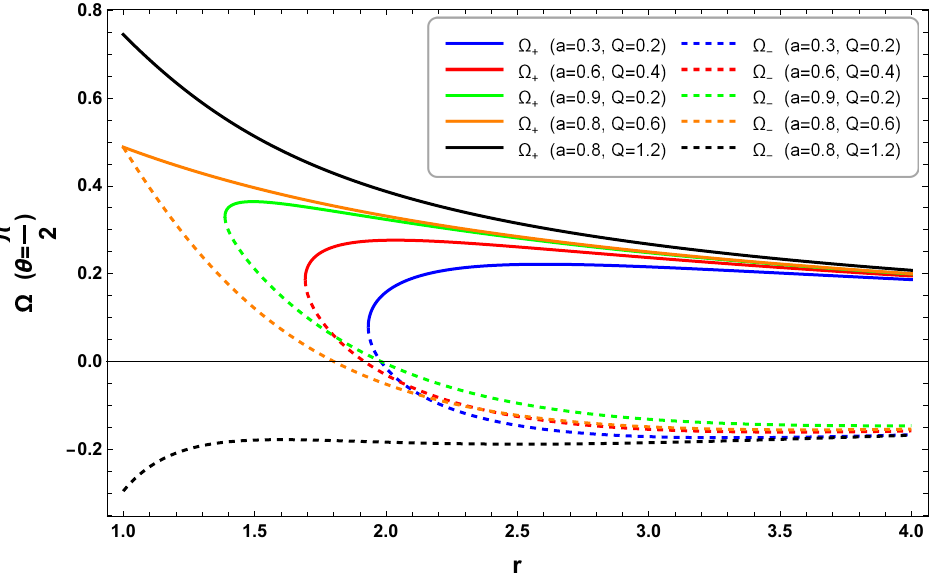}\label{bhr_1}} \hspace{0.4cm}
	\subfigure[]{\includegraphics[width=7.3cm,height=7.6cm]{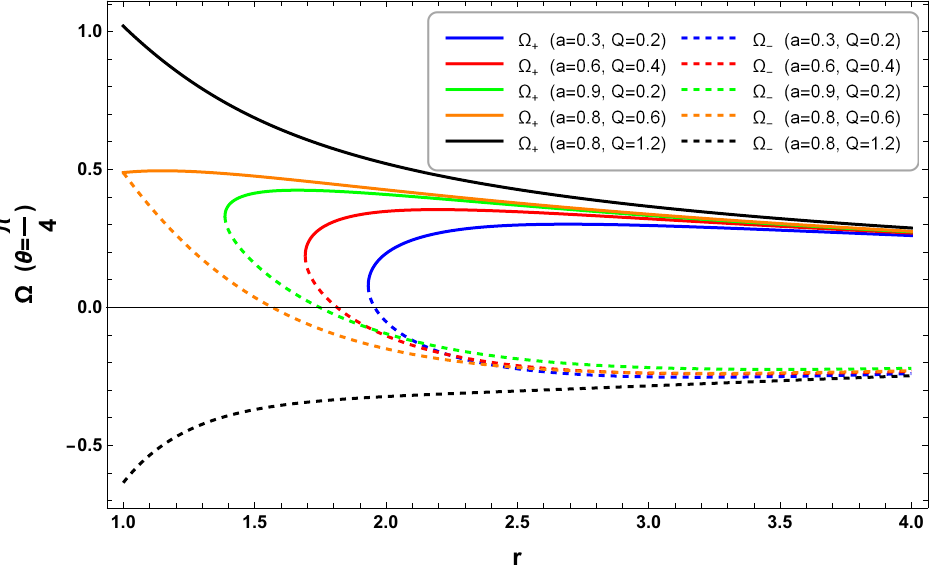}\label{bhr_2}}
    \subfigure[]{\includegraphics[width=7.3cm,height=7.7cm]{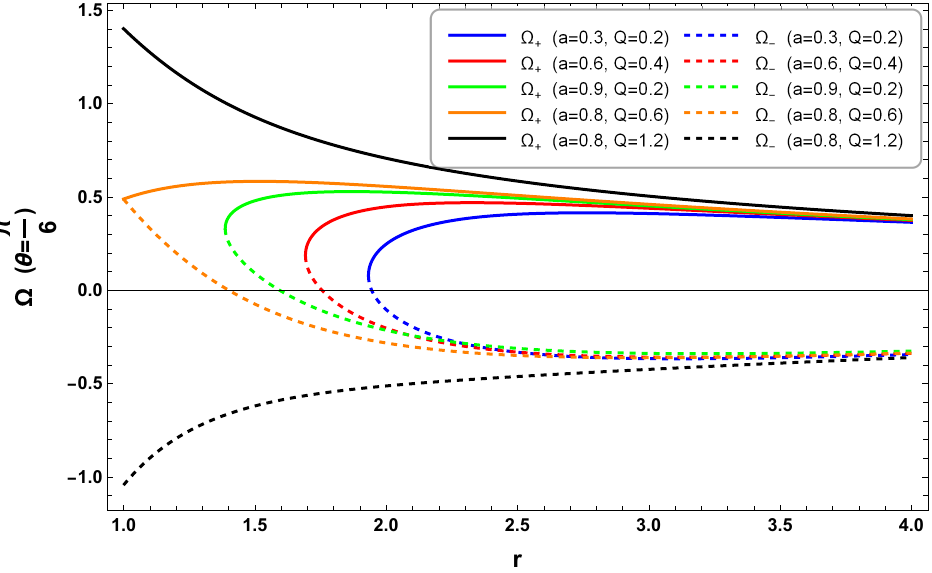}\label{bhr_3}} \hspace{0.4cm}
	\subfigure[]{\includegraphics[width=7.3cm,height=7.6cm]{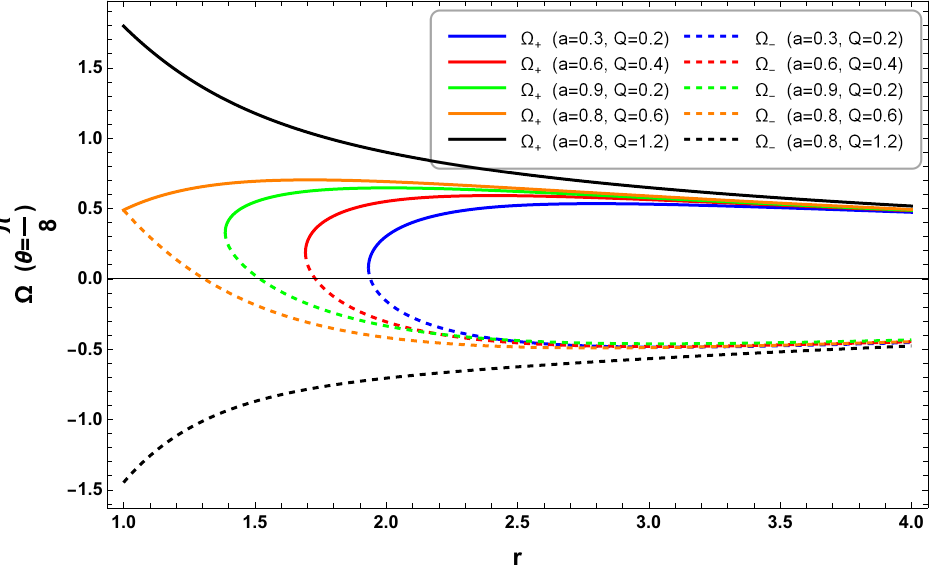}\label{bhr_4}}
    \caption{Illustration showing the variation of the orbital velocity $\Omega$ (in units of $M^{-1}$) of a test gyroscope versus the radial coordinate $r$ (in units of $M$) for different $Q$ values in the background of a rotating Kerr-Newman black hole, specifically within the ergoregion, restricted to the equatorial plane. The diverging red curve (solid and dotted) in all panels corresponds to the value of $\Omega_{\pm}$ in the naked singularity region. The behaviour of the curves clearly distinguishes the black hole configuration from the naked singularity case through the divergence of the orbital velocity near the central region.}\label{omega_pm}
\end{figure}

\vspace{0.2cm}
\noindent
This indicates that the admissible range of $\Omega$ shrinks as the observer moves closer to the horizon ($r \simeq r_{+}$), and in the horizon limit it collapses to a unique value depending on the charge parameter $Q$.

\begin{equation}
    \Omega_H=\frac{a}{2Mr_{+}-Q^2}
\end{equation}

As the black hole's horizon, or the ring singularity ($r=0$, $\theta = \pi/2$) is approached by the observer, the limiting angular velocities $\Omega_{+}$ and $\Omega_{-}$ become degenerate and coincide (see Fig.~\ref{omega_pm}), thereby eliminating the possibility of any timelike observer. However, its behaviour can still be examined arbitrarily close to the horizon in the near-horizon region, which retains significance for the study of precession in the neighbourhood of these limiting configurations. A more detailed discussion and analysis of this aspect will be presented in a later section of this paper. Next, we examine the behaviour of $\Omega_{s}$ in various limiting cases of $r$ and $\theta$.

\subsection{Behaviour of $\vec{\Omega}_{s}$ at $r=0$}

In this section, we examine various limiting cases of the generalised spin precession frequency to elucidate its underlying structure. We begin by considering the limit $r \rightarrow 0$, corresponding to the vicinity of the ring singularity located at $r = 0$ and $\theta = \pi/2$. It is important to note that the precession vector $\vec{\Omega}_{s}$ is not well defined exactly at the ring singularity. Nevertheless, its behaviour in the neighbourhood of this region, namely for $r \to 0$ and $0 \leq \theta < \pi/2$, remains physically meaningful and warrants investigation. We emphasise that this domain lies entirely outside the ergoregion. At the $r = 0$, the generalised spin precession frequency becomes 

\begin{equation}
	\vec{\Omega}_{s}|_{r=0}=\frac{\xi(\theta)\hat{r}+\eta(\theta)\hat{\theta}}{\zeta(\theta)}, \label{omega_r=0}
\end{equation}

where

\begin{equation}
\begin{aligned}
\xi(\theta) &= \sqrt{a^{2}+Q^{2}}\Bigg[
    -\,Q^{2} 
    - \frac{\Omega}{8}\Big\{(3+4\cos 2\theta+\cos 4\theta)a^{3}
    - 8aQ^{2}(1-\cos 2\theta)\Big\}
    - a^{2}Q^{2}\Omega^{2}\sin^{4}\theta
\Bigg], \\[10pt]
\eta(\theta) &= Ma^{2}\cos\theta\,\Big[
    -1 
    + a\Omega(1+\sin^{2}\theta)
    - a^{2}\Omega^{2}\sin^{2}\theta
\Big], \\[10pt]
\zeta(\theta) &= a^{2}\cos^{2}\theta\,\Big[
    Q^{2} 
    + a^{2}\cos^{2}\theta
    - 2a\Omega Q^{2}\sin^{2}\theta
    - a^{2}\Omega^{2}\sin^{2}\theta\left(a^{2}\cos^{2}\theta - Q^{2}\sin^{2}\theta\right)
\Big]. \notag
\end{aligned}
\end{equation}

\vspace{0.2cm}
The valid regime of the angular velocity $\Omega$ for the above equation is bounded by the two limiting branches $\Omega_{-}(\theta)$ and $\Omega_{+}(\theta)$. For a fixed polar angle $\theta$, physically admissible values of $\Omega$ satisfy

\begin{equation}
	\Omega_-(\theta)<\Omega<\Omega_+(\theta),
\end{equation}

where

\begin{equation}
\begin{aligned}
\Omega_{-}(\theta) &= 
 -\,\frac{
    Q^{2}\sin\theta
    + \sqrt{
        Q^{4}\sin^{2}\theta
        + \big[a^{2} - (a^{2}+Q^{2})\sin^{2}\theta\big]
          \big(Q^{2}+a^{2}\cos^{2}\theta\big)
      }
 }{
    a\sin\theta\,\big[a^{2} - (a^{2}+Q^{2})\sin^{2}\theta\big]
 }, \\[20pt]
\Omega_{+}(\theta) &= 
 \frac{
    -\,Q^{2}\sin\theta
    + \sqrt{
        Q^{4}\sin^{2}\theta
        + \big[a^{2} - (a^{2}+Q^{2})\sin^{2}\theta\big]
          \big(Q^{2}+a^{2}\cos^{2}\theta\big)
      }
 }{
    a\sin\theta\,\big[a^{2} - (a^{2}+Q^{2})\sin^{2}\theta\big]
 }.
\end{aligned}
\end{equation}

\vspace{0.3cm}
For values of $\Omega$ lying within this interval, the corresponding trajectories are timelike, whereas values outside this range lead to spacelike motion and are therefore unphysical. For a static observer outside the ergosphere, we have to put $\Omega = 0$; therefore, we get from Eq.~\eqref{omega_r=0}

\begin{equation}	|\vec{\Omega}_{s}|=\frac{\sqrt{Q^4(a^2+Q^2)+M^2a^4\sin^2\theta\cos^2\theta}}{a^2\cos^2\theta(Q^2+a^2\cos^2\theta)}.
\end{equation}

\vspace{0.3cm}
At $r = 0$, the precession frequency $\Omega_{s}$ varies within the range
\[
\frac{Q^2}{a^2\sqrt{a^2+Q^2}} \leq \Omega_{s} < \infty \quad \text{for } 0 \leq \theta < 90^\circ.
\]
This indicates that $\Omega_{s}$ remains finite everywhere except in the limit $\cos\theta \to 0$, where it diverges. This divergence occurs precisely at the ring singularity of the Kerr-Newman spacetime, located at
\[
x^2 + y^2 = a^2, \qquad z = 0
\]
in Cartesian Kerr-Schild coordinates. Inside the ring, where $x^2 + y^2 < a^2$, the precession frequency remains finite. Physically, this behaviour reflects that the frame-dragging and spin precession effects, encoded in $\Omega_{s}$, grow without bound only at the singular ring, while remaining well-defined and finite in the surrounding spacetime, highlighting the localized nature of the Kerr-Newman singularity and the influence of the charge parameter $Q$ on the precession dynamics.

\subsection{Behaviour of $\vec{\Omega}_{s}$ at $\theta=\frac{\pi}{2}$}

\noindent 
In the equatorial plane, the precession frequency is given by

\begin{equation}
	\vec{\Omega}_s|_{\theta=\frac{\pi}{2}}=\frac{\mathcal{N}(r)}{\mathcal{D}(r)}\hat{\theta},
\end{equation}

where

\begin{equation}
	\begin{array}{rcl}
    \mathcal{N}(r)&=&aMr^2-aQ^2r+\Omega[r^5-3Mr^4-2Ma^2r^2+2rQ^2(r^2+a^2)]+a\Omega^2[M(3r^4+a^2r^2)\\[6pt]&& \qquad\qquad\qquad\qquad\qquad\qquad\qquad\qquad\qquad\qquad-Q^2r(2r^2+a^2)],
		\vspace{0.2cm}\\[10pt] \nonumber
	\mathcal{D}(r)&=&r^3\left[r^2-2Mr+Q^2+2a\Omega(2Mr-Q^2)-\Omega^2\{r^2(r^2+a^2)+a^2(2Mr-Q^2)\}\right].
    \end{array}
\end{equation}

\vspace{0.2cm}
The range of $\Omega$ is determined by using Eq.~\eqref{sp_16}. Therefore, one can obtain the precession frequency at the static-limit surface (boundary of the ergoregion) $r=a+M$ of the extremal Kerr-Newman black hole as

\begin{equation}
\vec{\Omega}_{s}\big|_{\theta=\frac{\pi}{2}}^{r=a+M}
=
\frac{
\begin{aligned}
& a M (a+M)^2 - a (a+M) Q^2 \\
&\quad + (a+M)\Big[-2 a^2 M (a+M) - 3M (a+M)^3 + (a+M)^4
+ 2\big(a^2 + (a+M)^2\big)Q^2\Big]\Omega \\
&\quad + a\Big[M\big(a^2 (a+M)^2 + 3 (a+M)^4\big)
- (a+M)\big(a^2 + 2(a+M)^2\big)Q^2\Big]\Omega^2
\end{aligned}
}{
\begin{aligned}
&(a+M)^3\Big[
-2M(a+M) + (a+M)^2 + Q^2
+ 2a\big(2M(a+M) - Q^2\big)\Omega \\
&\qquad\qquad\qquad
-\Big((a+M)^2\big(a^2 + (a+M)^2\big)
+ a^2\big(2M(a+M) - Q^2\big)\Big)\Omega^2
\Big]
\end{aligned}
}.
\end{equation}

\vspace{0.2cm}
The value of $\Omega$ lies in the range

\begin{equation}
	0<\Omega<\frac{2a}{3a^2+M^2+2aM}.
\end{equation}

\vspace{0.2cm}
\noindent At the equatorial plane $(\theta=\pi/2)$, the allowed angular velocity of the observer is bounded by
$\Omega_{-}<\Omega<\Omega_{+}$. In particular, at the equatorial ergosurface of an extremal Kerr-Newman black hole,
$r=r_{\rm ergo}=M+a$ (for $a\ge 0$), one obtains $\Omega_{-}=0$ and $\Omega_{+}=\frac{2a}{M^{2}+2aM+3a^{2}}$. For fixed spin $a$, this upper bound decreases monotonically with increasing mass $M$. For fixed mass $M$, however, $\Omega_{+}$ is non-monotonic in $a$: it increases for $0<a<M/\sqrt{3}$, attains its maximum at $a=M/\sqrt{3}$, and decreases for $a>M/\sqrt{3}$. Thus, both in the equatorial plane $(\theta=\pi/2)$ and at the equatorial ergosurface, the admissible range of $\Omega$ is controlled by the pair $(M,a)$, with $M$ providing a strictly suppressing effect and $a$ contributing in a bounded, non-monotonic manner.

\vspace{0.2cm}
\noindent
Therefore, the precession frequency at the outer horizon of the extremal Kerr-Newman black hole is computed to be

\begin{equation}
	\vec{\Omega}_{s}|_{\theta=\frac{\pi}{2},r=M,a^{2}+Q^{2}=M^2}=-\frac{a}{M^2}.
\end{equation}

\subsection{Lense-Thirring Precession Frequency}

In the general case, the expression for the precession frequency presented in Eq.~(\ref{sp_13}) applies to all stationary observers situated either inside or outside the ergosphere, provided their angular velocity $\Omega$ lies within the restricted range defined by Eq.~(\ref{sp_14}). The total precession frequency incorporates the combined effects of spacetime rotation (LT precession) and curvature (geodetic precession). Setting $\Omega = 0$ in Eq.~(\ref{sp_13}) yields the specific form of the LT precession frequency for a Kerr-Newman black hole as

\begin{equation}
\vec{\Omega}_{LT} =
\frac{
a(2 M r - Q^2)\sqrt{r^2 - 2 M r + a^2 + Q^2}\cos\theta\,\hat{r}
+ a\left[M(r^2 - a^2\cos^2\theta) - Q^2r\right]\sin\theta\,\hat{\theta}
}{
(r^2 + a^2\cos^2\theta)^{3/2}\left[(r^2 + a^2\cos^2\theta) - 2Mr + Q^2\right]
}. \label{lt_vec}
\end{equation}

\vspace{0.2cm}
\begin{figure}[h!]
	\centering
	\subfigure[]{\includegraphics[width=7.4cm,height=8.1cm]{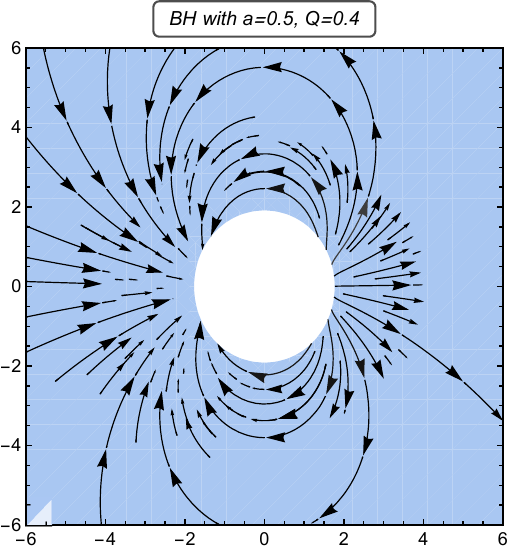}\label{vec_1}}
	\hspace{0.8cm}
	\subfigure[]{\includegraphics[width=7.4cm,height=8.1cm]{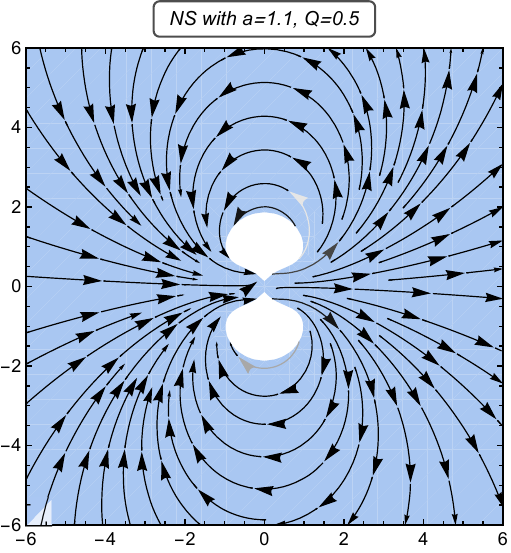}\label{vec_2}}\vspace{0.8em}
	\subfigure[]{\includegraphics[width=7.4cm,height=8.1cm]{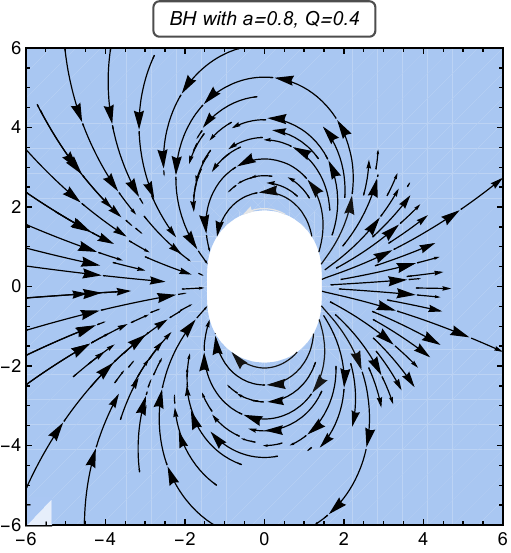}\label{Vec_3}}
	\hspace{0.8cm}
	\subfigure[]{\includegraphics[width=7.4cm,height=8.1cm]{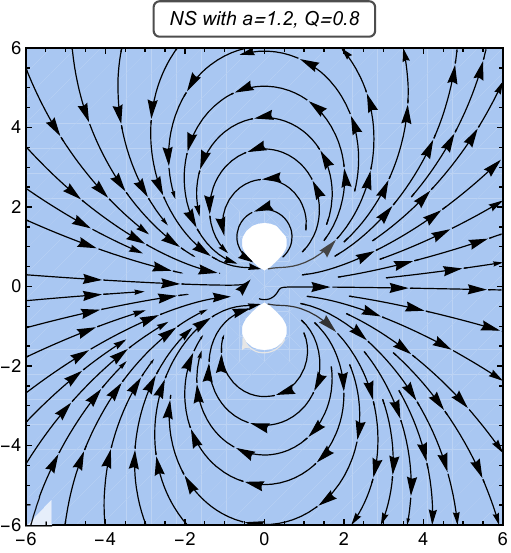}\label{vec_4}}
	\label{sup}
\end{figure}

\begin{figure}[h!]
	\centering
	\subfigure[]{\includegraphics[width=7.4cm,height=8.1cm]{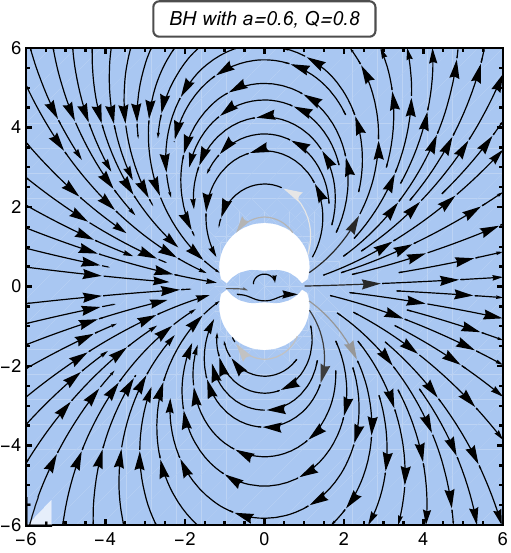}\label{Vec_5}}
	\hspace{0.8cm}
	\subfigure[]{\includegraphics[width=7.4cm,height=8.1cm]{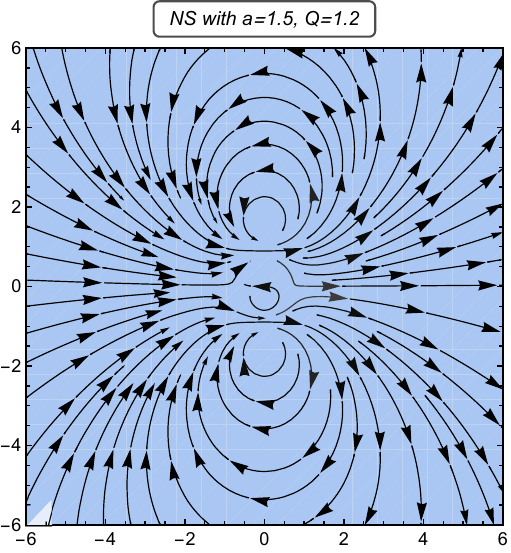}\label{vec_6}}
	\caption{Vector field representation of the $\vec{\Omega}_{LT}$ precession frequency in the Cartesian plane corresponding to $(r, \theta)$. Panels (a), (c), and (e) depict the vector field for black holes, whereas panels (b), (d), and (f) correspond to naked singularities. The field lines show that, for black holes, the vector field is confined outside the ergosphere, while for naked singularities it remains finite and extends up to the ring singularity. The vectors form closed or elliptical loops, illustrating the rotational influence of spacetime and the characteristic frame-dragging effect induced by the black hole.}
	\label{LT_fig}
\end{figure}

\vspace{0.2cm}
Figure~(\ref{LT_fig}) illustrates the vector field associated with the LT precession frequency \eqref{lt_vec} for black holes and naked singularities for several representative values of the parameter $Q$. The left column corresponds to black hole spacetimes, while the right one displays the behaviour for naked singularity, plotted in the Cartesian plane associated with $(r,\theta)$. For black holes, the LT precession frequency diverges as the observer approaches the ergosurface from any direction, whereas it remains finite everywhere outside the ergoregion. In contrast, for naked singularities the precession field is regular throughout the entire domain, except at the ring singularity $(r=0,\;\theta=\pi/2)$, where the denominator of Eq.~\eqref{lt_vec} vanishes, and the frequency becomes singular. Moreover, in black hole geometries, the field lines near the rotational poles precess in the same sense as the spin of the black hole, whereas those lying in the equatorial plane precess in the opposite direction. This behaviour is consistent with the case of the linearised gravitation field~\citep{Rindler:2006km}.

\begin{figure}[h!]
	\centering
	\subfigure[]{%
		\includegraphics[width=7.3cm,height=7.7cm]{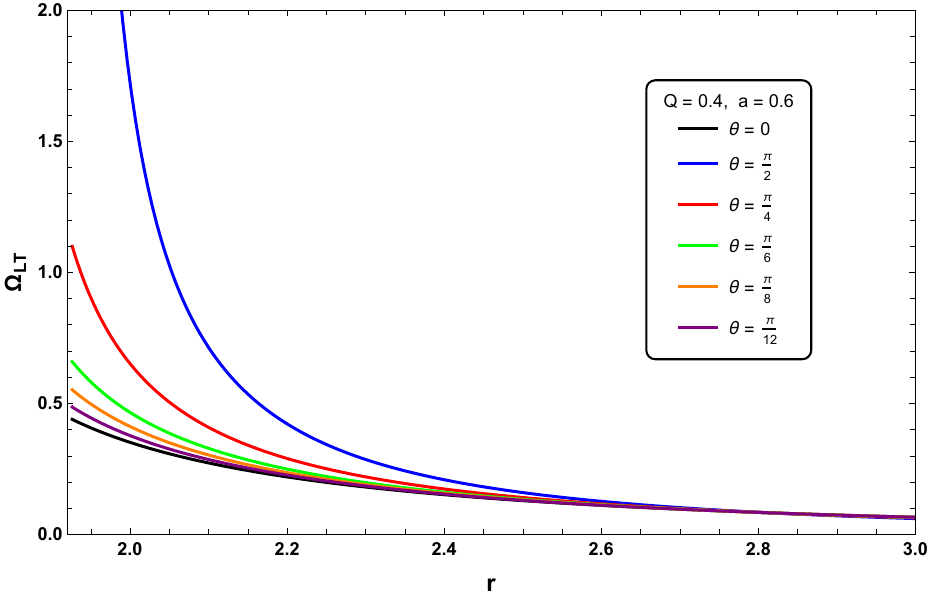}\label{lt_mag_fig_1}}\hspace{0.8cm}
	\subfigure[]{%
		\includegraphics[width=7.3cm,height=7.7cm]{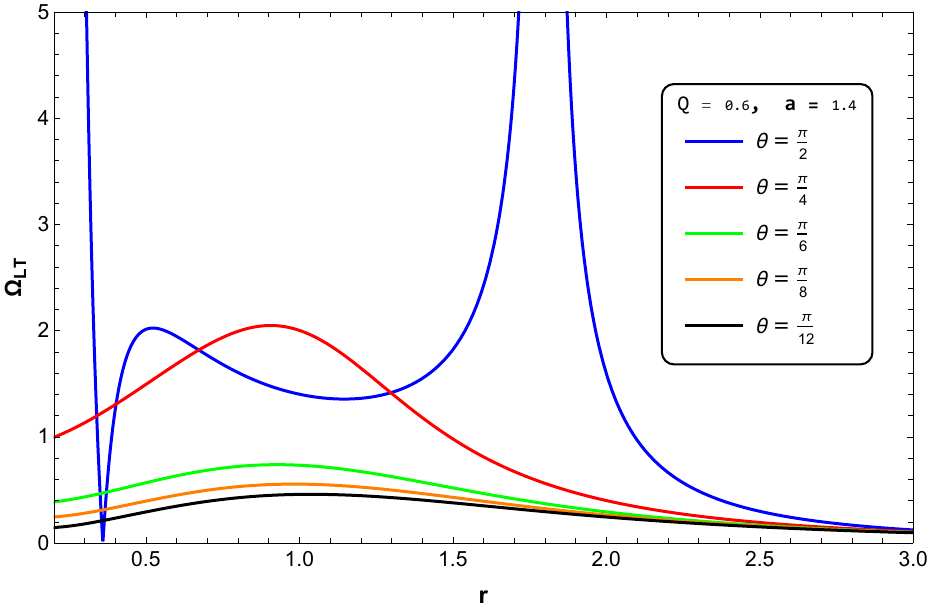}\label{lt_mag_fig_2}}\vspace{0.6em}
	\subfigure[]{%
		\includegraphics[width=7.3cm,height=7.7cm]{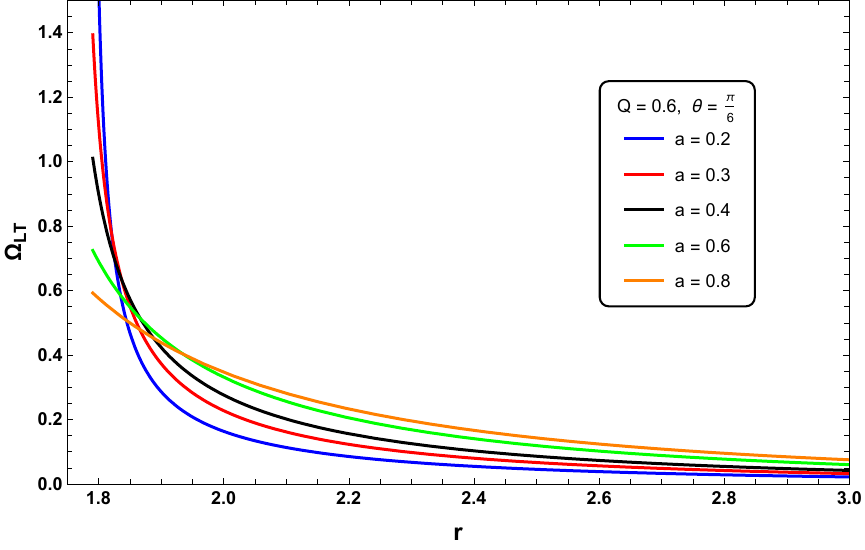}\label{lt_mag_fig_3}} \hspace{0.8cm}
	\subfigure[]{%
		\includegraphics[width=7.3cm,height=7.8cm]{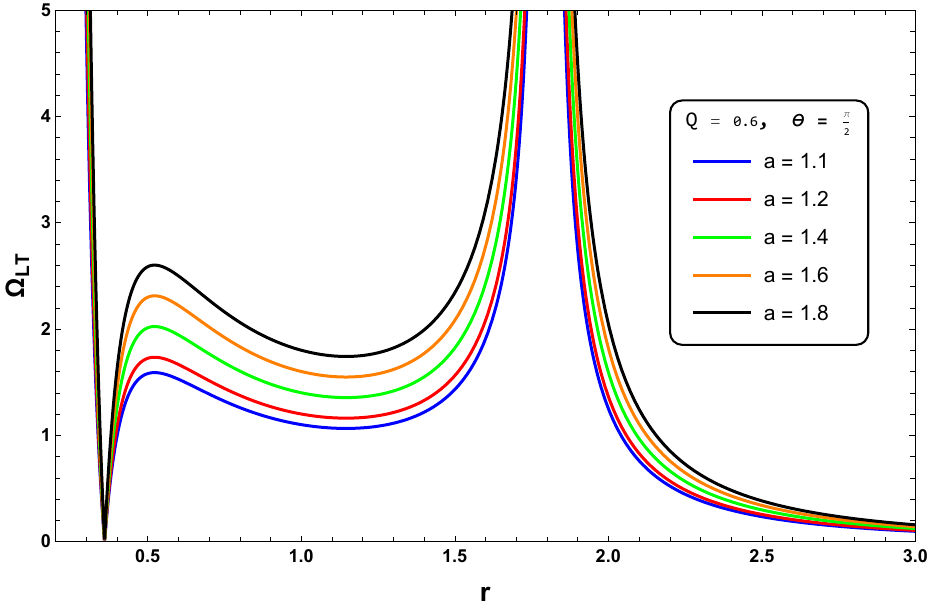}\label{lt_mag_fig_4}}
\end{figure}

\begin{figure}[h!]
	\centering
	\subfigure[]{\includegraphics[width=7.3cm,height=7.86cm]{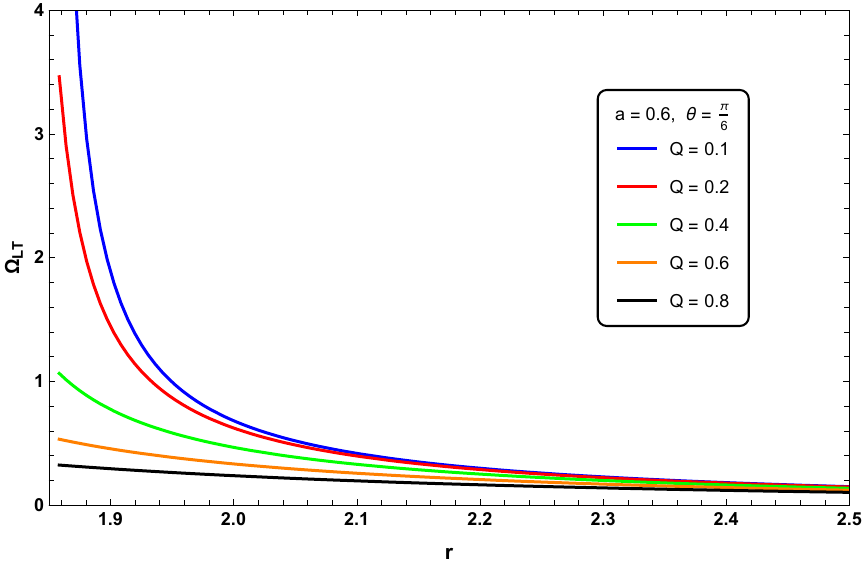}\label{lt_mag_fig_5}}\hspace{0.8cm}
	\subfigure[]{\includegraphics[width=7.3cm,height=7.7cm]{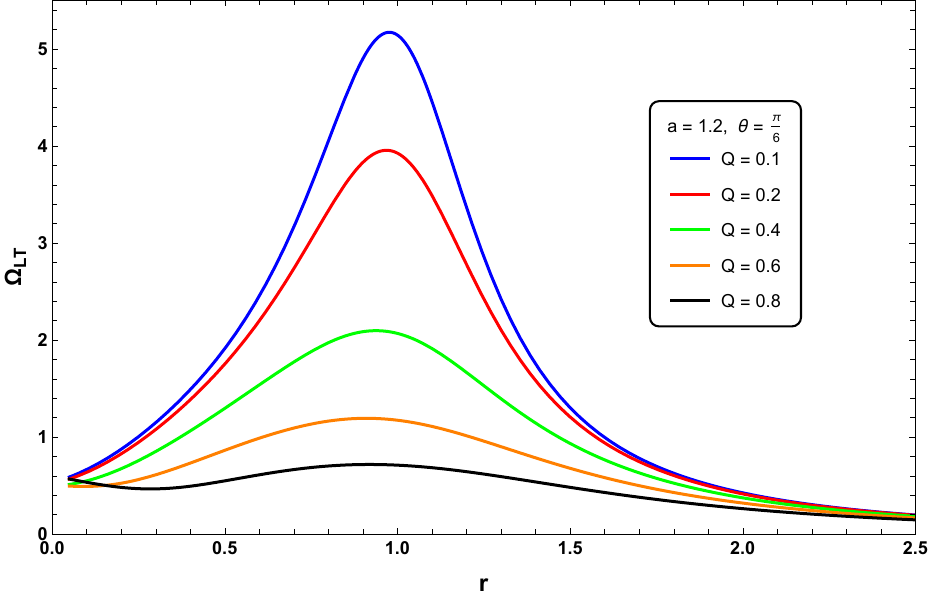}\label{lt_mag_fig_6}}
	\caption{Illustration showing the variation of the magnitude of $|\vec{\Omega}_{LT}|$ (in units of $M^{-1}$) precession frequency versus radial coordinate $r$ (in units of $M$) for different parameters. Panels (a), (c), and (e) depict the magnitude of Lense-Thirring precession frequency for black holes, whereas panels (b), (d), and (f) correspond to naked singularities.}\label{omega_pm_2}
\end{figure}

\vspace{0.3cm}
The magnitude of the LT precession frequency for the Kerr-Newman black hole is given by

\begin{equation}
|\vec{\Omega}_{LT}| =
\frac{
a
\sqrt{
(2Mr - Q^2)^2 (r^2 - 2Mr + a^2 + Q^2)\cos^2\theta
+
\left[M(r^2 - a^2\cos^2\theta) - Q^2r\right]^2 \sin^2\theta
}
}{
(r^2 + a^2\cos^2\theta)^{3/2}
\left[(r^2 + a^2\cos^2\theta) - 2Mr + Q^2\right]
}.\label{lt_mag}
\end{equation}

\vspace{0.2cm}
\noindent
Setting $(Q=0)$ in the above equation reproduces the standard Kerr result~\citep{PhysRevD.95.044006}. In Figures~(\ref{lt_mag_fig_1}, \ref{lt_mag_fig_3}, and \ref{lt_mag_fig_5}) we present the Kerr-Newman LT precession magnitude $|\vec{\Omega}_{LT}|$ for black hole configurations, while Figures~(\ref{lt_mag_fig_2}, \ref{lt_mag_fig_4}, and \ref{lt_mag_fig_6}) correspond to naked singularity configurations. As indicated by Eq.~\eqref{lt_mag}, the strong-field behaviour is mainly controlled by the denominator factors $\rho^{2}=r^{2}+a^{2}\cos^{2}\theta$ and $\rho^{2}-2Mr+Q^{2}$ (the latter defining the static-limit surface $g_{tt}=0$, distinct from the horizon $\Delta=0$). In the asymptotic region ($r\gg M,a,Q$), all curves exhibit a rapid falloff (of order $r^{-3}$), so black hole and naked singularity cases become nearly indistinguishable at large radii. For the black hole configurations shown in Figures~(\ref{lt_mag_fig_1}, \ref{lt_mag_fig_3}, and \ref{lt_mag_fig_5}), a pronounced enhancement of $|\vec{\Omega}_{LT}|$ is observed toward smaller radii, with a sharp growth as the static limit is approached from outside; this feature is strongest near the equatorial plane and weakens toward the rotation axis, reflecting the angular anisotropy of frame dragging. The dependence on parameters is systematic: increasing $a$ amplifies the overall magnitude, whereas increasing $Q$ generally suppresses $|\vec{\Omega}_{LT}|$ and shifts the strong-field features. By contrast, the naked singularity cases (Figures~(\ref{lt_mag_fig_2}, \ref{lt_mag_fig_4}, and \ref{lt_mag_fig_6}) show a strongly $\theta$ dependent inner-region behaviour. In Figure~(\ref{lt_mag_fig_2}), the equatorial curve ($\theta=\pi/2$) displays clear divergences at the static-limit radii, while the off-equatorial curves remain smooth. In Figure~(\ref{lt_mag_fig_4}) (equatorial plane), the static-limit structure is particularly prominent: two static-limit radii $(r_{ergo}=M\pm\sqrt{M^{2}-Q^{2}})$ generate divergences, and a characteristic dip occurs at $r=Q^{2}/M$. In Figure~(\ref{lt_mag_fig_6}) (off-equatorial plane), the profiles remain finite and show a smooth peak whose amplitude decreases with increasing $Q$. Finally, the curvature singularity corresponds to $\rho^{2}\to 0$, i.e.\ $(r,\theta)\to(0,\pi/2)$; thus a genuine $\rho^{-3}$ divergence is relevant only when the ring singularity is approached along the equatorial direction, which is accessible in the naked singularity case but hidden by the horizon for black holes. Therefore, the black hole case shows a clear static-limit divergence, while the naked singularity case lacks it for a wide range of $\theta$ and has a different inner-region behaviour. Hence, these figures allow us to clearly distinguish black hole and naked singularity profiles in charged rotating spacetimes.

\subsection{Geodetic Precession}

In the limiting case of $a = 0$, the metric presented in Eq.~(\ref{kn_1}) corresponds to the Reissner–Nordström black hole~\citep{1916AnP,chandrasekhar1983mathematical}. Since the spacetime in this case represents a non-rotating black hole, there is no precession arising from frame-dragging effects. Nevertheless, due to the inherent curvature of spacetime, the precession frequency $\Omega_s$ of a test gyroscope does not vanish. This residual precession, originating purely from spacetime curvature, is identified as the geodetic precession frequency \citep{deSitter:1916zz,Lense:1918zz}. Therefore, the spin precession frequency for the Kerr-Newman black hole at $a=0$ is expressed as 

\begin{equation}
\left. \vec{\Omega}_{s} \right|_{a=0}
= \frac{
\Omega\!\left(
-\,r\sqrt{\Delta}\,\cos\theta\,\hat{r}
+ \big(r^2 - 3 M r + 2 Q^2\big)\sin\theta\,\hat{\theta}
\right)
}{
\Delta - \Omega^2 r^4 \sin^2\theta
},
\label{GP_1}
\end{equation}



\begin{figure}[h!]
	\centering
	\subfigure[]{\includegraphics[width=7.3cm,height=7.7cm]{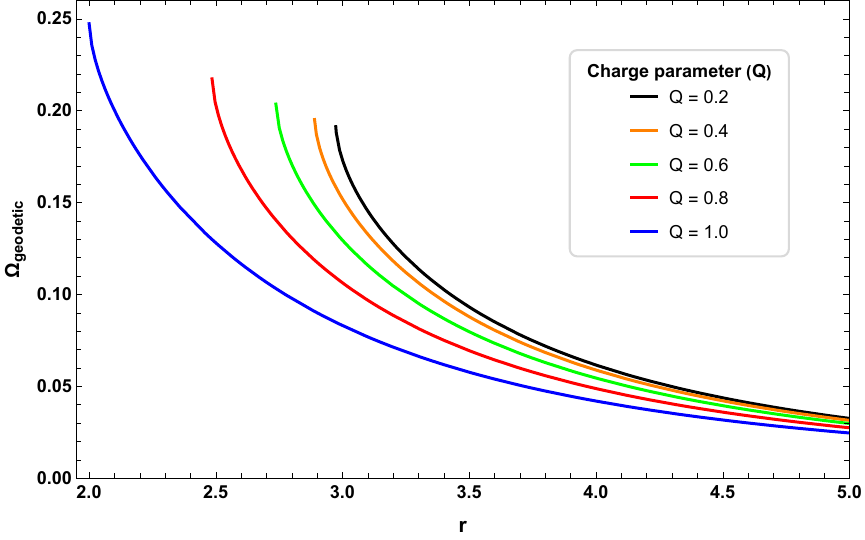}\label{omega_g_1}}\hspace{0.4cm}
	\subfigure[]{\includegraphics[width=7.3cm,height=7.7cm]{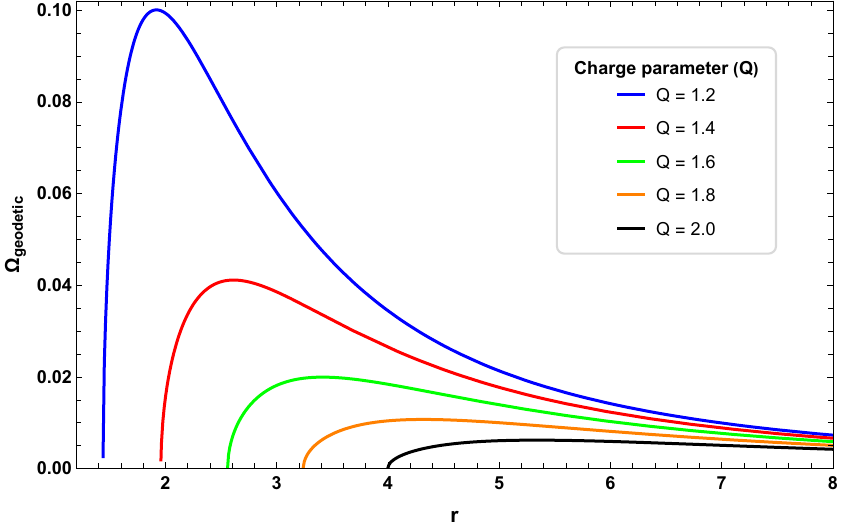}\label{omega_g_2}}\vspace{0.8em}
	\subfigure[]{\includegraphics[width=7.5cm,height=7.7cm]{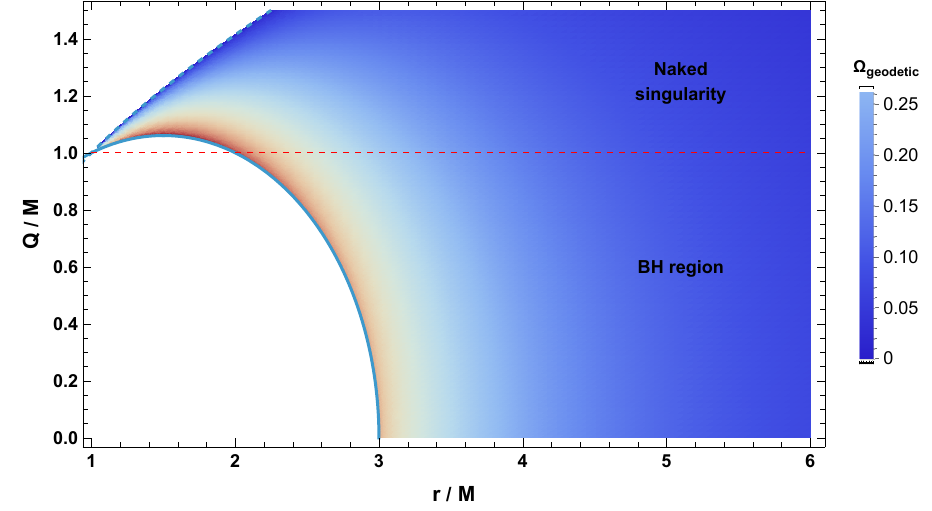}\label{omega_g_3}}\hspace{0.4cm}
	\subfigure[]{\includegraphics[width=7.3cm,height=7.7cm]{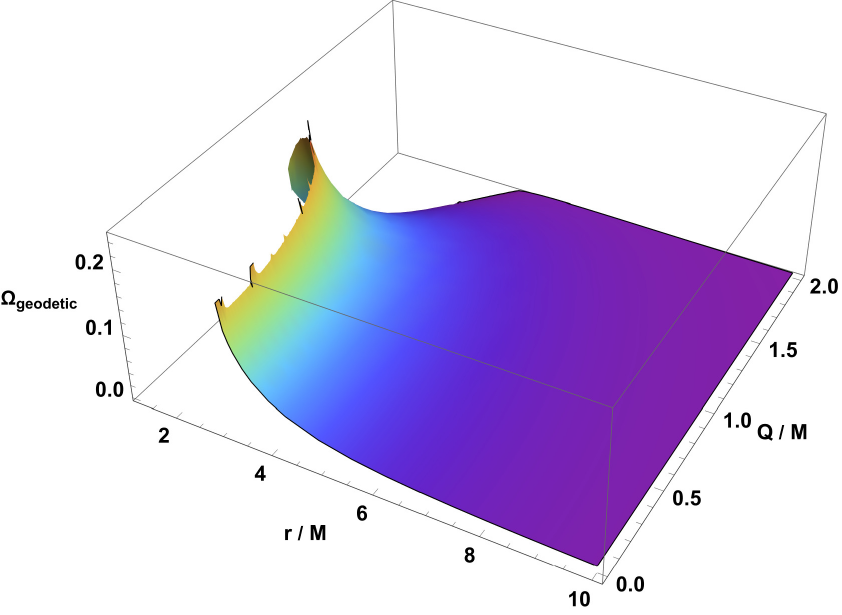}\label{omega_g_4}}
	\caption{Variation of the geodetic precession frequency,  $\Omega_{\mathrm{geodetic}}$ (in units of $M^{-1}$) versus the radial coordinate $r$ (in units of $M$) for different charge values $Q$ in a static charged black hole spacetime. Panel~(a) showing $\Omega_{\mathrm{geodetic}}$ decreases monotonically with increasing $r$, while increasing $Q$ shifts the curves to smaller radii and modifies the strong-field magnitude. Panel~(b) displays that $\Omega_{\mathrm{geodetic}}$ increases with $Q$ up to a threshold, beyond which it decreases. Panel~(c) displays the parametric plot highlighting the transition from a black hole spacetime to a naked singularity configuration as $Q$ surpasses the critical value defined by the extremality condition, while Panel~(d) provides a three-dimensional visualisation of the overall behaviour across the spacetime geometry.}\label{omega_g}
\end{figure}

\vspace{0.4cm}
\noindent
In the case of a static charged black hole, the spacetime geometry is spherically symmetric, implying that the geodetic precession frequency remains uniform over any spherical surface surrounding the black hole. Therefore, without loss of generality, we set out $\theta = \pi/2$ to analyse the geodetic precession in the equatorial plane. In this plane, for an observer in a circular orbit, the magnitude of the precession frequency coincides with the Keplerian frequency, expressed as

\begin{equation}
\Omega_{\mathrm{Kep}} \;\equiv\; \left.\Omega_{s}\right|_{a=0} = \Omega^{'}=
\sqrt{\frac{M}{r^3}-\frac{Q^2}{r^4}}. \label{GP_2}
\end{equation}

\vspace{0.2cm}
In the Copernican reference frame, i.e. (a frame that does not rotate relative to a test gyroscope and possesses a nonzero precession frequency due to the curvature of spacetime), the derived expression for the precession frequency remains valid when evaluated with respect to the proper time $\tau$. The proper time $\tau$ is related to the coordinate time $t$ through the corresponding relativistic transformation as 

\begin{equation}
    d\tau=\sqrt{1-\frac{3M}{r}+\frac{2Q^{2}}{r^{2}}}dt.\label{trans}
\end{equation}

\vspace{0.2cm}
Accordingly, the precession frequency expressed in the coordinate basis, $\Omega^{''}$, can be written as

\begin{equation}
	\Omega^{''}=\left(\dfrac{Mr-Q^2}{r^4}\right)^{\frac12}\sqrt{1-\dfrac{3M}{r}+\dfrac{2Q^2}{r^2}}. \label{geo_0_1}
\end{equation}

\vspace{0.2cm}
By applying the above transformation \eqref{trans}, the geodetic precession frequency ($\Omega^{'}-\Omega^{''}$), which characterises the variation in the orientation of the spin vector after one complete revolution of the observer around the black hole in the coordinate basis, is given by

\begin{equation}
	\Omega_{\text{geodetic}}=\left(\dfrac{Mr-Q^2}{r^4}\right)^{\frac12}\left(1-\sqrt{1-\dfrac{3M}{r}+\dfrac{2Q^2}{r^2}}\right). \label{geo_0}
\end{equation}

\vspace{0.2cm}
The above expression matches exactly the result reported in~\citep{Majumder:2025wsb}. In the absence of the charge parameter $Q$, the expression (\ref{geo_0}) correctly reduces to that of a Schwarzschild black hole~\citep{PhysRevD.19.2280,hartle2003gravity}. Figure~(\ref{omega_g}) illustrates the variation of the geodetic precession frequency versus radial coordinate $r$ for different values of the $Q$, respectively. From Figure~(\ref{omega_g_1}), it is evident that an increase in the charge parameter $Q$ leads to an enhancement in the magnitude of the geodetic precession frequency. However, beyond a certain threshold of $Q$, a distinct deviation in behaviour is observed; namely, the geodetic precession frequency begins to decrease with further increase in $Q$. This peculiar feature can be better understood from Fig.~(\ref{omega_g_3}), which clearly illustrates the transition of the spacetime from a black hole configuration to a naked singularity as $Q$ crosses the critical threshold value. Figure~(\ref{omega_g_4}) illustrates the three-dimensional representation of the overall configuration, providing a comprehensive visualisation of the behaviour of the precession frequency across the spacetime.

\section{Distinguishing Kerr-Newman Black Hole From Naked Singularity}\label{sec_4}

In this section, we utilise the spin precession behaviour of a test gyroscope to differentiate between a Kerr-Newman black hole and a naked singularity, thereby validating the consistency of our analysis. We calculate the values of $\Omega$ for an observer located close to the black hole's horizons. In the event horizon the angular velocity becomes $\Omega_+=\frac{a}{2Mr_{+}- Q^2}$ and in the Cauchy horizon the angular velocity becomes $\Omega_{-}=\frac{a}{2Mr_{-}- Q^{2}}$. In the equatorial plane, the expression reads as

\begin{equation}
	\Omega_{\pm}|_{\theta=\pi/2}=\frac{a(2Mr-Q^2)\pm r^2\sqrt{\Delta}}{r^2(r^2+a^2)+a^2(2Mr-Q^2)}. \label{NS_1}
\end{equation}

\vspace{0.3cm}
Following the approach adopted in \citep{Zahra_2025}, we examine the behaviour of the limiting angular velocities $\Omega_{\pm}$ in the Kerr-Newman spacetime in the vicinity of the ring singularity $r=0\mathrm{~and~}\theta=\frac{\pi}{2}$. As we previously discussed, the admissible angular velocity of a timelike observer is determined by the condition $\Omega_{-}(r,\theta) < \Omega(r,\theta) < \Omega_{+}(r,\theta)$, where $\Omega_{\pm}$ are obtained as the roots of the quadratic form built from the metric components. For the Kerr-Newman geometry, although the expressions for $\Omega_{\pm}$ become formally indeterminate at the singularity $(r=0,\theta=\pi/2)$, their limiting behaviour differs qualitatively from the rotating naked singularity case. In particular, in the joint limit $r\to 0^{+}$ and $\theta\to \pi/2$, the term proportional to $\rho^{2}$ in the numerator vanishes, and both branches approach the same finite value. 

\begin{equation}
	\Omega_{\pm}|_{r\to 0,\:\theta=\pi/2}=\frac{1}{a}.\label{NS_2}
\end{equation}

\vspace{0.2cm}
This confirms that both angular frequencies coincide precisely at the ring singularity, and it is independent of the approach path. Thus, unlike the rotating naked singularity, where the lower branch can exhibit path-dependent behaviour, a nonzero charge $Q$ regularises the limiting angular velocities in this spacetime geometry. Therefore, the admissible domain of angular velocities near the ring singularity remains well defined, providing a clear distinction between the rotating naked singularity and Kerr-Newman naked singularity scenarios. Now we analyse the behaviour of a test gyroscope within the ergosphere of a black hole in contrast to that of a naked singularity. To accomplish this, we first compute the magnitude of the precession frequency for various values of $\theta$. For this purpose, we introduce a dimensionless parameter $\delta$, which allows us to explore the entire range of the angular velocity for any timelike observer $\Omega$ as

\begin{equation}
\begin{aligned}
\Omega \: &= \: \delta\,\Omega_+ + (1-\delta)\,\Omega_- = \:\omega - (1-2\delta)\,\sqrt{\omega^2 - \frac{g_{tt}}{g_{\phi\phi}}}\:, \\[10pt]
       &= \frac{ a(2Mr - Q^2)\,\sin\theta \;-\; (1-2\delta)\,\rho^2\,\sqrt{\Delta} }
                { \sin\theta \,\bigl[(r^2 + a^2)\,\rho^2 + a^2\sin^2\theta\,(2Mr - Q^2)\bigr] }. \label{NS_3}
\end{aligned}
\end{equation}

\vspace{0.3cm}
\noindent
Here, $0 < \delta < 1$ and $\omega = -\,g_{t\phi}/g_{\phi\phi}$. 
It follows that the limiting values of $\delta$ correspond to the two extreme cases of the observer’s angular velocity, thereby constraining $\Omega$ within the range $\Omega_{-} < \Omega < \Omega_{+}$ governed by Eq.~\eqref{sp_16}. Using Eq.~\eqref{NS_3}, one obtains the spin precession frequency in a compact form as

\begin{equation}
	\vec{\Omega}_{s}=\dfrac{(r^2+a^2)^2-a^2\Delta\sin^2\theta}{4\delta(1-\delta)\Delta\rho^7}\Big(\xi(r)\sqrt{\Delta}\cos\theta\hat{r}+\eta(r)\sin\theta\hat{\theta}\Big),\label{dis_spin_vec}
\end{equation}

where
\begin{equation}
\begin{aligned}
\xi(r) =\;& a\left(2 M r - Q^2\right)
 - \frac{\Omega}{8}\Big[8 r^4 + 8 a^2 r^2 + 16 a^2 M r - 8 a^2 Q^2 + 3 a^4 
 + 4 a^2\left(2 \Delta - a^2\right)\cos 2\theta \\
 &+ a^4 \cos 4\theta\Big] 
 + \Omega^2 a^3\left(2 M r - Q^2\right)\sin^4\theta, \\[8pt]
\eta(r) =\;& a M\left(r^2 - a^2 \cos^2\theta\right) - a Q^2 r 
 + \Omega\Big[r^5 - 3 M r^4 + 2 a^2 r^3 \cos^2\theta - 2 M a^2 r^2 
 + a^4 r \cos^4\theta \\
 &+ M a^4 \cos^2\theta\left(1 + \sin^2\theta\right) 
 + 2 Q^2 r\left(r^2 + a^2\right)\Big] \\
 &+ \Omega^2 a \sin^2\theta\Big[
 M\!\left(3 r^4 + a^2 r^2 + a^2 \cos^2\theta (r^2 - a^2)\right) - Q^2 r\!\left(2 r^2 + a^2(1 + \cos^2\theta)\right)
 \Big]. \\[6pt]
\end{aligned} \notag
\end{equation}

Therefore, the magnitude of the spin–precession frequency is obtained as

\begin{equation}
\left|\vec{\Omega}_{s}\right|
= \frac{(r^{2}+a^{2})^{2}-a^{2}|\Delta| \sin^{2}\theta}
     {4\,\delta(1-\delta)\,|\Delta|\,\rho^{7}}
\sqrt{\xi(r)^{2}\,|\Delta|\,\cos^{2}\theta
      \;+\;
      \eta(r)^{2}\,\sin^{2}\theta}.
\end{equation}

\vspace{0.2cm}
\noindent
As evident from the structure of the above equation, the denominator becomes singular at both the ring singularity and the black hole's horizons. Furthermore, Eq.~\eqref{dis_spin_vec} shows that the numerator of the radial component of $\vec{\Omega}_{s}$ vanishes as the observer approaches the horizon. Thus, the limiting behaviour of the spin-precession frequency depends sensitively on the direction along which the observer approaches the horizon. In the subsequent analysis, we examine the spin precession for trajectories characterised by different values of $\delta$, $a$, and $Q$, and discuss the corresponding physical behaviour in the near-horizon regime.

\begin{figure}[h!]
	\centering
	\subfigure[]{%
		\includegraphics[width=7.3cm,height=7.7cm]{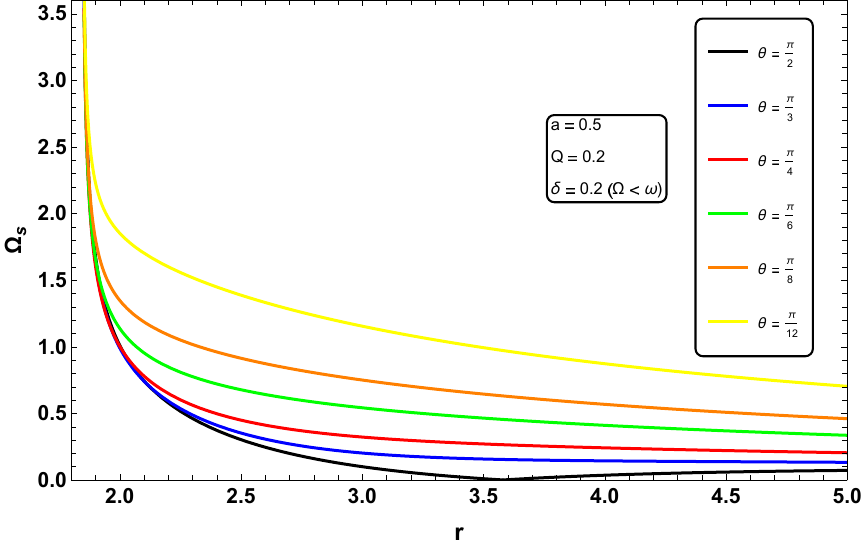}\label{g_1}}\hspace{0.8cm}
	\subfigure[]{%
		\includegraphics[width=7.3cm,height=7.7cm]{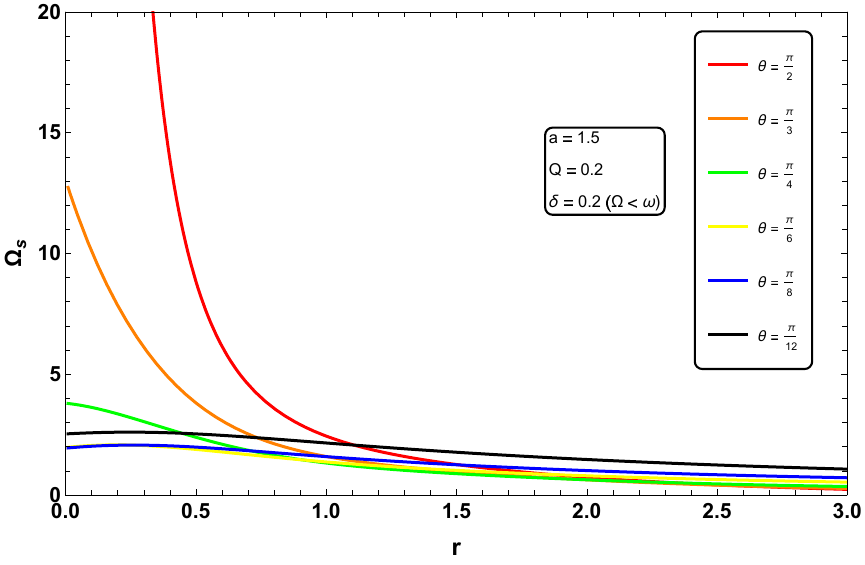}\label{g_2}}\vspace{0.8em}
	\subfigure[]{%
		\includegraphics[width=7.3cm,height=7.7cm]{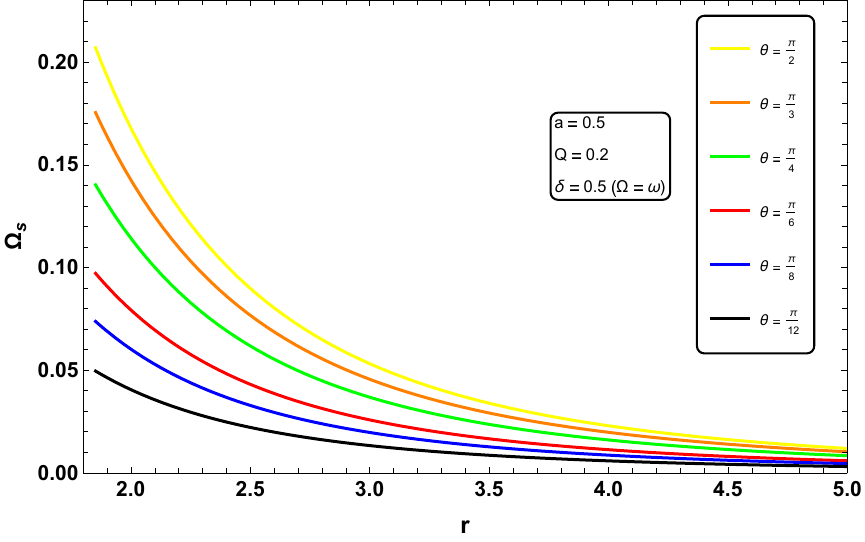}\label{g_3}}\hspace{0.8cm}
	\subfigure[]{%
		\includegraphics[width=7.3cm,height=7.7cm]{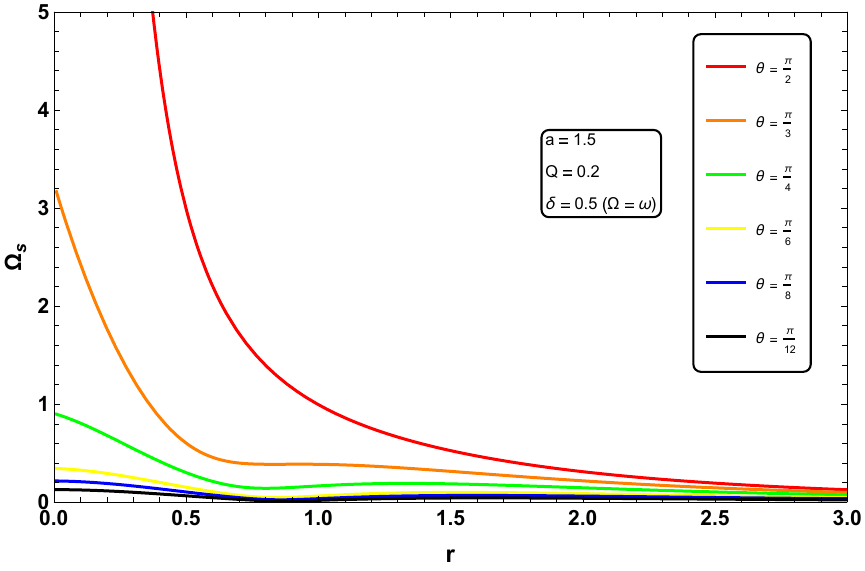}\label{g_4}}
\end{figure}

\begin{figure}[h!]
	\centering
	\subfigure[]{%
		\includegraphics[width=7.3cm,height=7.7cm]{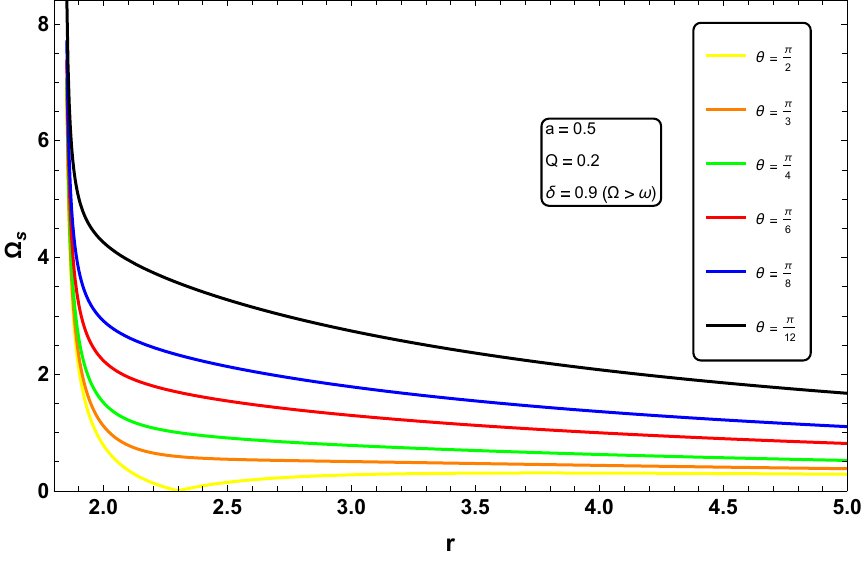}\label{g_4}}\hspace{0.8cm}
	\subfigure[]{%
		\includegraphics[width=7.3cm,height=7.7cm]{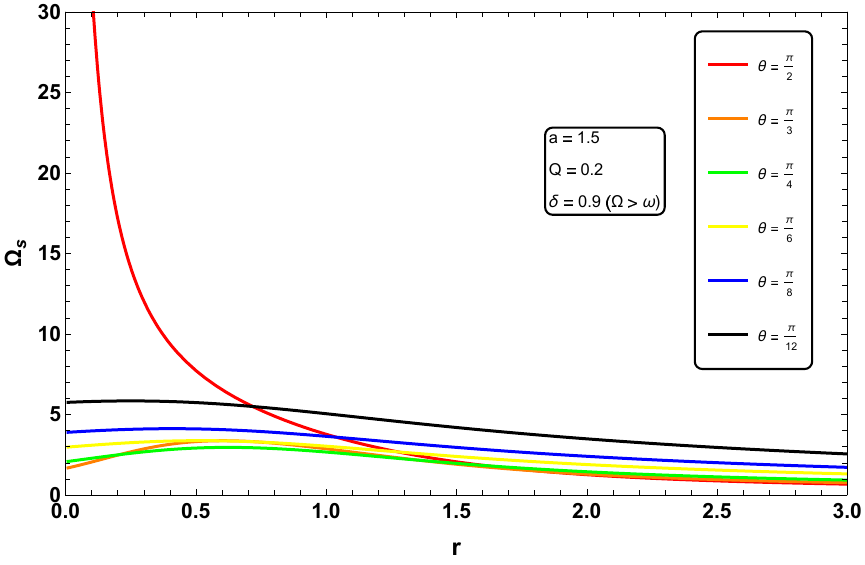}\label{g_4}}
	\caption{The magnitude of the spin–precession frequency \(\Omega_{s}\) (in units of \(M^{-1}\)) is plotted versus the radial coordinate \(r\) (in units of \(M\)). The left column corresponds to the black hole case, whereas the right column depicts the behaviour for a naked singularity. In the black hole geometry, the precession frequency $\Omega_{s}$ becomes unbounded for $\delta = 0.2$ and $\delta = 0.9$, whereas for $\delta = 0.5$ it remains finite as the observer approaches the event horizon from any direction. By contrast, in the naked singularity case, $\Omega_{s}$ stays finite for all directions of approach except at the ring singularity ($r = 0$,\, $\theta = \pi/2$), where it diverges.}\label{omega_BH_NS}
\end{figure}

\vspace{0.2cm}
\noindent Previously, we considered an observer moving with a four–velocity in a stationary and axisymmetric spacetime. For a particular class of observers whose angular momentum vanishes, the condition characterising such motion is given by

\begin{equation}
	\begin{array}{rcl}p_\phi&\equiv&\dfrac{\partial\mathcal{L}}{\partial\dot{\phi}}=g_{t\phi}\dot{t}+g_{\phi\phi}\dot{\phi}=\ell=0.\end{array}
\end{equation}

\begin{figure}[h!]
	\centering
	\subfigure[]{%
		\includegraphics[width=7.3cm,height=7.7cm]{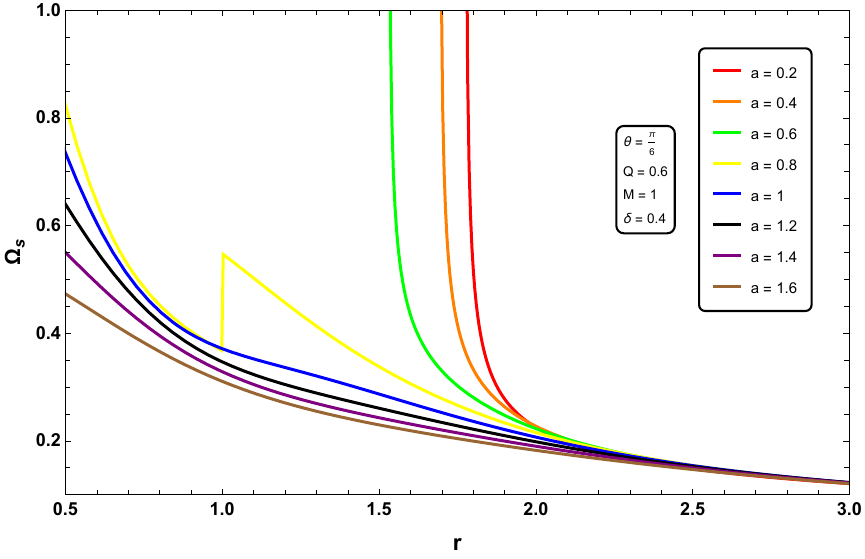}\label{g_1}}\hspace{0.8cm}
	\subfigure[]{%
		\includegraphics[width=7.3cm,height=7.7cm]{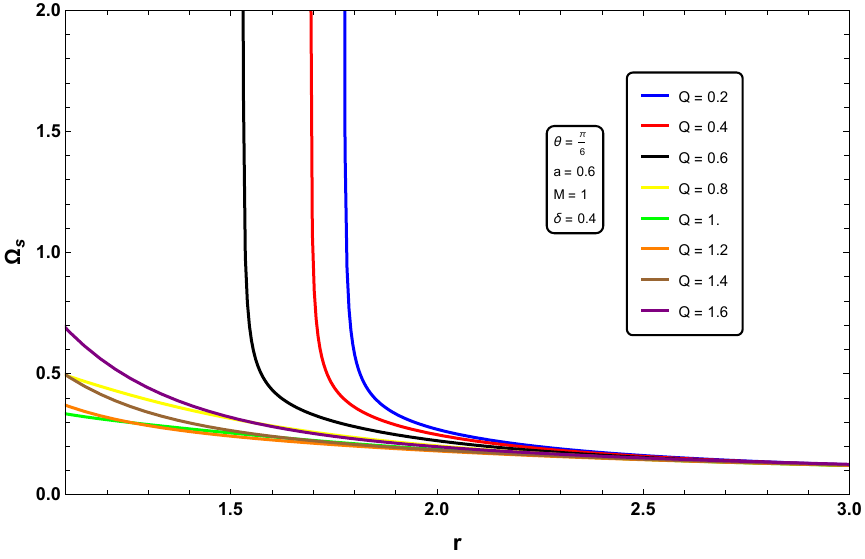}\label{g_2}}\vspace{0.8em}
	\subfigure[]{%
		\includegraphics[width=7.3cm,height=7.7cm]{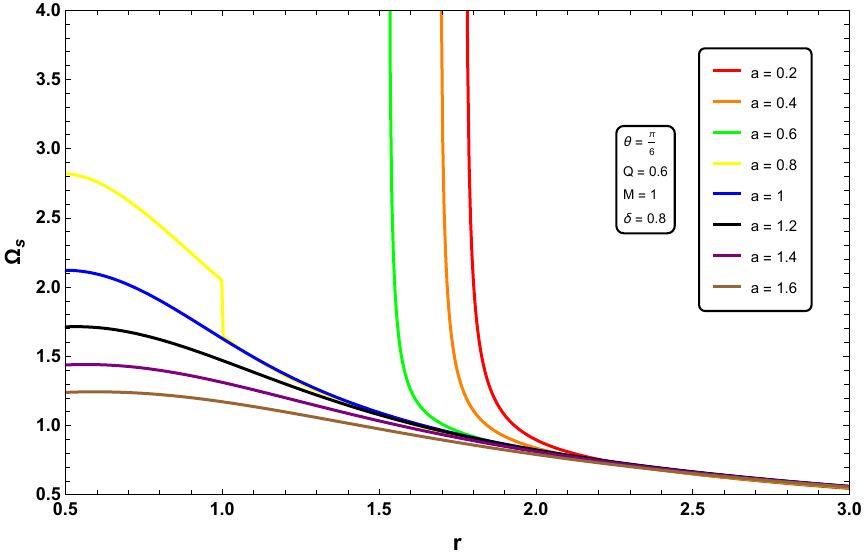}\label{g_3}}\hspace{0.8cm}
	\subfigure[]{%
		\includegraphics[width=7.3cm,height=7.7cm]{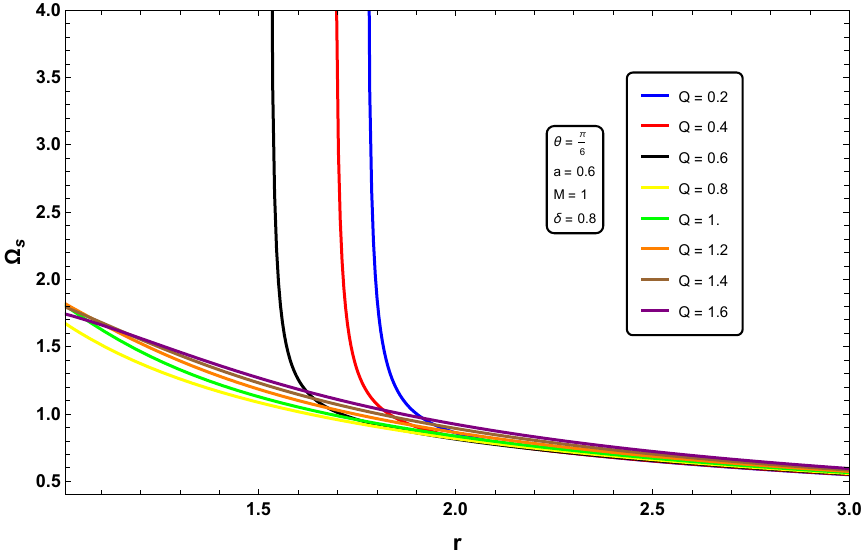}\label{g_4}}
\end{figure}

\begin{figure}[h!]
	\centering
	\subfigure[]{%
		\includegraphics[width=7.3cm,height=7.7cm]{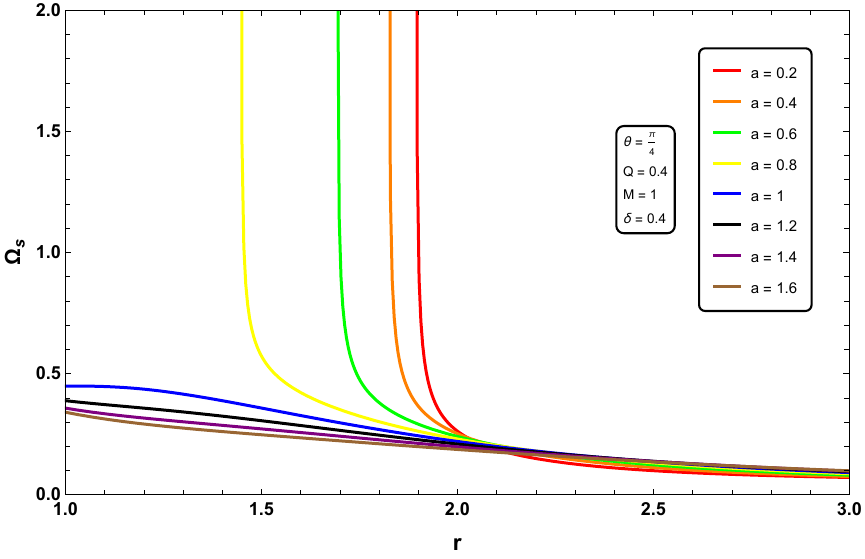}\label{g_1}}\hspace{0.8cm}
	\subfigure[]{%
		\includegraphics[width=7.3cm,height=7.7cm]{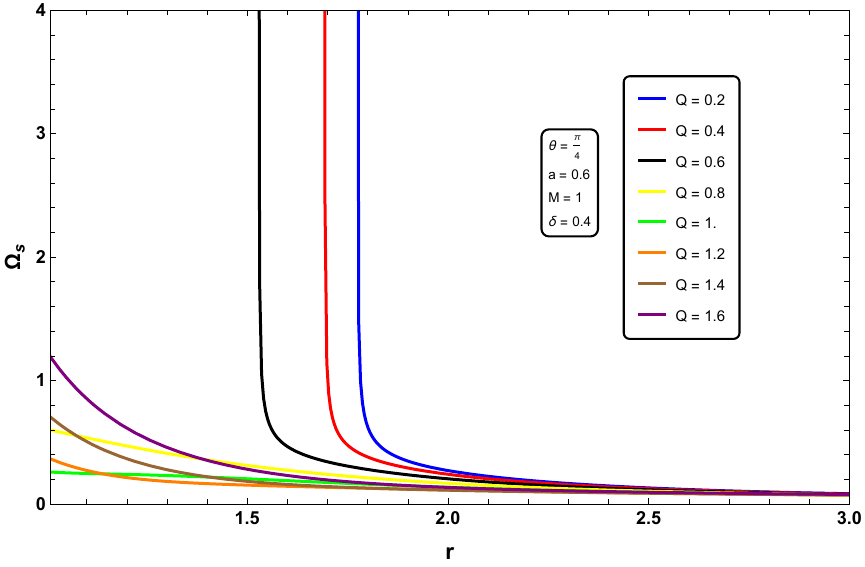}\label{g_2}}\vspace{0.8em}
	\subfigure[]{%
		\includegraphics[width=7.3cm,height=7.7cm]{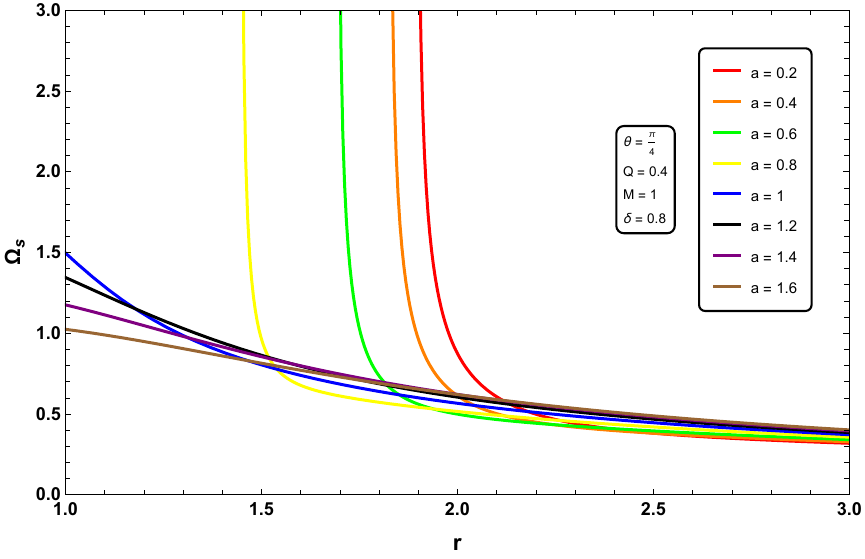}\label{g_3}}\hspace{0.8cm}
	\subfigure[]{%
		\includegraphics[width=7.3cm,height=7.7cm]{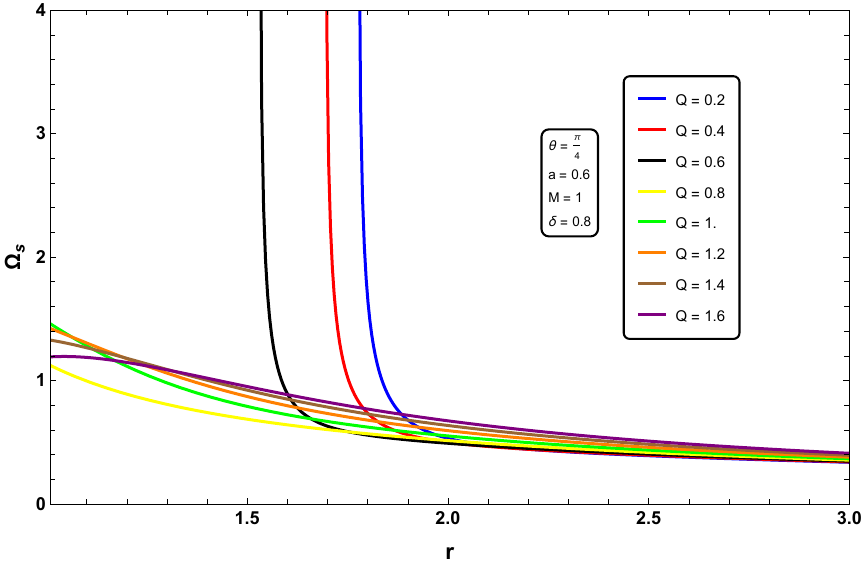}\label{g_4}}
	\caption{The variation of the precession frequency $\Omega_{s}$ (in unit of $M^{-1}$) versus the radial coordinate $r$ (in units of $M$) is shown for several parameter values. The plots indicate that \(\Omega_{s}\) diverges in the vicinity of the horizons for black hole spacetimes, whereas it remains finite for all accessible radii in the naked singularity configuration.}\label{omega_BH_NS_1}
\end{figure}

\vspace{0.2cm}
Such an observer is referred to as a zero-angular-momentum observer (ZAMO), a concept first introduced by Bardeen~\citep{misner2017gravitation,Bardeen:1970vja}. Later, Bardeen et al.~\citep{Bardeen:1972fi} established that the ZAMO framework offers a robust and effective approach for analysing physical processes in the vicinity of astrophysical compact objects. In the Newtonian regime, the specific angular momentum $\ell$ and the angular velocity $\Omega$ are related through $\ell = r^{2}\Omega$. This relation clearly indicates that no ambiguity arises when one considers a non-rotating frame, since the condition $\ell = 0$ immediately implies $\Omega = 0$. Thus, in Newtonian gravity, an observer with vanishing angular momentum naturally corresponds to one with zero angular velocity. The situation changes fundamentally when we consider Einstein’s theory of gravity. In this case, the specific angular momentum satisfies the relation $\ell \propto (\Omega - \omega)$, where $\omega = -\,\frac{g_{t\phi}}{g_{\phi\phi}}$ denotes the frame-dragging angular velocity. It is therefore clear that the condition $\ell = 0$ cannot be achieved simply by imposing $\Omega = 0$. This leads naturally to the distinction between two physically different classes of observers, i.e., the ZAMO, for whom $\ell = 0$, and the zero-angular-velocity observer (ZAVO), for whom $\Omega = 0$. For both frames, the associated frame-dragging angular velocity is $\omega = -\,\frac{g_{t\phi}}{g_{\phi\phi}}$. In the ZAMO frame, one may introduce a gravitational potential defined by $\Phi = -\frac{1}{2}\ln |g^{tt}|$, where $g^{tt}$ is the contravariant time—time component of a stationary and axisymmetric spacetime metric. This potential is related to the gravitational acceleration experienced by the observer through $g_{\mu} = (a_{\mu})_{\text{ZAMO}} = \nabla_{\mu}\Phi$. The acceleration $a_{\mu}$ is a purely kinematical invariant. Using Eq.~\eqref{NS_3}, one can readily show that for the special choice $\delta = \tfrac{1}{2}$, the angular velocity of a Kerr-Newman black hole in the ZAMO frame takes the form

\begin{equation}
	\Omega \:=\: \frac{a(2Mr-Q^{2})\sin\theta}{\sin\theta[(r^{2}+a^{2})\rho^{2}+a^{2}\sin^{2}\theta(2Mr-Q^{2})]}=-\,\frac{g_{t\phi}}{g_{\phi\phi}}=\omega.
\end{equation}

\vspace{0.2cm}
\noindent 
This implies that the ZAMOs angular velocity is a function of the spin parameter and charge parameter. In the $\theta = \frac{\pi}{2}$ plane, the ZAMOs angular velocity is given by

\begin{equation}
	\Omega|_{\theta=\frac\pi2}=\omega|_{\theta=\frac\pi2}=\frac{a(2Mr-Q^2)}{(r^2+a^2)r^2+a^2(2Mr-Q^2)}.
\end{equation}

\vspace{0.2cm}
\noindent
In Fig.~(\ref{omega_BH_NS}), we illustrated the behaviour of the precession frequency $\Omega_{s}$ for the black hole with $a = 0.5$ and $Q = 0.2$ (left column) and for the naked singularity with $a = 1.5$ and $Q = 0.2$ (right column). The three rows correspond to observers parametrised by $\delta = 0.2,\,0.5,$ and $0.9$, respectively. For the black hole, $\Omega_{s}$ diverges for all observers except the ZAMO as the event horizon is approached along any direction. At the same time, the ZAMO continues to exhibit a finite precession frequency at the horizon. For the naked singularity, $\Omega_{s}$ remains finite as $r \rightarrow 0$ for all directions except along the equatorial plane ($\theta = \pi/2$), where it becomes substantial due to the ring singularity. Fig.~(\ref{omega_BH_NS_1}) confirms that this qualitative behaviour persists for other choices of parameters. In particular, for any black hole configuration with arbitrary $a$ and $Q$, $\Omega_{s}$ diverges near the horizon for every observer except the ZAMO, whereas in all naked singularity cases it remains finite up to $r = 0$ along any direction except $\theta = \pi/2$, where the divergence is again associated with the ring singularity.

\vspace{0.2cm}
\noindent
Finally, the limiting behaviour of the spin-precession frequency provides an effective diagnostic for distinguishing Kerr-Newman black holes from naked singularities. Consider a gyroscope carried by stationary observers whose four-velocity includes a nonvanishing azimuthal component $\Omega$. These observers move along circular orbits at fixed values of $(r,\theta)$ with constant angular velocity, and the admissible range of $\Omega$ is determined by demanding that their motion remain timelike; within this interval, $\Omega$ may be parametrised in terms of a convenient auxiliary parameter $\delta$. When stationary observers are located along two different polar directions, $\theta_{1}$ and $\theta_{2}$, are examined, the behaviour of their spin-precession frequency $\Omega_{s}$ clearly reveals the nature of the underlying geometry. In the Kerr-Newman black hole spacetime, $\Omega_{s}$ diverges as the observer approaches the outer event horizon, which completely encloses the curvature singularity. Consequently, observers approaching the central object from any direction encounter an arbitrarily large precession rate. In contrast, for a Kerr-Newman naked singularity, the absence of an event horizon exposes the ring singularity at $(r=0,\ \theta=\pi/2)$, and the divergence of $\Omega_{s}$ occurs only along this equatorial direction. Observers approaching the central region away from the equatorial plane measure a finite precession frequency. Thus, an isotropic divergence of $\Omega_{s}$ characterises the Kerr-Newman black hole, whereas a divergence confined to the equatorial direction alone signals the presence of a naked singularity.

\subsection{Acceleration scalar for stationary observers in Kerr–Newman Spacetime}

In the framework of a stationary and axisymmetric spacetime, the motion of a test gyroscope can be characterised by its four-velocity $u^{\mu}$, which, in general, corresponds to a non-zero acceleration. These gyroscopes move along helical trajectories that are tangent to the timelike Killing vector field of the spacetime. Such motion indicates that the gyroscopes are not freely falling; rather, they are subjected to a continuous acceleration to maintain their stationary orbits. This acceleration is not arbitrary but must be externally sustained, i.e., either through continuous external thrust or by some equivalent physical mechanism that counterbalances the gravitational and inertial effects acting on the gyroscope. Therefore, the required acceleration scalar corresponding to the motion of the gyroscopes can be determined from the expression of the four-acceleration, given by

\begin{equation}
\alpha_\beta =\tfrac12\,\nabla_\beta\ln|K^2|. 
\end{equation}

\vspace{0.2cm}
The scalar magnitude of the acceleration associated with the four-velocity is obtained as follows:

\begin{multline}
	\alpha = \sqrt{g^{\beta\gamma}\alpha_\beta\alpha_\gamma} = \frac{1}{\rho^{5}\,|K^{2}|}
	\left\{
	\Delta
	\left[
	\bigl(M(2r^{2}-\rho^{2})-rQ^{2}\bigr)\,\mathcal{B}^{2}
	- r\,\rho^{4}\,\Omega^{2}\sin^{2}\theta
	\right]^{2}
	\right. \\[6pt]
	\left.
	+\,\sin^{2}\theta\,\cos^{2}\theta\,
	\left[
	(2Mr-Q^{2})\,\mathcal{C}^{2}
	+ \Omega^{2}\rho^{4}\Delta
	\right]^{2}
	\right\}^{1/2},
	\label{lt_a_1}
\end{multline}

where 

\begin{equation*}
\mathcal{B}\equiv 1-a\Omega\sin^2\theta,
\qquad
\mathcal{C}\equiv a-\Omega(r^2+a^2).
\end{equation*}

\vspace{0.2cm}
Now $K^2$ is expressed as 

\begin{equation}
K^2
=-\,\frac{4\,\delta(1-\delta)\,\rho^2\,\Delta}
{(r^2+a^2)^2-a^2\Delta\sin^2\theta}. \label{lt_a_2}
\end{equation}

\vspace{0.3cm}
Finally, substituting Eq.~\eqref{lt_a_2} into Eq.~\eqref{lt_a_1} gives the reduced form

\begin{multline}
	\alpha
	= \frac{(r^{2}+a^{2})^{2}-a^{2}\Delta\sin^{2}\theta}
	{4\,\delta(1-\delta)\,\rho^{7}\,\Delta}
	\left\{
	\Delta\left[\bigl(M(2r^{2}-\rho^{2})-rQ^2\bigr)\mathcal{B}^{2}
	- r\,\rho^{4}\Omega^{2}\sin^{2}\theta\right]^{2}
	\right. \\[8pt]
	\left.
	+ \sin^{2}\theta\,\cos^{2}\theta\,
	\left[(2Mr-Q^{2})\,\mathcal{C}^{2}+\Omega^{2}\rho^{4}\Delta\right]^{2}
	\right\}^{1/2}. \label{KN_acc_red}
\end{multline}

\vspace{0.2cm}
\noindent
The above equation specifies the proper acceleration required to transport a test gyroscope along a stationary trajectory in the Kerr-Newman spacetime. In the limit \(Q\to 0\), it reduces exactly to the corresponding Kerr expression reported in \citep{Chakraborty:2016mhx}. Although the electric charge introduces additional \(Q^{2}\)-dependent contributions, the qualitative behaviour of the acceleration remains similar to that in the Kerr geometry. For the black hole case, the acceleration diverges as the event horizon is approached from any direction. This reflects the fact that an observer attempting to remain stationary arbitrarily close to the horizon must undergo an increasingly large proper acceleration. Hence, a stationary timelike worldline cannot be maintained on the horizon with finite acceleration. On the other hand, in the naked singularity case, where no horizon exists, the acceleration remains finite throughout the spacetime except at the ring singularity. In particular, it stays finite even at \(r=0\) for \(\theta\neq \pi/2\). The divergence appears only at the ring singularity \(r=0\) and \(\theta=\pi/2\), where \(\rho^{2}=0\). Thus, the behaviour of the gyroscope's acceleration clearly distinguishes the Kerr-Newman black hole from the Kerr-Newman naked singularity. 

\vspace{0.2cm}
\noindent
It is also instructive to note that the proper acceleration vanishes when the gyroscope follows a geodesic trajectory. In the equatorial plane \((\theta=\pi/2)\), Eq.~\eqref{KN_acc_red} reduces to

\begin{equation}
	\alpha
	=
	\frac{(r^{2}+a^{2})^{2}-a^{2}\Delta}{4\delta(1-\delta)r^{7}\Delta}
	\left\{
	\Delta\left[r(Mr-Q^{2})(1-a\Omega)^{2}-r^{5}\Omega^{2}\right]^{2}
	\right\}^{1/2}.
\end{equation}

\vspace{0.2cm}
Therefore, the condition \(\alpha=0\) yields

\begin{equation}
	(Mr-Q^{2})(1-a\Omega)^{2}-r^{4}\Omega^{2}=0,
\end{equation}

\vspace{0.2cm}
whose solutions are

\begin{equation}
	\Omega_{kep}= \Omega_{\rm \phi} = \frac{\pm\sqrt{Mr-Q^{2}}}{r^{2}\pm a\sqrt{Mr-Q^{2}}},
\end{equation}

\vspace{0.2cm}
which is precisely the Keplerian frequency (see Eq.~\eqref{omeg_phi}) for equatorial circular geodesics in the Kerr-Newman spacetime. Hence, when the gyroscope moves along a circular geodesic with angular velocity $\Omega=\Omega_{kep}= \Omega_{\rm \phi}$, no external force is required to maintain its motion, and the proper acceleration vanishes identically. Thus, Eq.~\eqref{KN_acc_red} not only characterises the acceleration needed to hold a gyroscope on a stationary orbit in the Kerr-Newman spacetime but also consistently reproduces the geodesic limit in which the motion becomes force-free.

\section{Role of Quasi-Periodic Oscillations in Shaping Accretion Flows in Kerr–Newman Geometry}\label{sec_5}

To investigate the properties of an accretion disk surrounding a charged rotating black hole, it is essential to analyse the stable circular geodesics in the vicinity of the spacetime under a strong-gravity regime. The innermost stable circular orbit (ISCO) defines the smallest radius at which such orbits remain stable and therefore determines the inner edge of the accretion disk. In this section, we investigate the characterisation of these orbits through the three fundamental frequencies as follows: (i) the orbital (Keplerian) frequency $(\nu_\phi)$, i.e., the inverse of the orbital period; (ii) the radial epicyclic frequency $(\nu_r)$, describing small radial oscillations about the mean circular orbit; and (iii) the vertical epicyclic frequency $(\nu_\theta)$, describing small latitudinal (polar) oscillations about the equilibrium plane. They are derived from the conditions for circular geodesics and their linear perturbations, together with the conserved specific energy and angular momentum. Moreover, we employ the effective potential formalism. These frequencies play a central role in accretion-disk physics and in modelling quasiperiodic oscillations (QPOs), which can serve as probes of the strong-gravity regime near black holes~\citep{Abramowicz_2013}. Their characteristic values depend on the structure of the background metric and on the orbital radius. Analytic expressions for the epicyclic frequencies have been derived repeatedly for Schwarzschild, Kerr, Kerr–MOG, and Hartle–Thorne spacetimes \citep{Abramowicz_2005,Pradhan_2019,2005ragt.meet..315T,Urbancov__2019}. To the best of our knowledge, however, closed-form formulae for the Kerr–Newman case have not been reported. Motivated by the results of the preceding studies that established analytic expressions for the epicyclic frequencies in various black hole geometries, we extend the analysis to the Kerr-Newman black hole and derive the corresponding closed-form expressions.

\vspace{0.2cm}
\noindent
In Newtonian gravity, the three characteristic frequencies have the same value when the potential is $\Phi=-\frac{M}{r}$, i.e.,
\begin{equation}
\nu_{\phi}=\nu_{\theta}=\nu_{r}=\frac{M}{r^{3/2}}\,.
\end{equation}

\vspace{0.2cm}
\noindent 
The equality of these three characteristic frequencies implies that, in the potential $\Phi=-\tfrac{M}{r}$, the orbits are periodic and closed. To derive the corresponding frequencies for the Kerr-Newman black hole, we first recall the geodesic motion of a test particle. Thus, the Lagrangian of the test particle becomes
\begin{equation}
\mathcal{L}=\frac{1}{2}\,g_{\mu\nu}(x)\,\dot{x}^{\mu}\dot{x}^{\nu},
\end{equation}

\vspace{0.2cm}
\noindent
where an overdot denotes differentiation with respect to the proper time $\tau$ for timelike geodesics and with respect to an affine parameter $\lambda$ for null geodesics. Now, the general stationary and axisymmetric spacetime metric can be expressed as

\begin{equation}
	ds^2=g_{tt}dt^2+g_{rr}dr^2+g_{\theta\theta}d\theta^2+g_{\phi\phi}d\phi^2+2g_{t\phi}d\phi dt,
\end{equation}

\vspace{0.2cm}
\noindent
The metric coefficients are independent of $t$ and $\phi$, i.e.\ the spacetime is stationary and axisymmetric in the canonical form. Hence, it admits the Killing vectors $\partial_t$ and $\partial_\phi$, and along any geodesic there are two conserved quantities: the specific energy $e \equiv -p_t$ and the specific (azimuthal) angular momentum $\ell \equiv p_\phi$. Thus, the Lagrange equation of motion gives

\begin{equation}
	\begin{array}{rcl}p_t&\equiv&\frac{\partial\mathcal{L}}{\partial\dot{t}}=g_{tt}\dot{t}+g_{t\phi}\dot{\phi}=-e\end{array}.
\end{equation}

\begin{equation}
	\begin{matrix}
    p_\phi&\equiv&\dfrac{\partial\mathcal{L}}{\partial\dot{\phi}}=g_{t\phi}\dot{t}+g_{\phi\phi}\dot{\phi}=\ell
    \end{matrix}.
\end{equation}

Solving the above equations, we have the four-velocity components of $t$ and $\phi$ are

\begin{equation}
	\begin{aligned}
		\dot{\phi}&=&-\frac{(e\:g_{t\phi}+\ell\:g_{tt})}{z}, \\[6pt]
		\dot{t}&=&\dfrac{(e\:g_{\phi\phi}+\ell\:g_{t\phi})}{z},
	\end{aligned}
\end{equation}

where $z=g_{t\phi}^2-g_{tt}g_{\phi\phi}$ From the conservation of rest mass $g_{\mu\nu}u^{\mu}u^{\nu}=-1$, we have

\begin{equation}
	\begin{array}{rcl}g_{rr}\dot{r}^2+g_{\theta\theta}\dot{\theta}^2&=&\mathcal{V}_{eff}(r,\theta,e,\ell).
    \end{array}
\end{equation}

Therefore, the effective potential can be written as

\begin{equation}
	\mathcal{V}_{eff}=\frac{(e^2+g_{tt})g_{\phi\phi}+(2e\ell-g_{t\phi})g_{t\phi}+\ell^2g_{tt}}{z}.
\end{equation}

For circular orbits in the equatorial plane one has $\dot{r}=\dot{\theta}=0$, which directly implies $\nu_{eff}=0$, and $\ddot{r}=\ddot{\theta}=0$ which gives $\partial_{r}\mathcal{V}_{eff}=0$ and $\partial_{\theta}\mathcal{V}_{eff}=0$ respectively. From these conditions, one can obtain the energy and angular momentum \citep{RiffertHerold1995} as

\begin{equation}
	e=-\frac{g_{tt}+\Omega_{\phi}g_{t\phi}}{\sqrt{-g_{tt}-2g_{t\phi}\Omega_{\phi}-g_{\phi\phi}\Omega_{\phi}^{2}}},
\end{equation}

and

\begin{equation}
	\begin{array}{rcl}\ell&=&\dfrac{g_{t\phi}+\Omega_\phi g_{\phi\phi}}{\sqrt{-g_{tt}-2g_{t\phi}\Omega_\phi-g_{\phi\phi}\Omega_\phi^2}}.\end{array}
\end{equation}

\vspace{0.2cm}
The orbital frequency of a test particle is defined as

\begin{equation}
	\Omega_\phi\equiv2\pi\nu_\phi=\frac{\dot{\phi}}{\dot{t}}=\frac{(\frac{d\phi}{d\tau})}{(\frac{dt}{d\tau})}=\frac{d\phi}{dt}=\frac{-\partial_rg_{t\phi}\pm\sqrt{(\partial_rg_{t\phi})^2-(\partial_rg_{tt})(\partial_rg_{\phi\phi})}}{\partial_rg_{\phi\phi}}.
\end{equation}

\vspace{0.2cm}
\noindent 
The $+$ sign corresponds to a corotating orbit, and the $-$ sign corresponds to a counterrotating orbit, respectively. If $\partial_{r}^{2}\mathcal{V}_{eff}\leq0$ $\partial_{\theta}^{2}\mathcal{V}_{eff}\leq0$ then the orbits are stable under small perturbations. The radial and vertical frequencies can be easily computed by considering small perturbations around equatorial circular orbits along the radial and vertical directions, respectively. If $\delta_{r}$ and $\delta_{\theta}$ are the small displacements around the mean orbit $(\mathrm{i.e.}r=r_0+\delta_r\mathrm{~and~}\theta=\frac{\pi}{2}+\delta_\theta)$, then we get the following second-order differential equations~\citep{Rayimbaev_2022,PhysRevD.108.044063} as 

\begin{equation}
	\frac{d^2\delta_r}{dt^2}+\Omega_r^2\delta_r=0,
\end{equation}

\begin{equation}
	\frac{d^2\delta_\theta}{dt^2}+\Omega_\theta^2\delta_\theta=0.
\end{equation}

\vspace{0.2cm}
Where the squared epicyclic frequencies associated with the radial and polar ($\theta$) motions~\citep{Bambi:2017khi} are given as 

\begin{equation}
	\Omega_{r}^{2}=-\frac{1}{2\dot{t}^{2}g_{rr}}\partial_{r}^{2}\mathcal{V}_{eff},
\end{equation}

\begin{equation}
	\Omega_{\theta}^{2}=-\frac{1}{2\dot{t}^{2}g_{\theta \theta}}\partial_{\theta}^{2}\mathcal{V}_{eff}.
\end{equation}

\vspace{0.3cm}
\noindent 
The radial epicyclic frequency is therefore $\nu_{r}=\frac{\Omega_{r}}{2\pi}$ and the vertical one is $\nu_{\theta}=\frac{\Omega_{\theta}}{2\pi}$~\citep{PhysRevD.90.044004}. For the Kerr-Newman black hole, the Kepler frequency is derived to be

\begin{equation}
\Omega_\phi=\pm \frac{\sqrt{M-\dfrac{Q^{2}}{r}}}{r^{3/2}\pm a\sqrt{M-\dfrac{Q^{2}}{r}}}. \label{omeg_phi}
\end{equation}

\vspace{0.2cm}
After some long algebraic steps, one can obtain the radial epicyclic frequencies $\Omega_{r}$ for the prograde and retrograde rotation as

\begin{equation}
\Omega_r=\Omega_\phi\left(
\frac{1-\dfrac{6M}{r}+\dfrac{9Q^2}{r^2}-\dfrac{3a^2}{r^2}+\dfrac{4Q^2\!\left(a^2-Q^2\right)}{M r^3}}{1-\dfrac{Q^2}{Mr}}\;\pm\;\frac{8a}{r^{3/2}}\sqrt{\,M-\frac{Q^2}{r}\,}\right)^{1/2}. \label{omeg_r}
\end{equation}

\vspace{0.2cm}
\noindent
Furthermore, the conditions $\Omega_{r}^{2}\geq0$ and $\Omega_{\theta}^{2}\geq0$ imply the stability of the circular geodesic motions against small oscillations. A non-negative value of $\Omega_{\theta}$ implies stability of the geodesic motion against
small vertical (out-of-plane) perturbations. By imposing the marginal stability condition for equatorial circular motion, namely $\Omega_{r}^{2}=0$, we obtain the defining relation for the ISCO in the Kerr-Newman black hole. For example, it is well known that for a Schwarzschild black hole, the ISCO is located at
$r_{\mathrm{ISCO}}=6M$, whereas for an extremal Kerr black hole, it is located at $r_{\mathrm{ISCO}}=M$ for prograde (direct) orbits and at $r_{\mathrm{ISCO}}=9M$ for retrograde orbits \citep{Bardeen1972}. Therefore, the corresponding ISCO equation for the Kerr-Newman black hole can be expressed as

\begin{equation}
   r^{2}\!\left(r^{2}-6Mr+9Q^{2}\right)+4Q^{2}\!\left(Q^{2}-2Mr\right)
\mp 8a\left(Mr-Q^{2}\right)^{3/2}-3a^{2}\!\left(Mr-Q^{2}\right)=0.
\end{equation}

\begin{figure}[h!]
    \centering
    \subfigure[]{%
        \includegraphics[width=7.3cm,height=7.7cm]{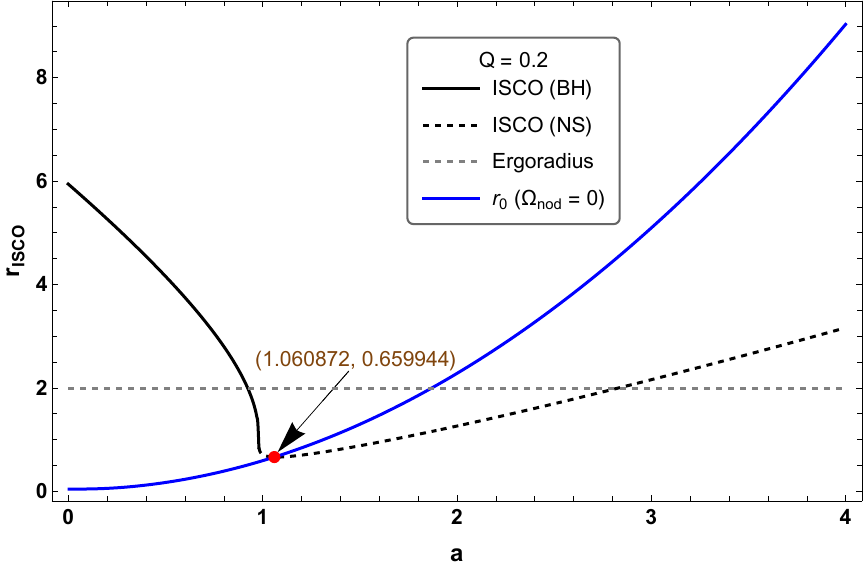}%
        \label{isco_1}}\hspace{0.8cm}
    \subfigure[]{%
        \includegraphics[width=7.3cm,height=7.7cm]{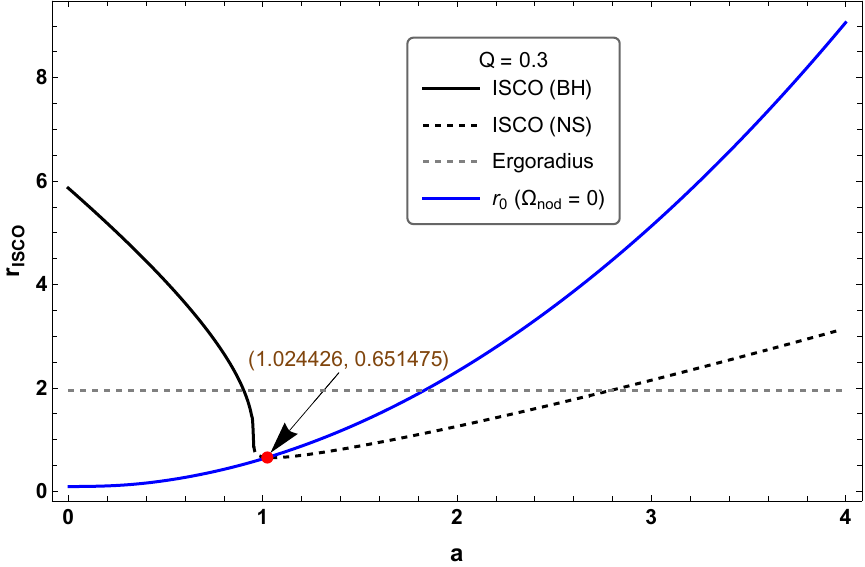}%
        \label{isco_2}}\vspace{0.8em}
    \subfigure[]{%
        \includegraphics[width=7.3cm,height=7.7cm]{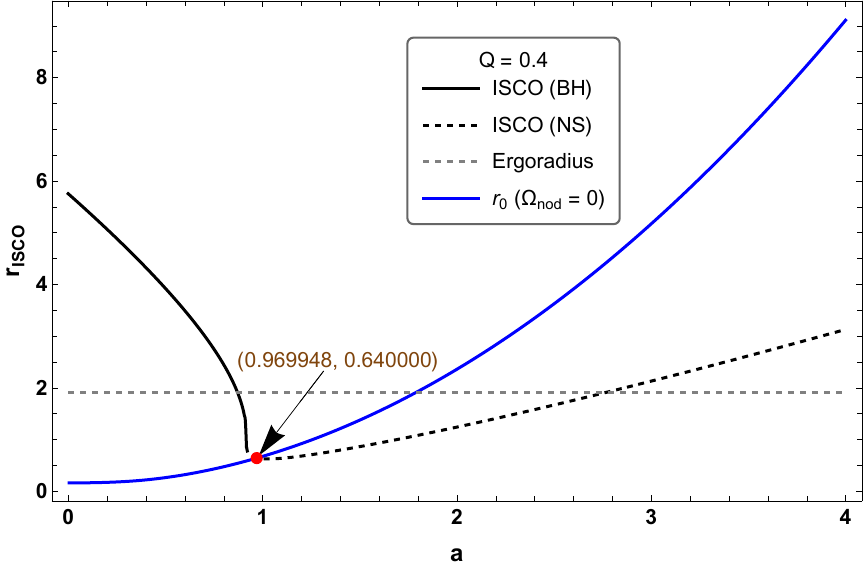}%
        \label{isco_3}}\hspace{0.8cm}
    \subfigure[]{%
        \includegraphics[width=7.3cm,height=7.7cm]{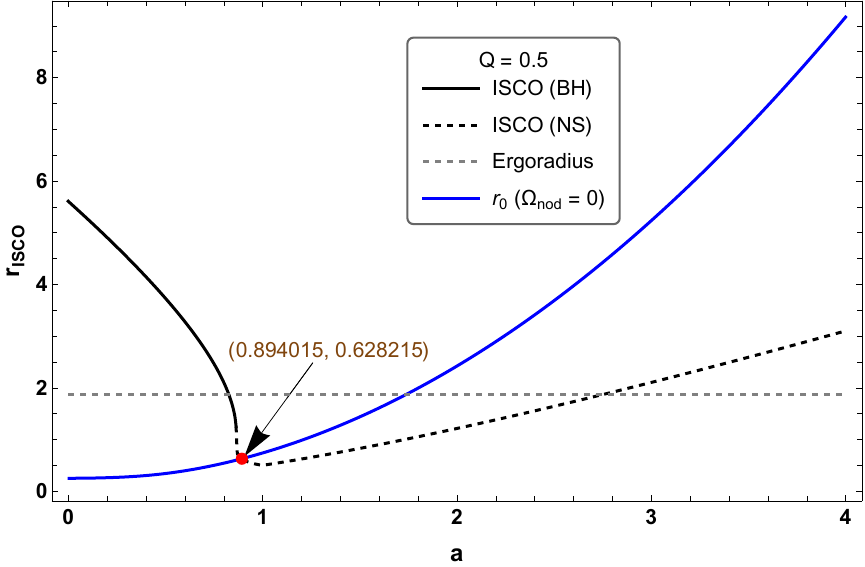}%
        \label{isco_4}}
    \caption{Illustration of Kerr-Newman ISCO (in units of $M$) for prograde orbits versus the spin parameter $a$ (in units of $M$), for black hole (solid curves) and naked singularity (dashed curves) configurations corresponding to different values of the charge parameter $Q$. The horizontal dotted lines (grey) represent the ergoradius (in units of $M$) for different values of $Q$. The blue curve represents the radius $r_0$ at which the nodal precession frequency, $\Omega_{\mathrm{nod}}$, vanishes.}
    \label{r_isco}
\end{figure}

\vspace{0.3cm}
\noindent
Since we restrict our analysis to the equatorial plane, i.e. $\theta=\pi/2$, the periastron precession frequency~\citep{Bambi:2017khi} for prograde and retrograde orbits can be defined as

\begin{equation}
	\Omega_{per}=\Omega_\phi-\Omega_r,
\end{equation}

\vspace{0.2cm}
which can be further expressed as

\begin{align}
\Omega_{per}
=\pm \frac{\sqrt{M-\dfrac{Q^{2}}{r}}}{\,r^{3/2}\pm a\sqrt{M-\dfrac{Q^{2}}{r}}\,}
\Bigg\{
1-&\Bigg[
\frac{
1-\dfrac{6M}{r}+\dfrac{9Q^{2}}{r^{2}}-\dfrac{3a^{2}}{r^{2}}
+\dfrac{4Q^{2}(a^{2}-Q^{2})}{Mr^{3}}
}{
1-\dfrac{Q^{2}}{Mr}
}
\nonumber\\[8pt]
&\hspace{3.5cm}
\pm \frac{8a}{r^{3/2}}\sqrt{M-\frac{Q^{2}}{r}}
\Bigg]^{1/2}
\Bigg\}.\label{omeg_par}
\end{align}

\vspace{0.2cm}
In an analogous manner, the vertical epicyclic frequency $\Omega_{\theta}$ corresponding to prograde and retrograde motion is given by

\begin{equation}
\Omega_\theta
=\Omega_\phi\left(
1
\mp \frac{4a}{r^{3/2}}\sqrt{M-\frac{Q^{2}}{r}}
+\frac{3a^{2}}{r^{2}}
\mp \frac{2aQ^{2}}{r^{5/2}\sqrt{M-\frac{Q^{2}}{r}}}
+\frac{a^{2}Q^{2}}{r^{3}\left(M-\frac{Q^{2}}{r}\right)}
\right)^{1/2}. \label{omeg_the}
\end{equation}

\vspace{0.4cm}
\noindent 
Here, the $\pm$ sign corresponds to a prograde and retrograde orbit, respectively. Having obtained the Keplerian frequency and the vertical epicyclic frequency above, the nodal-plane precession frequency (NPPF) follows straightforwardly from their difference \citep{Bambi:2017khi}. This frequency is also referred to as the orbital-plane precession frequency, or equivalently, the LT precession frequency of a test particle. The expression can be read as

\begin{equation}
    \Omega_{nod} = \Omega_{\phi} - \Omega_{\theta}.
\end{equation}

\vspace{0.2cm}
After some algebraic manipulations, the above expression can be further expressed as

\begin{align}
\Omega_{nod}
=\pm \frac{\sqrt{\,M-\dfrac{Q^{2}}{r}\,}}
{\,r^{3/2}\pm a\sqrt{\,M-\dfrac{Q^{2}}{r}\,}\,}
\Bigg\{
1-\Bigg[
1
&\mp \frac{4a}{r^{3/2}}\sqrt{\,M-\frac{Q^{2}}{r}\,}
+\frac{3a^{2}}{r^{2}} \mp \frac{2aQ^{2}}{r^{5/2}\sqrt{\,M-\frac{Q^{2}}{r}\,}}\nonumber\\[8pt]
&\hspace{3.6cm} +\frac{a^{2}Q^{2}}{r^{3}\left(M-\dfrac{Q^{2}}{r}\right)}
\Bigg]^{1/2}
\Bigg\},\label{omeg_nod}
\end{align}

where $\pm$ represents the prograde and retrograde rotation, respectively.

\vspace{0.2cm}
\noindent
Orbital-plane precession is an intrinsically rotational effect, i.e., in the nonrotating limit, the azimuthal and vertical frequencies coincide ($\Omega_{\phi}=\Omega_{\theta}$), so LT (nodal) precession disappears. In contrast, periastron precession is not exclusive to rotation and persists even in non-rotating spacetimes. A key stability indicator is the radial epicyclic frequency, for which $\Omega_r^{2}$ falls to zero at the ISCO and turns negative for smaller radii, signaling the onset of radial instability for circular orbits. For $r>r_{\mathrm{ISCO}}$, $\Omega_r^{2}$ remains positive, while $\Omega_{\theta}^{2}$ stays positive and finite in rotating geometries. The NPPF $\Omega_{\mathrm{nod}}$, however, does not follow an equally uniform trend and can display nontrivial radial structure, including the possibility of sign reversal depending on the black hole parameters.

\begin{figure}[h!]
\centering
    \subfigure[]{\includegraphics[width=7.3cm,height=7.7cm]{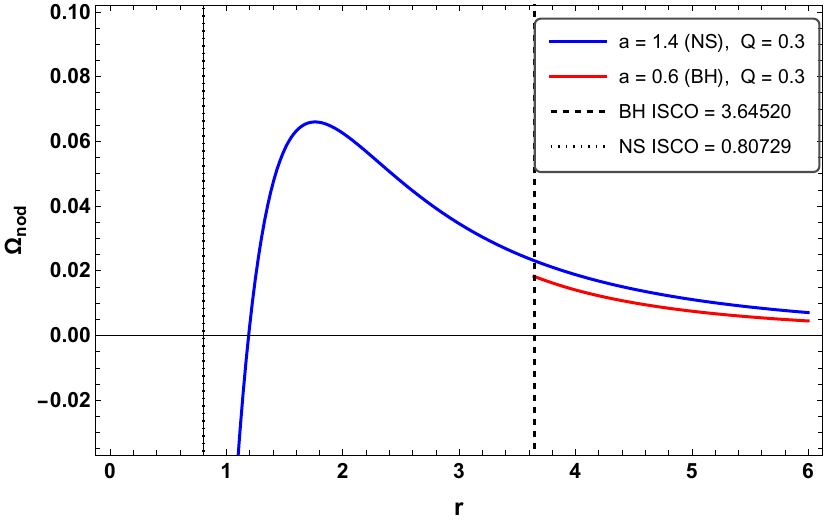}\label{f_1}}\hspace{0.8cm}
	\subfigure[]{\includegraphics[width=7.3cm,height=7.7cm]{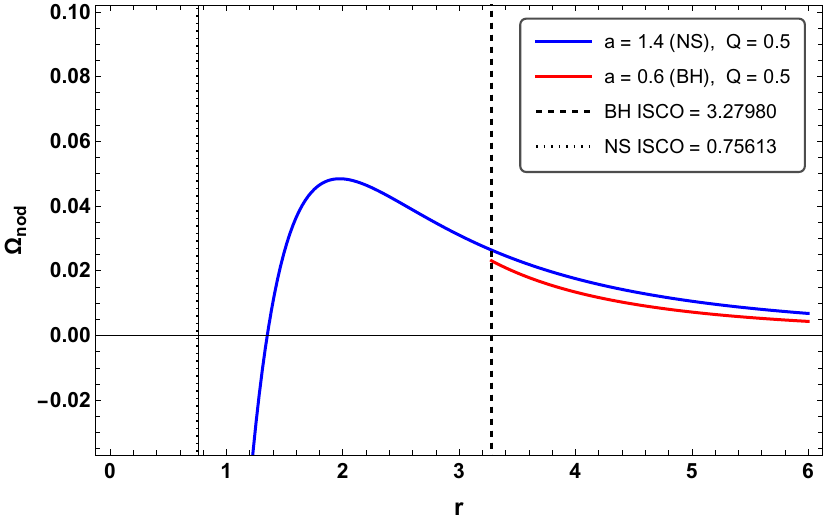}\label{f_2}}
    \caption{Illustration showing the variation of the NPPF $\Omega_{\mathrm{nod}}$ (in units of $M^{-1}$) versus the radial coordinate $r$ (in units of $M$) for different values of charge parameters $Q = 0.3$ and $Q = 0.5$. In the black-hole regime, $\Omega_{\mathrm{nod}}$ decreases monotonically with increasing $r$. By contrast, in the naked-singularity regime, it typically exhibits a peak and then falls, vanishing at a finite radius $r_{0}$. Negative values of $\Omega_{\mathrm{nod}}$ indicate a reversal in the direction of precession.}\label{nod_1}
\end{figure}

\begin{figure}[h!]
\centering
    \subfigure[]{\includegraphics[width=7.3cm,height=7.7cm]{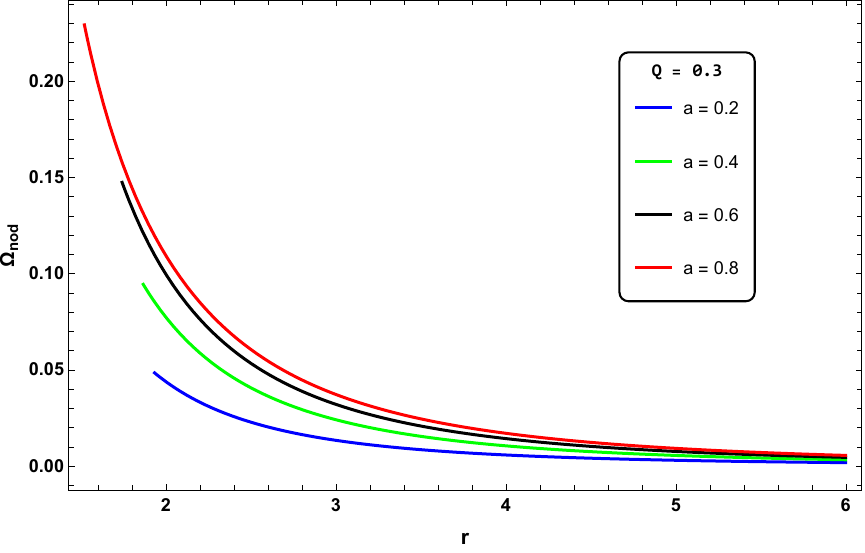}\label{r_1}}\hspace{0.8cm}
	\subfigure[]{\includegraphics[width=7.3cm,height=7.7cm]{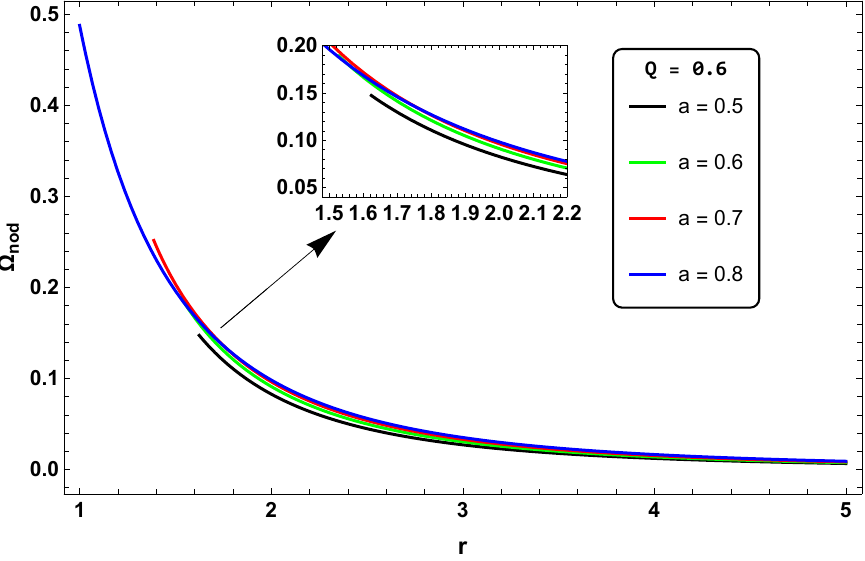}\label{r_2}}
    \caption{Variation of the NPPF $\Omega_{\mathrm{nod}}$ (in units of $M^{-1}$) versus the radial coordinate $r$ (in units of $M$) for the black hole case with charge parameters $Q = 0.3$ and $Q = 0.6$. In both cases, $\Omega_{\mathrm{nod}}$ exhibits a monotonic decrease with increasing $r$ for both values of the charge parameter, indicating a systematic weakening of nodal precession effects at larger radial distances.}\label{nod_2}
\end{figure}

\vspace{0.2cm}
Figure~(\ref{r_isco}) shows that, for prograde orbits, the ISCO radius decreases with increasing spin parameter $a$ for both black hole and naked singularity configurations up to the (critical) values $a_{c}=1.060872,\: 1.024426,\: 0.969948,$ and $0.894015$ for $Q=0.2, 0.3, 0.4,$ and $0.5$, respectively; beyond these values, it reverses its trend and increases with further increase in $a$~\citep{Chakraborty:2016mhx,10.1007/978-3-319-06761-2_61,Stuchlik1980}. Therefore, the ISCO reaches its minimum radius, $r_{ISCO} = (0.659944,\: 0.651475,\; 0.640000$, and $0.628215)$, which occurs for the parameter value $a_{c} = 1.060872,\: 1.024426,\: 0.969948,$ and $0.894015$ corresponds to $Q=0.2, 0.3, 0.4,$ and $0.5$. Further, it is shown in Figure~(\ref{r_isco}) that the ISCO lies on or within the ergosurface. For each value of $a$ and $Q$, there exists a threshold radius $r_0$ at which the frame-dragging effect vanishes, and consequently, the LT precession becomes zero. In the black hole regime, the LT precession frequency increases as the orbital radius $r$ decreases, rising continuously down to the inner boundary of the accretion disc (see Figures~(\ref{nod_1}) and (\ref{nod_2})). In contrast, for a naked singularity configurations the LT frequency exhibits a non-monotonic behaviour: it reaches a maximum at $r=r_{p}$, with $r_{p}>r_{ISCO}$ (see Table~(\ref{tab:KN_Q0p2} - \ref{tab:KN_Q0p4})), and subsequently decreases as $r$ is reduced further (see Figures~(\ref{nod_1}) and (\ref{nod_2})). As illustrated in Figures~(\ref{r_isco}) and (\ref{nod_1})), for \( a > a_c \) the LT precession frequency becomes negative in the radial range $r_{\mathrm{ISCO}} \leq r < r_0$, indicating a reversal in the direction of LT precession. Therefore, these characteristics provide important diagnostics for distinguishing between black hole and naked singularity spacetimes in the Kerr–Newman geometry.

\begin{figure}[h!]
    \centering
    \subfigure[]{\includegraphics[width=7.3cm,height=7.7cm]{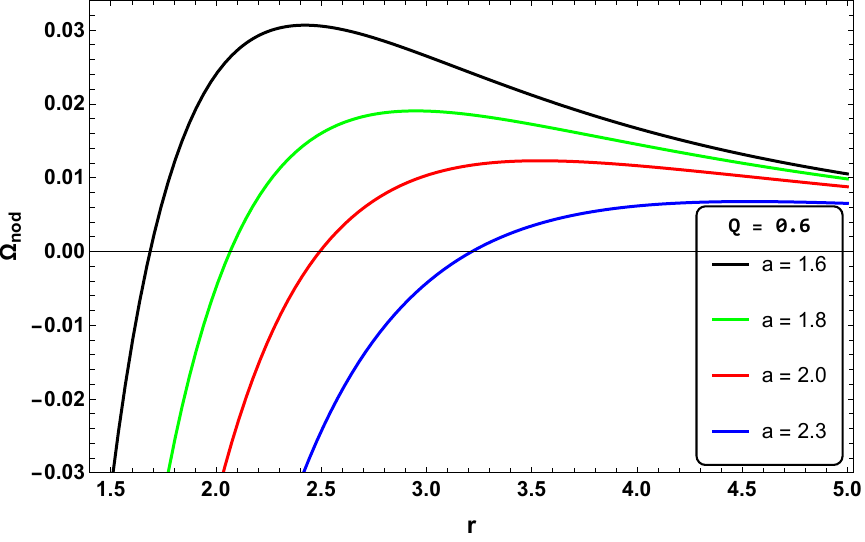}\label{r_2}}\hspace{0.8cm}
    \subfigure[]{\includegraphics[width=7.3cm,height=7.7cm]{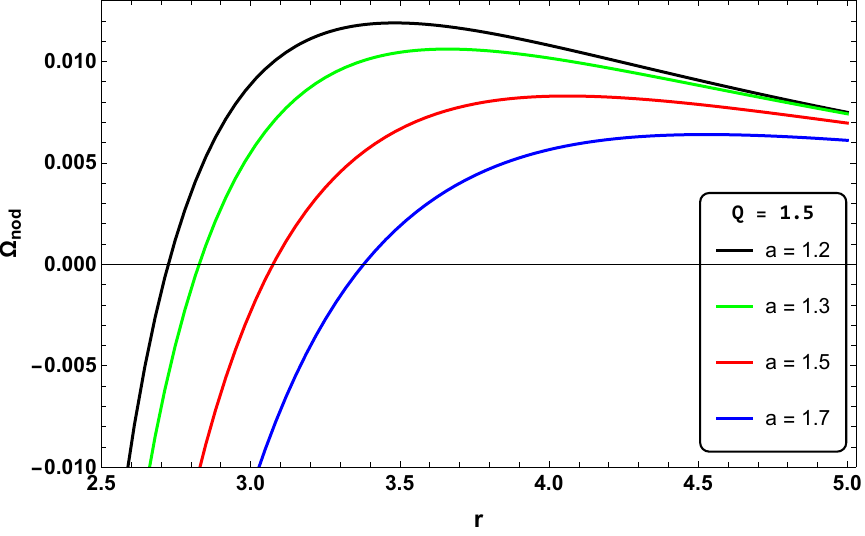}\label{r_2}}
    \caption{Variation of the NPPF $\Omega_{\mathrm{nod}}$ (in units of $M^{-1}$) versus the radial coordinate $r$ (in units of $M$) for naked singularity configurations with $Q=0.6$ and $Q=1.5$. The frequency initially increases with $r$, attains a characteristic maximum, and subsequently decreases for larger radii. The emergence of negative values of $\Omega_{\mathrm{nod}}$ reflects a reversal in the direction of nodal precession.}\label{nod_3}
\end{figure}

\vspace{0.2cm}
\noindent
One can readily verify that, in the limiting case $Q\rightarrow 0$, the characteristic frequencies~\cref{omeg_phi,omeg_r,omeg_the} smoothly reduce to their standard Kerr counterparts~\citep{Chakraborty:2016mhx,Motta2013wga,Bambi_2015}. From Figures~(\ref{nod_1} - \ref{nod_3}), it is apparent that, in the black hole regime, the NPPF $\Omega_{nod}$ exhibits a strictly monotonic decay with increasing orbital radius $r$. This behaviour can be summarised by the following condition:

\begin{equation}
\left.\frac{d\Omega_{\mathrm{nod}}}{dr}\right|_{r_{\mathrm{ISCO}}} < 0
\qquad \text{(BH regime)}.
\end{equation}

\vspace{0.2cm}
\noindent
In contrast to the black hole case, naked singularity configurations exhibit a non-monotonic NPPF profile: $\Omega_{\mathrm{nod}}(r)$ rises from its inner-orbit value, reaches a peak at $r=r_{\max}$, and then decays with increasing $r$. Thus, we can have the following condition in the case of a naked singularity.

\begin{equation}
\left.\frac{d\Omega_{\mathrm{nod}}}{dr}\right|_{r_{\mathrm{ISCO}}} > 0
\qquad \text{(NS regime)}.
\end{equation}

\vspace{0.2cm}
\noindent
Finally, we note that negative values of $\Omega_{nod}$ carry a clear physical meaning: they signal a reversal of the nodal precession direction with respect to the adopted reference orientation.

\begin{figure}[h!]
\centering
    \subfigure[]{\includegraphics[width=7.3cm,height=7.7cm]{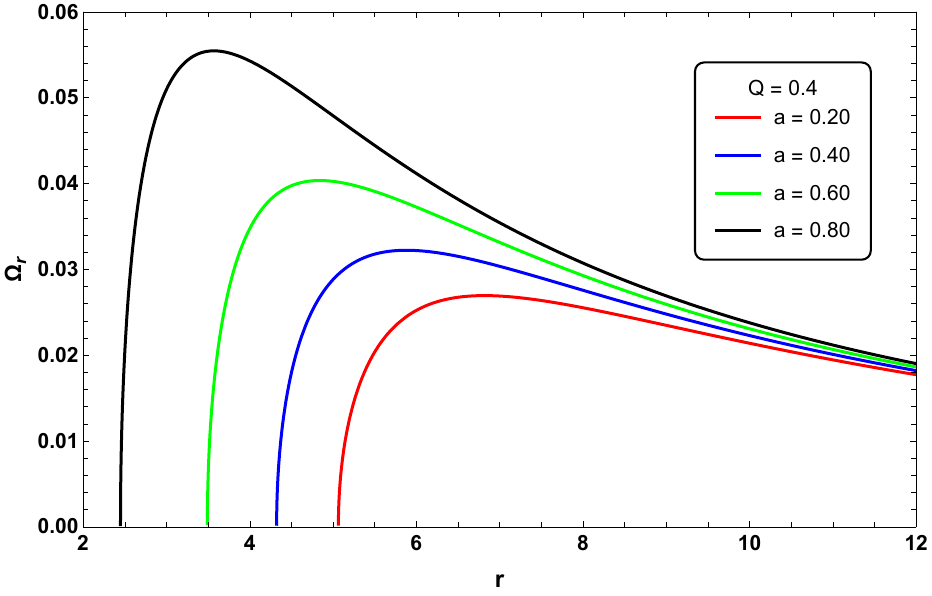}\label{r_1}}\hspace{0.8cm}
	\subfigure[]{\includegraphics[width=7.3cm,height=7.7cm]{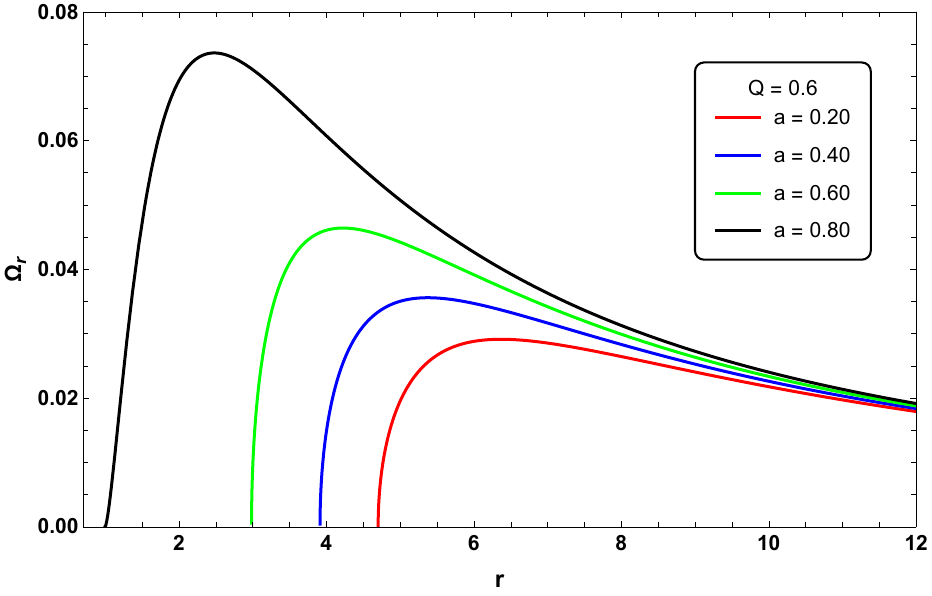}\label{r_2}}\vspace{0.8em}
    \subfigure[]{\includegraphics[width=7.3cm,height=7.8cm]{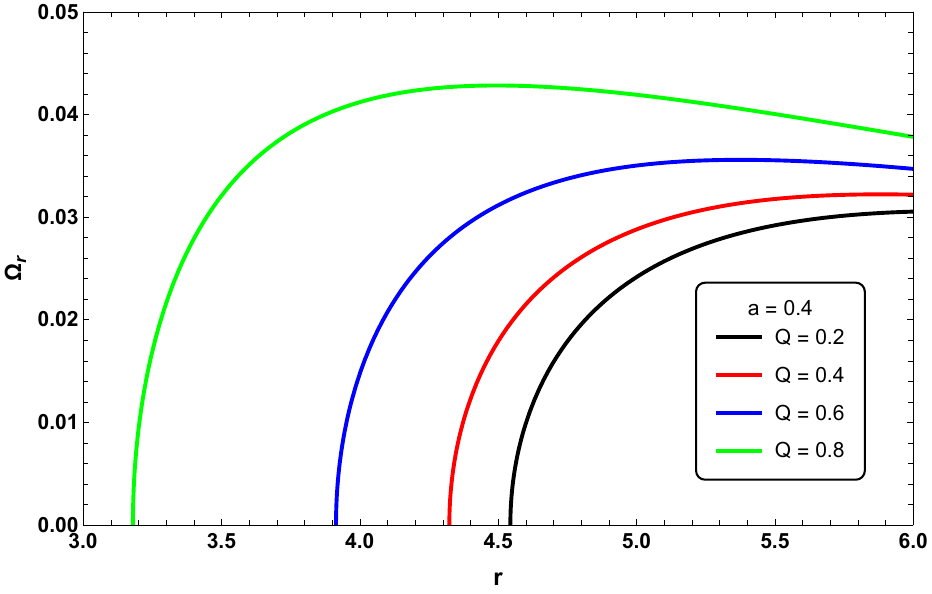}\label{r_3}}\hspace{0.8cm}
    \subfigure[]{\includegraphics[width=7.3cm,height=7.7cm]{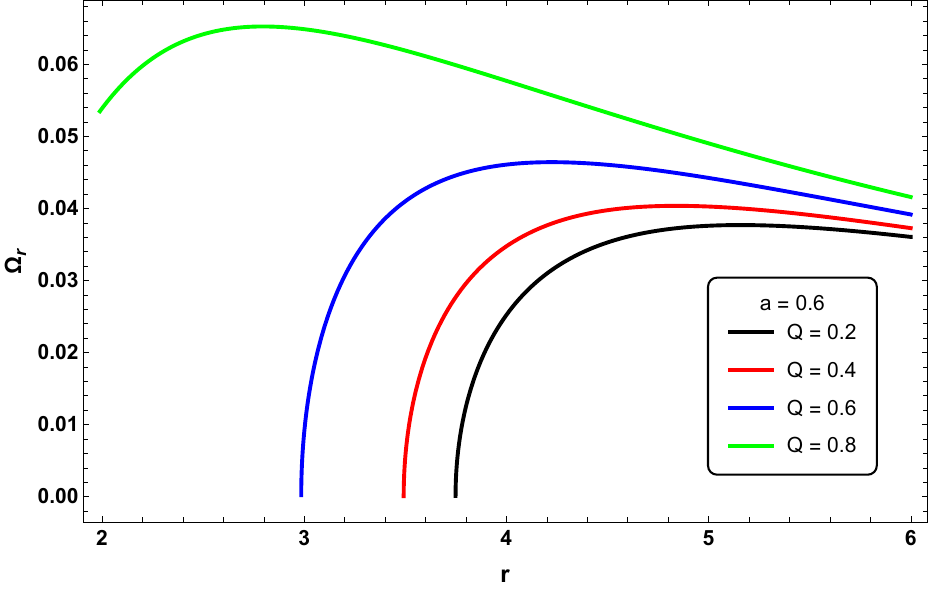}\label{r_4}}
    \caption{Variation of the radial epicyclic frequency $\Omega_{r}$ (in unit of $M^{-1}$) versus the radial coordinate $r$ (in unit of $M$) for different choices of the spin parameter $a$ and charge parameter $Q$ in the black hole configuration. For each parameter set, $\Omega_{r}$ vanishes at a characteristic radius, signalling the onset of radial instability, and then rises to a maximum before decreasing with increasing $r$.}\label{nod_4}
\end{figure}

\begin{figure}[h!]
\centering
    \subfigure[]{\includegraphics[width=7.3cm,height=7.7cm]{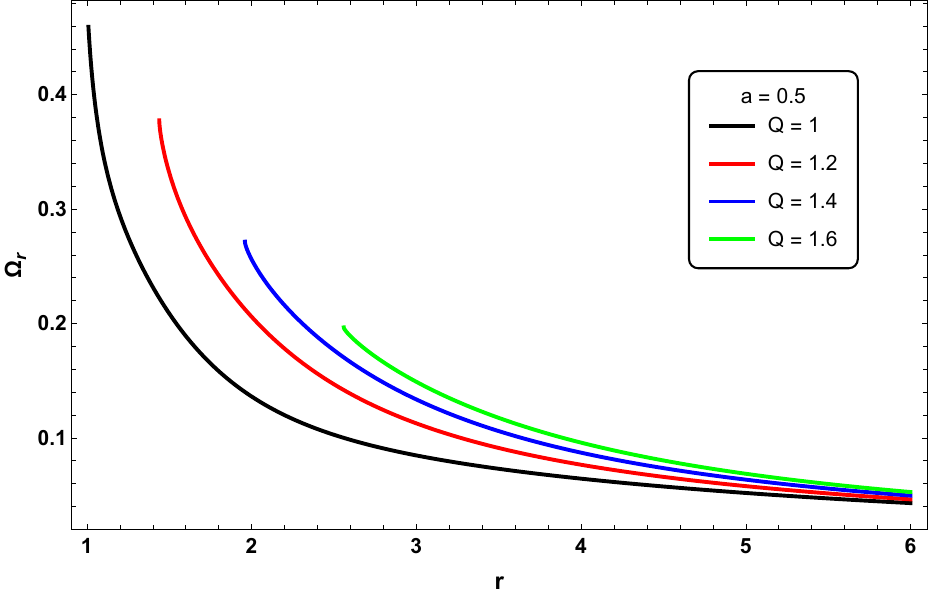}\label{r_1}}\hspace{0.8cm}
	\subfigure[]{\includegraphics[width=7.3cm,height=7.7cm]{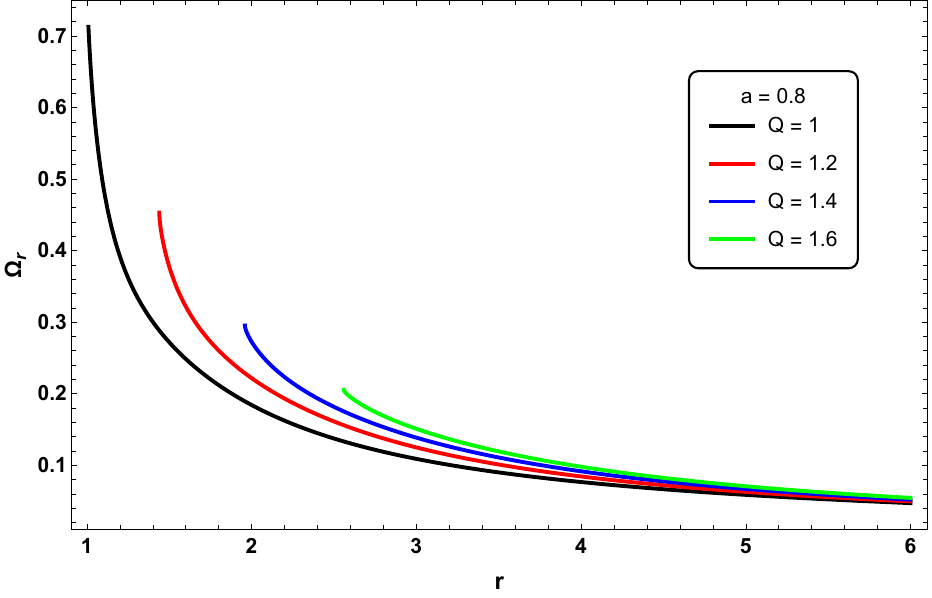}\label{r_2}}\vspace{0.8em}
    \subfigure[]{\includegraphics[width=7.3cm,height=7.8cm]{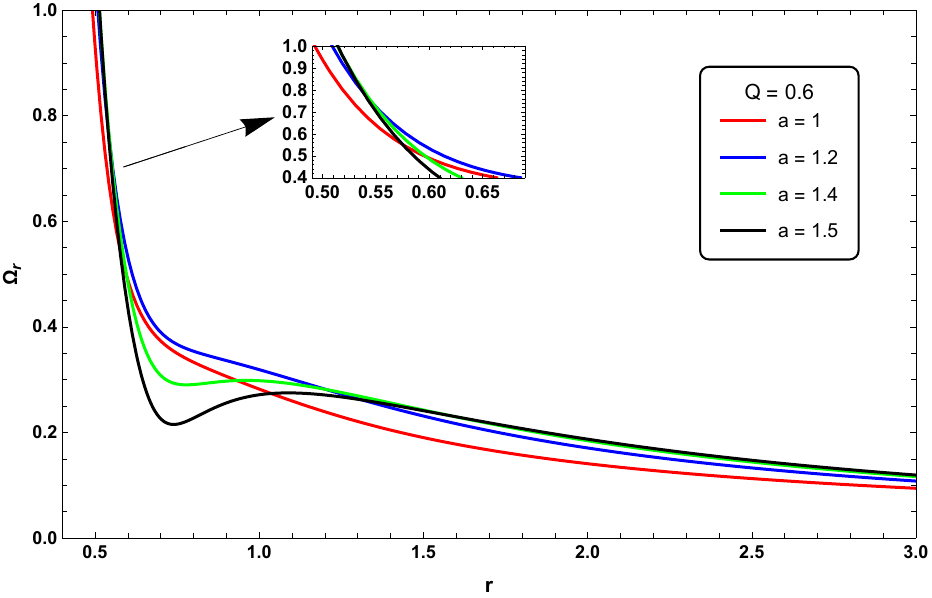}\label{r_3}}\hspace{0.8cm}
    \subfigure[]{\includegraphics[width=7.3cm,height=7.7cm]{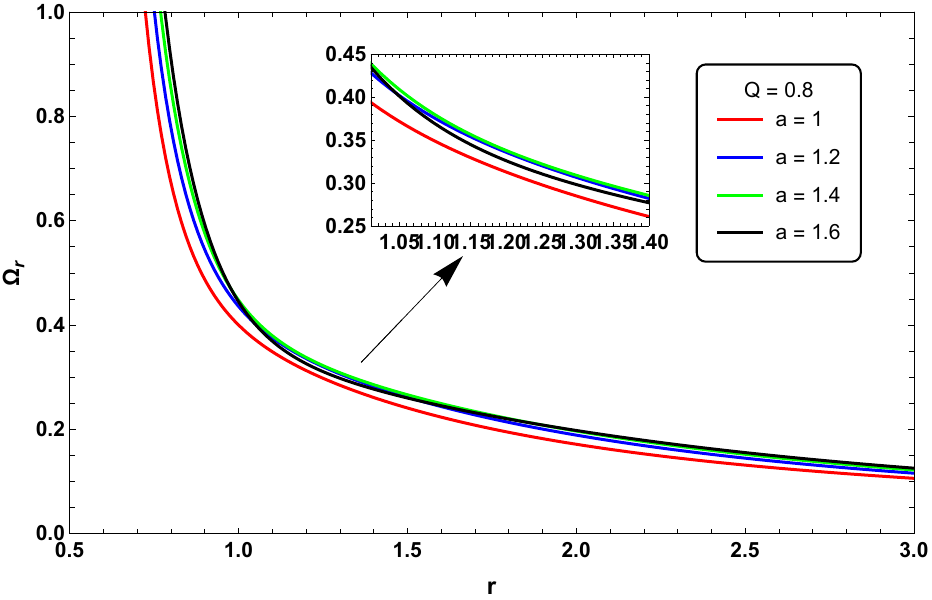}\label{r_4}}
    \caption{Variation of the radial epicyclic frequency $\Omega_{r}$ (in unit of $M^{-1}$) versus the radial coordinate $r$ (in unit of $M$) for different choices of the spin parameter $a$ and charge parameter $Q$ for the naked singularity configuration.}\label{nod_5}
\end{figure}

\begin{figure}[h!]
	\centering
	\subfigure[]{\includegraphics[width=7.3cm,height=7.7cm]{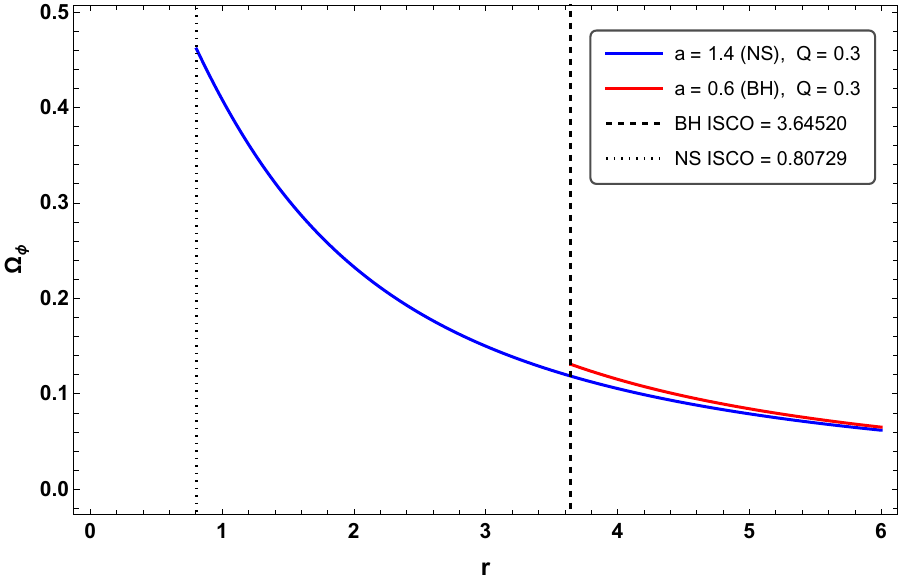}\label{r_2}}\hspace{0.8cm}
	\subfigure[]{\includegraphics[width=7.3cm,height=7.7cm]{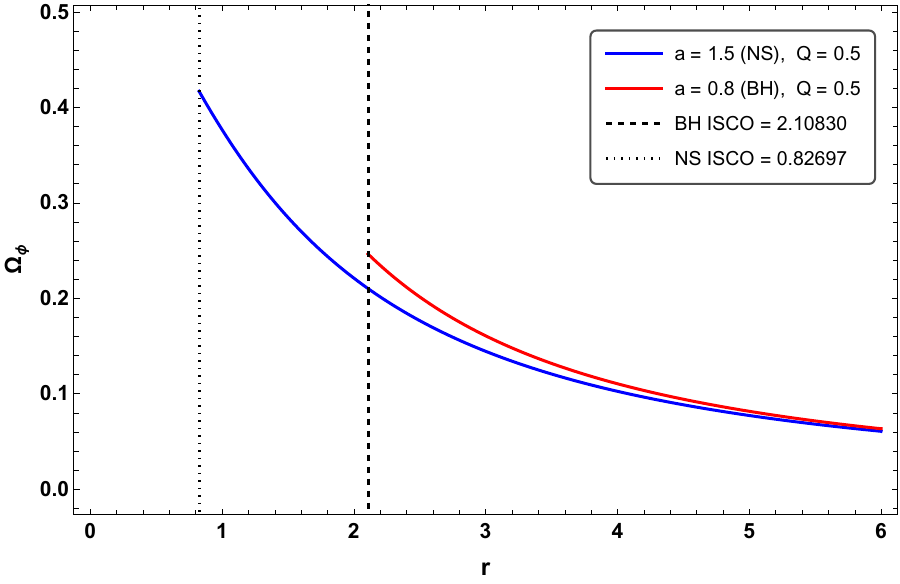}\label{r_2}}
	\caption{Variation of the $\Omega_{\phi}$ (in units of $M^{-1}$) versus the radial coordinate $r$ (in units of $M$) for different choices of the spin parameter $a$ and charge parameter $Q$. The plots clearly indicate that the Keplerian frequency ($\Omega_{\phi}$) at the ISCO is significantly higher for a naked singularity than for a black hole.}\label{phi_1}
\end{figure}

\vspace{0.2cm}
\noindent
The remaining fundamental frequencies, namely $\Omega_{\phi}$, $\Omega_{r}$, $\Omega_{\theta}$ and the periastron-precession frequency $\Omega_{per}$, likewise exhibit characteristic trends that can differ between black hole and naked singularity configurations, thereby providing additional discriminants between the two spacetime geometries. In particular, from the $\Omega_r(r)$ profiles, one finds that the charge parameter $Q$ leaves a qualitatively different imprint on the black hole and naked singularity branches, primarily through its control of the innermost admissible radius for circular motion (hence the effective inner edge of the accretion flow). For the black hole configuration (see Figure~\ref{nod_4}), increasing $Q$ shifts the onset of $\Omega_r$ (where it vanishes at the marginally stable orbit) to smaller radii, i.e., the ISCO moves inward; simultaneously, the peak of $\Omega_r$ becomes higher and occurs closer to the compact object, indicating stronger radial restoring oscillations in the inner disk and thus the possibility of higher-frequency epicyclic activity sourced from deeper regions. However, for Kerr-Newman naked singularities (see Figure~\ref{nod_5}), the trend is reversed in an operational sense: larger $Q$ pushes the inner cutoff of the physically allowed circular-orbit region outward, so the disk cannot probe the smallest radii where the strong-field behaviour is most extreme; while $\Omega_r$ at a fixed radius can be modestly enhanced, the dominant effect of increasing $Q$ is to truncate the inner disk and to shift the locations of any sharp features (rapid turnover) in $\Omega_r$ to larger radii, thereby reducing the contribution of ultra-compact radii to the disk's dynamical variability. More importantly, Figure~(\ref{phi_1}) indicates that the Keplerian (azimuthal) frequency $\Omega_{\phi}$ follows a broadly similar qualitative trend for both black hole and naked singularity configurations. The effect of the charge parameter becomes increasingly pronounced as ($Q$) is raised, leading to a substantial modification of the corresponding frequency values and their radial profiles. The charge parameter ($Q$) primarily affects the strong-field region by shifting the ISCO and suppressing the inner-disc orbital frequency $\Omega_\phi$. As ($Q$) increases (compare $Q=0.3$ to $Q=0.5$), the accessible maximum $\Omega_\phi$ near the inner edge is reduced, while the large-radius behaviour remains nearly unchanged, indicating that the effect of charge is predominantly a near-ISCO (near-horizon) modification. Nevertheless, the qualitative black hole-naked singularity contrast persists because it is controlled mainly by how far the disc can extend inward: naked singularity configurations can probe smaller radii and hence achieve larger $\Omega_\phi$ than black holes even in the presence of nonzero ($Q$). 

\begin{figure}[h!]
	\centering
	\subfigure[]{\includegraphics[width=7.3cm,height=7.7cm]{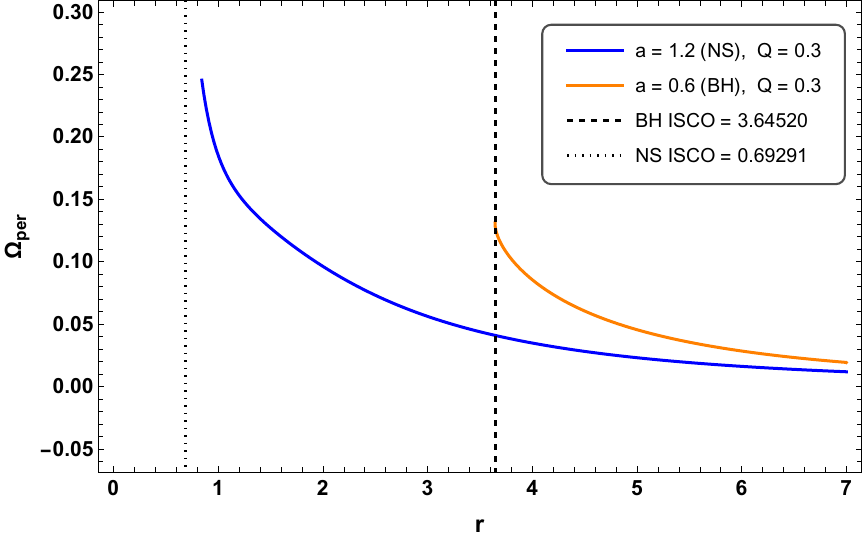}\label{r_2}}\hspace{0.8cm}
	\subfigure[]{\includegraphics[width=7.3cm,height=7.7cm]{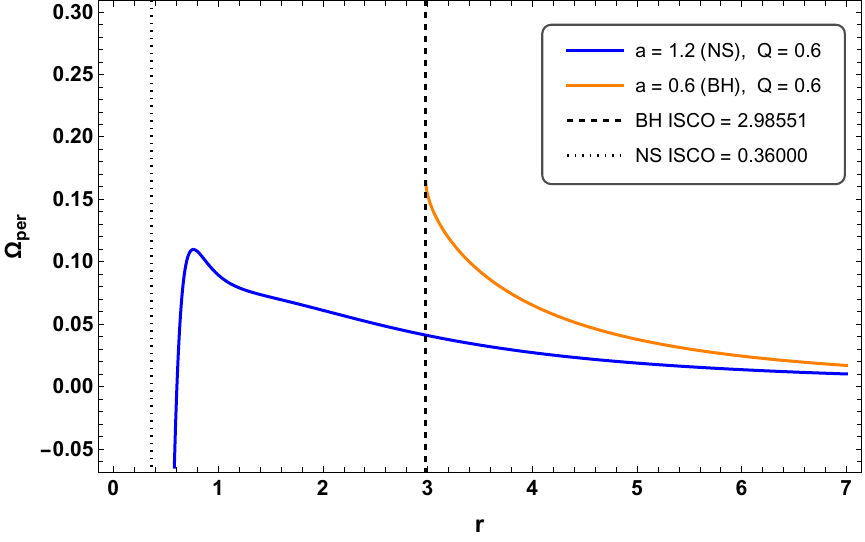}\label{r_2}}
	\caption{Variation of the $\Omega_{per}$ (in units of $M^{-1}$) versus the radial coordinate $r$ (in units of $M$) for different choices of the spin parameter $a$ and charge parameter $Q$. The emergence of negative values of $\Omega_{per}$ reflects a reversal in the direction of nodal precession.}\label{omega_r1}
\end{figure}

\begin{figure}[h!]
\centering
    \subfigure[]{\includegraphics[width=7.3cm,height=7.7cm]{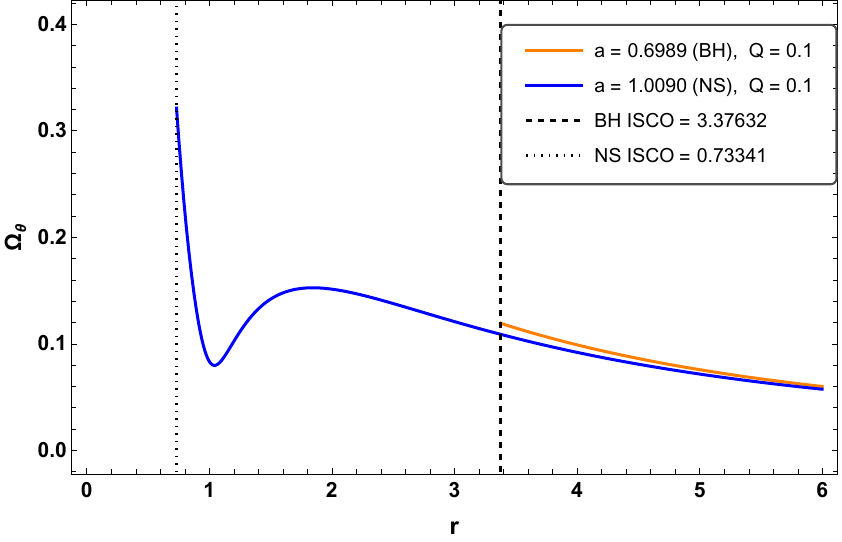}\label{r_1}}\hspace{0.8cm}
	\subfigure[]{\includegraphics[width=7.3cm,height=7.7cm]{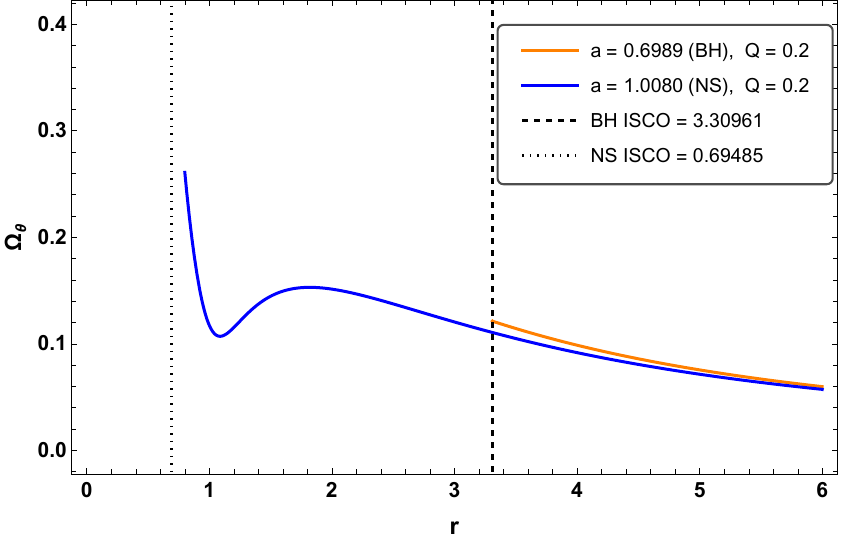}\label{r_2}}\vspace{0.8em}
    \subfigure[]{\includegraphics[width=7.3cm,height=7.7cm]{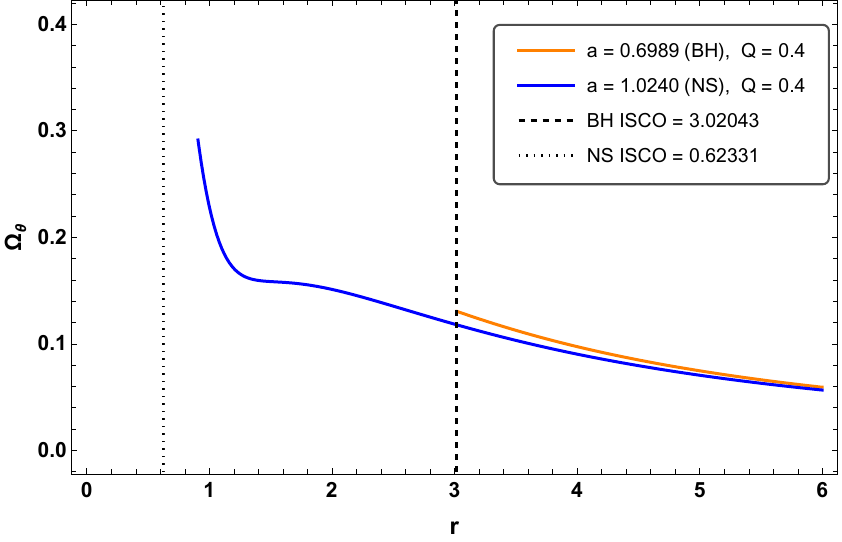}\label{r_3}}\hspace{0.8cm}
    \subfigure[]{\includegraphics[width=7.3cm,height=7.7cm]{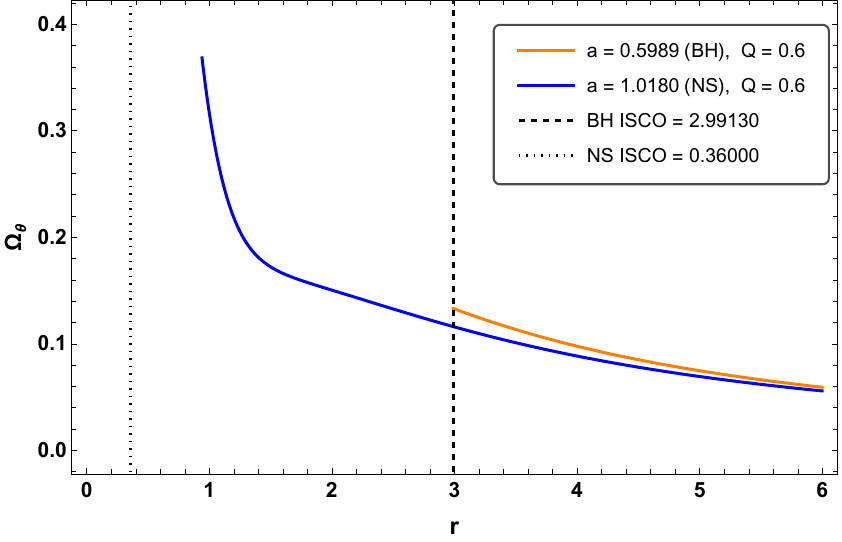}\label{r_4}}
    \caption{Variation of the radial epicyclic frequency $\Omega_{\theta}$ (in units of $M^{-1}$) versus the radial coordinate $r$ (in units of $M$) for different choices of the spin parameter $a$ and charge parameter $Q$.}\label{omega_phi}
\end{figure}

\vspace{0.2cm}
\noindent
Interestingly, from the $\Omega_{\rm per}(r)$ profiles (see Figure~\ref{omega_r1}), the charge parameter ($Q$) chiefly reshapes the inner-disk behaviour of the accretion disk, and it does so differently for black holes and naked singularities. In the case of black hole, $(\Omega_{\rm per})$ is only relevant outside the black hole ISCO (marked by the dashed line), and it decreases monotonically with radius; increasing ($Q$) shifts the inner edge inward (smaller $(r_{\rm ISCO})$) and correspondingly raises $(\Omega_{\rm per})$ near the disk's inner rim, but without introducing any qualitative feature such as a turning point or sign change. In contrast, for the naked singularity, the accretion disk can extend to much smaller radii (ISCO of the naked singularity lies much closer in), so $(\Omega_{\rm per})$ reaches significantly larger strong-field values. Moreover, increasing $(Q)$ produces a pronounced inner-region modification, i.e., the naked singularity curve develops a non-monotonic segment and can even exhibit a sign reversal (a narrow region of $(\Omega_{\rm per}<0)$) close to the inner edge, indicating that the periastron-precession sense can flip in the deepest part of the naked singularity disk. Across the panels from Figure~(\ref{omega_phi}), the vertical epicyclic (polar) frequency $\Omega_\theta(r)$ exhibits a qualitative behaviour that cleanly separates the two geometries. For the black hole configurations (orange curves), ($\Omega_\theta$) remains monotonic over the disk domain shown, i.e., it decreases smoothly as ($r$) increases and shows no turning points outside the black hole's ISCO. In contrast, the naked singularity configurations (blue dashed curves) display a non-monotonic "dip–rise–decay" profile at low $Q$ value, i.e., a local minimum appears at radii larger than the naked singularity's ISCO, followed by a mild recovery to a local maximum before joining the common outer fall-off. This "external minimum" is therefore an intrinsic strong-field signature accessible to the NS disk but absent in the black hole case. Nonetheless, the charge parameter ($Q$) systematically weakens this naked singularity-specific minimum in the displayed profiles. As ($Q$) increases from ($0.1$) to ($0.2$), the dip becomes shallower, and by $Q\simeq (0.4)--(0.6)$ the non-monotonic feature is largely washed out, leaving an almost monotonic naked singularity curve. From the accretion-disk perspective, this means that at small $(Q)$ the naked singularity disk (whose inner edge sits at the ISCO) can extend into a radial band where ($\Omega_\theta$) develops the minimum, imprinting a potentially observable spectral/timing feature. Yet at larger ($Q$), that diagnostic becomes progressively less pronounced, since the ($\Omega_\theta$) profile smooths out even though the disk still reaches deeper radii than in the black hole case.  

\subsection*{$\bullet$ Observational Aspects}

Black hole X-ray binaries (BHXBs) exhibit a wide variety of X-ray timing signatures that encode the dynamics of the accretion flow in the strong-gravity~\citep{Belloni_2014,1998ApJ...492L..59S,Stella_1999,2018MNRAS.473..431M,Remillard_2006,Zhang_2013}. Among the most prominent features are QPOs, which are broadly classified into HF QPOs and LF QPOs. HF QPOs are typically detected in the range of several tens to several hundreds of hertz, and in some observations, two HF QPOs are simultaneously present as a pair of peaks. For instance, XTE~J1650-500 has shown HF QPOs in the interval $\sim 50$-$270~\mathrm{Hz}$, while 4U~1630-47 has exhibited HF QPOs in the range $\sim 150$--$450~\mathrm{Hz}$~\citep{Belloni_2012}.

\par\medskip
LF QPOs are commonly grouped into three phenomenological subclasses, denoted as types A, B, and C, with characteristic frequency bands $\sim 6.5$--$8~\mathrm{Hz}$, $\sim 0.8$--$6.4~\mathrm{Hz}$, and $\sim 0.01$--$30~\mathrm{Hz}$, respectively. Although several theoretical models have been proposed to explain these oscillations, a widely explored interpretation relates them to the relativistic precession (RP) of matter in the inner accretion disk. Within the RP framework, the observed QPO frequencies are associated with the fundamental frequencies of (nearly) circular motion, namely the azimuthal (Keplerian) frequency $\Omega_{\phi}$~$\eqref{omeg_phi}$, the radial epicyclic frequency $\Omega_{r}$~$\eqref{omeg_r}$, and the vertical epicyclic frequency $\Omega_{\theta}$~\eqref{omeg_the}, together with the derived precession frequencies, i.e., $\Omega_{per}$~\eqref{omeg_par} and $\Omega_{nod}$~\eqref{omeg_nod}. The RP model was first introduced to interpret the simultaneous occurrence of the twin kilohertz QPOs together with an accompanying LF QPO in neutron-star low-mass X-ray binaries \citep{1998ApJ...492L..59S,Stella_1999}. Motivated by the same physical picture, its application to BHXBs proceeds by associating the C-type LF QPO with the nodal LT precession frequency $\Omega_{nod}$, the lower-frequency HF QPO with the periastron precession frequency $\Omega_{per}$, and the higher-frequency HF QPO with the azimuthal (Keplerian) orbital frequency $\Omega_{\phi}$ \citep{Ingram_2014}. Since these characteristic frequencies are governed by the spacetime parameters of the central object, this identification scheme provides a practical route to constrain both the mass $M$ and the dimensionless spin parameter $a$, as demonstrated in~\citep{Motta2013wga}.

\vspace{0.2cm}
\noindent
To assess the robustness of our results, we construct a simple theoretical model by considering a compact object of mass \(M = 10\,M_{\odot}\). A detailed analysis of the results is presented in Tables~(\ref{tab:KN_Q0p2} - \ref{tab:KN_Q0p4}), where we emphasize the observational implications by quantifying how the charge parameter $Q$ modifies the characteristic frequencies, namely the Keplerian frequency $\nu_{\phi}$, the vertical epicyclic frequency $\nu_{\theta}$, and the nodal precession frequency $\nu_{\mathrm{nod}}$. We find that even moderate (astrophysically relevant) values of $Q$ can produce appreciable shifts in these frequencies, indicating that the presence of charge leaves a measurable imprint on the precession spectrum. Notably, in the limiting case where the charge parameter vanishes ($Q=0$), Table~(\ref{tab:KN_Q0p0}) reduces to the corresponding Kerr black hole results as reported in \citep{Chakraborty:2016mhx}. Most BHXBs are transient systems in which an accretion disk forms only during outbursts, while even persistent sources undergo spectral state changes that reflect variations in accretion components. Consequently, the inner disk radius is expected to advance toward or recede from the compact object depending on the source state. If QPOs arise from characteristic frequencies of the accretion flow, changes in the disk radius should lead to corresponding evolution in QPO frequencies, as is commonly observed. Since black holes and naked singularities exhibit distinct radial profiles of these frequencies, tracking QPO evolution as the disk moves inward or outward may allow the nature of the compact object to be identified. In particular, the nodal precession frequency, $\Omega_{\mathrm{nod}}$, increases monotonically toward the ISCO in black hole systems, reaching a maximum at that radius. In contrast, in the naked singularity regime, $\Omega_{\mathrm{nod}}$ increases to a maximum and then decreases as the disk approaches the ISCO, potentially reaching zero before increasing again; this behaviour depends on the spin parameter $a$. Thus, LT precession provides a potential diagnostic for distinguishing black holes from naked singularities. Similarly, the maximum azimuthal (orbital) frequency $\Omega_\phi$ exhibits a clear dependence on the spin parameter $a$. In contrast, the radial epicyclic frequency $\Omega_r$, which in the RP framework may be identified with the separation between HF QPOs, develops qualitatively distinct radial profiles in the naked singularity regime. This behaviour is governed by the combined influence of $a$ and $Q$, and thus provides an additional observational discriminant between black hole and naked singularity geometries.

\vspace{0.2cm}
\noindent
Our results indicate that, for increasing values of the charge parameter ($Q$), the characteristic frequencies are modified appreciably. From Tables~(\ref{tab:KN_Q0p2} - \ref{tab:KN_Q0p4}) it is evident that increasing the Kerr–Newman charge parameter $Q$ reshapes both the location of the innermost stable circular orbit and the strong-field values of the characteristic frequencies, $\nu_\phi$, $\nu_\theta$, and $\nu_{\rm nod}$. For fixed spin $a$, the ISCO radius decreases monotonically with $Q$, i.e., $r_{\rm ISCO}(a;Q)$ is a decreasing function in the explored range; for example, at $a=0.9$ one finds $r_{\rm ISCO}=2.32,\,2.18,\,1.98,$ and $1.58$ for $Q=0,\,0.2,\,0.3,$ and $0.4$, respectively. Since $\nu_\phi$ grows rapidly at smaller radii, evaluating the orbital frequency at this inward-shifted ISCO yields a systematic enhancement $\nu_\phi(r_{\rm ISCO})\uparrow$ with increasing $Q$ (for the same case: $717\to 766\to 848\to 1065$~Hz), while $\nu_\theta(r_{\rm ISCO})$ also increases, with a markedly stronger sensitivity as $a\to 1$. The nodal frequency, being the difference $\nu_{\rm nod}(r_{\rm ISCO};a,Q)=\nu_\phi(r_{\rm ISCO})-\nu_\theta(r_{\rm ISCO})$, therefore exhibits two distinct regimes: for moderate spins it increases with $Q$ (e.g.\ $a=0.7:\ 79\to 90\to 94\to 109$~Hz), whereas in the near-extremal sector the rapid charge-driven growth of $\nu_\theta$ dominates and suppresses $\nu_{\rm nod}$, leading ultimately to a sign reversal as $Q$ is increased (e.g.\ at $a=1:\ \nu_{\rm nod}=1592,\,758,\,274,\,-343$~Hz for $Q=0,\,0.2,\,0.3,\,0.4$). In this sense, the inclusion of charge introduces a qualitatively important deviation from the Kerr limit: defining $\Delta\nu_{\rm nod}\equiv \nu^{Q}_{\rm nod}(r_{\rm ISCO})-\nu^{0}_{\rm nod}(r_{\rm ISCO})$, the tables show that $\Delta\nu_{\rm nod}$ is comparatively small in the moderate-spin domain so that $\nu^{Q}_{\rm nod}(r_{\rm ISCO})\sim 10^{1}\!-\!10^{2}$~Hz and remains naturally associated with the LF QPO band within the RP framework. However, as the spin approaches the rapidly rotating regime, the combined action of the inward ISCO shift and the enhanced vertical epicyclic response substantially alters the balance between $\nu_\phi$ and $\nu_\theta$, making $\Delta\nu_{\rm nod}$ large and driving $\nu^{Q}_{\rm nod}$ into the hundreds-of-Hz range (and even negative for sufficiently large $Q$, thereby signalling a strong-field restructuring of nodal precession. Moreover, for $Q\neq 0$ the nodal profile develops a finite maximum $\nu^{Q,\mathrm{peak}}_{\rm nod}=\nu^{Q}_{\rm nod}(r_p)$ at radii $r_p=\mathcal{O}(M)$; while this peak can reach $\sim 10^{3}$~Hz in near-extremal cases, its amplitude decreases monotonically with increasing $Q$ (e.g.\ at $a=0.98:\ 1565\to 1216\to 986$~Hz for $(Q=0.2,\,0.3,\,0.4)$, indicating that charge both sculpts the near-horizon nodal-precession profile and damps its maximum even as $\nu_\phi(r_{\rm ISCO})$ is enhanced. Taken together, these trends imply a consistent phenomenological picture for QPOs identification: $\nu_{\rm nod}$ retains its conventional LF QPO interpretation at moderate spins, whereas in the high-spin sector the charged nodal precession, particularly near its peak, can migrate into the HF domain, making it plausible to associate $\nu^{Q}_{\rm nod}$ with HF QPO phenomenology rather than strictly with the LF branch when $a$ is sufficiently large and $Q\neq 0$.



\newpage
\begin{table}[tbp!]
\centering
\small
\renewcommand{\arraystretch}{1.15}
\caption{We consider a compact object of mass \(M = 10\,M_{\odot}\simeq 15\,\mathrm{km}\) for the evaluation of the kepler frequency  \(\nu^{\alpha}_{\theta}\), the vertical epicyclic frequency \(\nu^{\alpha}_{\theta}\), and nodal plane precession frequency \(\nu^{\alpha}_{\mathrm{nod}}\). Conversion between the azimuthal angular frequency and the corresponding linear frequency: 
$\nu_{\phi}$ (in kHz) is obtained from $\Omega_{\phi}$ (in km$^{-1}$) via 
$\nu_{\phi} = \Omega_{\phi} \times \frac{300}{2\pi}$. The calculations are carried out at the $r_{ISCO}$ for $\bf Q=0$, with the ISCO radius chosen in the range \(M \leq r_{\mathrm{ISCO}} \leq 6M\).}
\label{tab:KN_Q0p0}
\vspace{0.5cm}
\begin{tabular*}{\textwidth}{@{\extracolsep{\fill}} c c c c c >{\centering\arraybackslash}p{3.8cm} @{}}
\hline\hline
$\bf a$ &
\begin{tabular}[c]{@{}c@{}}$\bf r_{\rm ISCO}$\\$(M)$\end{tabular} &
\begin{tabular}[c]{@{}c@{}}$\bf \nu_\phi$\\(Hz)\end{tabular} &
\begin{tabular}[c]{@{}c@{}}$\bf \nu_\theta$\\(Hz)\end{tabular} &
\begin{tabular}[c]{@{}c@{}}$\bf \nu_{\rm nod}$ at\\$r_{\rm ISCO}$\\(Hz)\end{tabular} &
\begin{tabular}[c]{@{}c@{}}$\bf \nu_{\rm nod}$ (Hz) at $\bf r_p$\\(radius in parentheses)\end{tabular}\\
\hline
0.1 & 5.67 & 234 & 231 & 3 & Not applicable \\
0.2 & 5.33 & 255 & 247 & 8 & Not applicable \\
0.3 & 4.98 & 279 & 265 & 14 & Not applicable \\
0.4 & 4.61 & 309 & 286 & 22 & Not applicable \\
0.5 & 4.23 & 346 & 312 & 34 & Not applicable \\
0.6 & 3.83 & 393 & 341 & 52 & Not applicable \\
0.7 & 3.39 & 457 & 378 & 79 & Not applicable \\
0.8 & 2.91 & 552 & 421 & 131 & Not applicable \\
0.9 & 2.32 & 717 & 472 & 245 & Not applicable \\
0.98 & 1.61 & 1050 & 463 & 587 & Not applicable \\
0.99 & 1.45 & 1160 & 422 & 738 & Not applicable \\
0.999 & 1.18 & 1394 & 254 & 1139 & Not applicable \\
0.9999 & 1.08 & 1501 & 132 & 1370 & Not applicable \\
0.99999 & 1.04 & 1550 & 64 & 1486 & Not applicable \\
0.999999 & 1.02 & 1572 & 30 & 1542 & Not applicable \\
1 & 1.00 & 1592 & 0 & 1592 & Not applicable \\
1.000001 & 0.984 & 1610 & 31 & 1579 & 1590 (0.999) \\
1.00001 & 0.967 & 1632 & 69 & 1563 & 1586 (0.997) \\
1.0001 & 0.931 & 1677 & 155 & 1521 & 1573 (0.992) \\
1.001 & 0.862 & 1767 & 363 & 1404 & 1530 (0.978) \\
1.01 & 0.752 & 1915 & 891 & 1024 & 1371 (0.961) \\
1.02 & 0.715 & 1959 & 1169 & 790 & 1264 (0.965) \\
1.04 & 0.684 & 1983 & 1513 & 470 & 1107 (0.985) \\
1.06 & 0.671 & 1977 & 1735 & 241 & 987 (1.012) \\
1.08 & 0.667 & 1959 & 1894 & 65 & 888 (1.041) \\
$\sqrt{32/27}\approx 1.089^*$ & 0.667 & 1949 & 1949 & 0 & 851 (1.054) \\
1.1 & 0.667 & 1935 & 2012 & -77 & 805 (1.072) \\
2 & 1.26 & 931 & 1581 & -650 & 49 (3.222) \\
4 & 3.17 & 330 & 566 & -236 & 2 (12.4) \\
6 & 5.38 & 172 & 289 & -116 & 0.25 (27.7) \\
\hline\hline
\end{tabular*}
\end{table}


\newpage
\begin{table}[tbp!]
\centering
\small
\renewcommand{\arraystretch}{1.15}
\caption{We consider a compact object of mass \(M = 10\,M_{\odot}\simeq 15\,\mathrm{km}\) for the evaluation of the kepler frequency  \(\nu^{\alpha}_{\theta}\), the vertical epicyclic frequency \(\nu^{\alpha}_{\theta}\), and nodal plane precession frequency \(\nu^{\alpha}_{\mathrm{nod}}\). Conversion between the azimuthal angular frequency and the corresponding linear frequency: 
$\nu_{\phi}$ (in kHz) is obtained from $\Omega_{\phi}$ (in km$^{-1}$) via 
$\nu_{\phi} = \Omega_{\phi} \times \frac{300}{2\pi}$. The calculations are carried out at the $r_{ISCO}$ for $\bf Q=0.2$, with the ISCO radius chosen in the range \(M \leq r_{\mathrm{ISCO}} \leq 6M\).}
\label{tab:KN_Q0p2}
\vspace{0.5cm}
\begin{tabular*}{\textwidth}{@{\extracolsep{\fill}} c c c c c >{\centering\arraybackslash}p{3.8cm} @{}}
\hline\hline
$\bf a$ &
\begin{tabular}[c]{@{}c@{}}$\bf r_{\rm ISCO}$\\$(M)$\end{tabular} &
\begin{tabular}[c]{@{}c@{}}$\bf \nu_\phi$\\(Hz)\end{tabular} &
\begin{tabular}[c]{@{}c@{}}$\bf \nu_\theta$\\(Hz)\end{tabular} &
\begin{tabular}[c]{@{}c@{}}$\bf \nu_{\rm nod}$ at\\$\bf r_{\rm ISCO}$\\(Hz)\end{tabular} &
\begin{tabular}[c]{@{}c@{}}$\bf \nu_{\rm nod}$ (Hz) at $\bf r_p$\\(radius in parentheses)\end{tabular} \\
\hline
0.1 & 5.607 & 237 & 233 & 3 & Not applicable \\
0.2 & 5.265 & 258 & 250 & 8 & Not applicable \\
0.3 & 4.912 & 283 & 269 & 14 & Not applicable \\
0.4 & 4.544 & 314 & 291 & 23 & Not applicable \\
0.5 & 4.159 & 353 & 317 & 36 & Not applicable \\
0.6 & 3.749 & 404 & 346 & 58 & Not applicable \\
0.7 & 3.304 & 480 & 383 & 97 & Not applicable \\
0.8 & 2.803 & 579 & 432 & 146 & Not applicable \\
0.9 & 2.182 & 766 & 482 & 284 & Not applicable \\
0.98 & 0.913 & 1922 & 924 & 998 & 1564 (0.988) \\
0.99 & 0.748 & 1985 & 1142 & 843 & 1369 (0.962) \\
0.999 & 0.711 & 1967 & 1208 & 759 & 1266 (0.966) \\
0.9999 & 0.711 & 1967 & 1210 & 757 & 1264 (0.966) \\
0.99999 & 0.711 & 1967 & 1210 & 757 & 1264 (0.966) \\
0.999999 & 0.711 & 1967 & 1210 & 757 & 1264 (0.966) \\
1 & 0.711 & 1967 & 1210 & 757 & 1264 (0.966) \\
1.000001 & 0.711 & 1967 & 1210 & 757 & 1264 (0.966) \\
1.00001 & 0.711 & 1967 & 1210 & 757 & 1264 (0.966) \\
1.0001 & 0.711 & 1967 & 1210 & 757 & 1264 (0.966) \\
1.001 & 0.708 & 1966 & 1225 & 740 & 1257 (0.967) \\
1.01 & 0.691 & 1974 & 1357 & 617 & 1199 (0.975) \\
1.02 & 0.678 & 1968 & 1501 & 467 & 1127 (0.984) \\
1.04 & 0.665 & 1989 & 1799 & 190 & 990 (1.013) \\
$1.060872^*$ & 0.660 & 1971 & 1971 & 0 & 888 (1.041) \\
1.08 & 0.660 & 1948 & 2087 & -139 & 809 (1.071) \\
$1.089$ & 0.660 & 1937 & 2131 & -194 & 777 (1.085) \\
1.1 & 0.662 & 1923 & 2182 & -259 & 738 (1.103) \\
2 & 1.228 & 930 & 1580 & -650 & 48 (3.259) \\
4 & 3.170 & 330 & 566 & -236 & 1.82 (12.305) \\
6 & 5.380 & 172 & 288 & -116 & 0.25 (27.781) \\
\hline\hline
\end{tabular*}
\end{table}


\newpage
\begin{table}[tbp!]
\centering
\small
\renewcommand{\arraystretch}{1.15}
\caption{We consider a compact object of mass \(M = 10\,M_{\odot}\simeq 15\,\mathrm{km}\) for the evaluation of the kepler frequency  \(\nu^{\alpha}_{\theta}\), the vertical epicyclic frequency \(\nu^{\alpha}_{\theta}\), and nodal plane precession frequency \(\nu^{\alpha}_{\mathrm{nod}}\). Conversion between the azimuthal angular frequency and the corresponding linear frequency: 
$\nu_{\phi}$ (in kHz) is obtained from $\Omega_{\phi}$ (in km$^{-1}$) via 
$\nu_{\phi} = \Omega_{\phi} \times \frac{300}{2\pi}$. The calculations are carried out at the $r_{ISCO}$ for $\bf Q=0.3$, with the ISCO radius chosen in the range \(M \leq r_{\mathrm{ISCO}} \leq 6M\).}
\label{tab:KN_Q0p3}
\vspace{0.5cm}
\begin{tabular*}{\textwidth}{@{\extracolsep{\fill}} c c c c c >{\centering\arraybackslash}p{3.6cm} @{}}
\hline\hline
$\bf a$ &
\begin{tabular}[c]{@{}c@{}}$\bf r_{\rm ISCO}$\\$(M)$\end{tabular} &
\begin{tabular}[c]{@{}c@{}}$\bf \nu_\phi$\\(Hz)\end{tabular} &
\begin{tabular}[c]{@{}c@{}}$\bf \nu_\theta$\\(Hz)\end{tabular} &
\begin{tabular}[c]{@{}c@{}}$\bf \nu_{\rm nod}$ at\\$\bf r_{\rm ISCO}$\\(Hz)\end{tabular} &
\begin{tabular}[c]{@{}c@{}}$\bf \nu_{\rm nod}$ (Hz) at $\bf r_p$\\(radius in parentheses)\end{tabular} \\
\hline
0.1 & 5.53 & 245 & 241 & 4 & Not applicable \\
0.2 & 5.18 & 267 & 258 & 9 & Not applicable \\
0.3 & 4.83 & 294 & 278 & 15 & Not applicable \\
0.4 & 4.45 & 326 & 302 & 25 & Not applicable \\
0.5 & 4.06 & 368 & 329 & 39 & Not applicable \\
0.6 & 3.65 & 423 & 363 & 60 & Not applicable \\
0.7 & 3.19 & 499 & 404 & 95 & Not applicable \\
0.8 & 2.67 & 618 & 453 & 165 & Not applicable \\
0.9 & 1.98 & 861 & 500 & 361 & Not applicable \\
0.98 & 0.691 & 2025 & 1424 & 601 & 1236 (0.974) \\
0.99 & 0.675 & 2038 & 1611 & 426 & 1159 (0.980) \\
0.999 & 0.665 & 2042 & 1748 & 294 & 1099 (0.992) \\
0.9999 & 0.664 & 2042 & 1767 & 275 & 1092 (0.993) \\
0.99999 & 0.664 & 2042 & 1768 & 274 & 1092 (0.993) \\
0.999999 & 0.664 & 2042 & 1768 & 274 & 1092 (0.993) \\
1.0 & 0.664 & 2042 & 1768 & 274 & 1092 (0.993) \\
1.000001 & 0.664 & 2042 & 1768 & 274 & 1092 (0.993) \\
1.00001 & 0.664 & 2042 & 1768 & 274 & 1092 (0.993) \\
1.0001 & 0.664 & 2042 & 1768 & 274 & 1091 (0.994) \\
1.001 & 0.664 & 2042 & 1780 & 262 & 1086 (0.995) \\
1.01 & 0.658 & 2040 & 1886 & 154 & 1034 (1.007) \\
$1.024426^*$ & 0.651 & 2031 & 2031 & 0 & 958 (1.024) \\
1.04 & 0.648 & 2016 & 2156 & -139 & 886 (1.045) \\
1.06 & 0.648 & 1992 & 2279 & -287 & 806 (1.077) \\
1.08 & 0.650 & 1964 & 2371 & -407 & 737 (1.113) \\
$1.089$ & 0.652 & 1952 & 2404 & -452 & 709 (1.123) \\
1.1 & 0.655 & 1935 & 2441 & -506 & 675 (1.143) \\
2 & 1.25 & 935 & 1655 & -720 & 47 (3.3) \\
4 & 3.15 & 334 & 582 & -247 & 1.83 (12.5) \\
6 & 5.35 & 175 & 295 & -121 & 0.25 (27.8) \\
\hline\hline
\end{tabular*}
\end{table}


\newpage
\begin{table}[tbp!]
\centering
\small
\renewcommand{\arraystretch}{1.15}
\caption{We consider a compact object of mass \(M = 10\,M_{\odot}\simeq 15\,\mathrm{km}\) for the evaluation of the kepler frequency  \(\nu^{\alpha}_{\theta}\), the vertical epicyclic frequency \(\nu^{\alpha}_{\theta}\), and nodal plane precession frequency \(\nu^{\alpha}_{\mathrm{nod}}\). Conversion between the azimuthal angular frequency and the corresponding linear frequency: 
$\nu_{\phi}$ (in kHz) is obtained from $\Omega_{\phi}$ (in km$^{-1}$) via 
$\nu_{\phi} = \Omega_{\phi} \times \frac{300}{2\pi}$. The calculations are carried out at the $r_{ISCO}$ for $\bf Q=0.4$, with the ISCO radius chosen in the range \(M \leq r_{\mathrm{ISCO}} \leq 6M\).}
\label{tab:KN_Q0p4}
\vspace{0.5cm}
\begin{tabular*}{\textwidth}{@{\extracolsep{\fill}} c c c c c >{\centering\arraybackslash}p{3.8cm} @{}}
\hline\hline
$\bf a$ &
\begin{tabular}[c]{@{}c@{}}$\bf r_{\rm ISCO}$\\$(M)$\end{tabular} &
\begin{tabular}[c]{@{}c@{}}$\bf \nu_\phi$\\(Hz)\end{tabular} &
\begin{tabular}[c]{@{}c@{}}$\bf \nu_\theta$\\(Hz)\end{tabular} &
\begin{tabular}[c]{@{}c@{}}$\bf \nu_{\rm nod}$ at\\$\bf r_{\rm ISCO}$\\(Hz)\end{tabular} &
\begin{tabular}[c]{@{}c@{}}$\bf \nu_{\rm nod}$ (Hz) at $\bf r_p$\\(radius in parentheses)\end{tabular} \\
\hline
0.1 & 5.41 & 247 & 243 & 4 & Not applicable \\
0.2 & 5.06 & 270 & 261 & 9 & Not applicable \\
0.3 & 4.70 & 298 & 282 & 16 & Not applicable \\
0.4 & 4.32 & 333 & 307 & 26 & Not applicable \\
0.5 & 3.92 & 377 & 336 & 41 & Not applicable \\
0.6 & 3.49 & 437 & 371 & 66 & Not applicable \\
0.7 & 3.01 & 524 & 415 & 109 & Not applicable \\
0.8 & 2.45 & 668 & 469 & 199 & Not applicable \\
0.9 & 1.58 & 1065 & 468 & 597 & Not applicable \\
$0.969948^*$ & 0.640 & 2070 & 2070 & 0 & 1055 (1.004) \\
0.98 & 0.633 & 2036 & 2166 & -130 & 986 (1.016) \\
0.99 & 0.629 & 2030 & 2273 & -243 & 937 (1.030) \\
0.999 & 0.626 & 2023 & 2356 & -333 & 897 (1.042) \\
0.9999 & 0.626 & 2022 & 2364 & -342 & 893 (1.044) \\
0.99999 & 0.626 & 2022 & 2364 & -342 & 892 (1.044) \\
0.999999 & 0.626 & 2022 & 2365 & -343 & 892 (1.044) \\
1 & 0.626 & 2022 & 2365 & -343 & 892 (1.044) \\
1.000001 & 0.626 & 2022 & 2365 & -343 & 892 (1.044) \\
1.00001 & 0.626 & 2022 & 2365 & -343 & 892 (1.044) \\
1.0001 & 0.626 & 2022 & 2365 & -343 & 892 (1.044) \\
1.001 & 0.626 & 2020 & 2375 & -355 & 888 (1.046) \\
1.01 & 0.624 & 2012 & 2443 & -431 & 851 (1.058) \\
1.02 & 0.623 & 2001 & 2511 & -510 & 813 (1.073) \\
1.04 & 0.624 & 1975 & 2620 & -645 & 743 (1.103) \\
1.06 & 0.627 & 1947 & 2700 & -753 & 682 (1.135) \\
1.08 & 0.631 & 1917 & 2757 & -840 & 628 (1.167) \\
$1.089$ & 0.634 & 1904 & 2776 & -872 & 607 (1.181) \\
1.1 & 0.637 & 1886 & 2797 & -911 & 580 (1.200) \\
2 & 1.24 & 915 & 1684 & -769 & 44 (3.4) \\
4 & 3.13 & 329 & 579 & -250 & 1.78 (12.6) \\
6 & 5.33 & 172 & 293 & -121 & 0.25 (27.9) \\
\hline\hline
\end{tabular*}
\end{table}

\clearpage
\section{Discussion \& Summary}\label{sec_6}

We have employed the behaviour of the spin-precession frequency as a diagnostic tool to distinguish between black holes and naked singularities within the Kerr–Newman spacetime. By parametrising the angular velocity of a stationary observer, the spin-precession frequency of an attached gyroscope can be systematically analysed for different choices of this parameter along various spatial directions. This approach provides a physically transparent criterion rooted in local observables, thereby offering an operational distinction between the two classes of compact objects.

\vspace{0.2cm}
\noindent
For the Kerr-Newman black hole, the spin-precession frequency diverges isotropically as the event horizon is approached, except for the special case of ZAMO observers, for whom the precession remains finite. In contrast, Kerr-Newman naked singularities exhibit a finite precession frequency throughout the spacetime, with divergences confined solely to the ring singularity along the equatorial plane. This sharp qualitative distinction provides a clear and robust criterion for differentiating between the two classes of compact objects. We have further demonstrated how the charge parameter modifies both LT and geodetic precession, revealing a nontrivial interplay between electromagnetic fields and frame-dragging effects. The transition from black hole to naked singularity is consistently accompanied by characteristic changes in the precession behaviour, reinforcing the diagnostic value of gyroscopic observables. Our results suggest that spin-precession effects, in principle, offer a viable theoretical framework for probing the nature of compact objects and testing the cosmic censorship conjecture. With future high-precision observations of relativistic precession phenomena near astrophysical compact objects, such effects may provide valuable insights into the true structure of spacetime in the strong-field regime. Specifically, if the spin–precession frequency corresponding to at least one of two observers, each characterised by a distinct angular velocity, diverges as the observer approaches the central object along any direction in the Kerr–Newman spacetime, the geometry is identified as that of a black hole. In contrast, if the spin–precession frequency associated with all admissible observers remains finite except possibly along at most one specific direction in the approach towards the centre, the spacetime corresponds to a naked singularity. This criterion highlights the crucial role of horizon-induced divergences in black hole spacetimes and underscores the absence of such universal directional divergences in naked singularity geometries.

\vspace{0.2cm}
The essential conclusions emerging from our investigations can be outlined as follows.

\begin{itemize}

    \item [(i)] In the Kerr-Newman geometry, the notion of "rest" is observer dependent: vanishing angular momentum does not, in general, imply vanishing angular velocity. This naturally motivates the use of distinct observer families (e.g., ZAMOs) and fixes the physically admissible range of angular velocities. These kinematical restrictions, inherited from the metric structure, play a central role in determining the allowed precession behaviour in the strong-field region.

    \item [(ii)] The limiting behaviour of the spin-precession frequency of a stationary gyroscope provides a sharp diagnostic to distinguish Kerr-Newman black holes from Kerr-Newman naked singularities. In the Kerr-Newman black hole, the spin-precession frequency generically diverges as the horizon is approached, whereas in the Kerr-Newman naked singularity, it remains finite throughout the spacetime except at the ring singularity on the equatorial plane. For black hole configurations, the precession divergence near the horizon occurs for all admissible stationary observers except the ZAMO family, for which the precession can remain finite at the horizon. This highlights the privileged role of ZAMOs in strong-field rotational spacetimes and clarifies how observer choice affects near-horizon precession measurements.

    \item [(iii)] The LT precession field in Kerr-Newman spacetime exhibits qualitatively distinct structures for black hole and naked singularity regimes. For black holes, the LT precession becomes unbounded as the observer approaches the static-limit surface (boundary of the ergoregion) while remaining finite outside the ergoregion. For naked singularities, the LT precession field is regular over the domain except at the ring singularity, where the frequency diverges. The direction of the LT-induced precession also shows a characteristic polar versus equatorial behaviour consistent with the expected rotational influence of spacetime.

    \item [(iv)] In the limit of vanishing spin parameter, the precession is purely geodetic and arises solely due to spacetime curvature. We obtain an explicit closed-form expression for the geodetic precession frequency, which can be used to build parameter-space maps and contour/density plots in the $(r,\:Q)$ plane, thereby isolating curvature-driven precession effects from rotational ones.
    
    \item[(v)] The acceleration scalar provides a clear physical characterisation of stationary motion in the Kerr--Newman spacetime. The derived expression extends the Kerr result by incorporating the effect of charge and reduces smoothly to the known Kerr form in the limit $Q\to 0$. Its behaviour sharply distinguishes the black hole and naked singularity cases: the acceleration diverges at the event horizon for a Kerr--Newman black hole, whereas for a Kerr--Newman naked singularity it remains finite everywhere except at the ring singularity. Moreover, in the equatorial plane, the vanishing of the acceleration reproduces the Keplerian angular velocity of neutral circular geodesics, thereby confirming the consistency of the formalism in the geodesic limit. Hence, the acceleration scalar acts as an effective probe of both the causal structure and the dynamical properties of the Kerr--Newman spacetime.
  
    \item [(vi)] The tabulated results obtained from our theoretical model demonstrate that the fundamental orbital frequencies and derived precession frequencies display systematic dependence on the charge parameter $Q$. The Charge parameter modifies the location of the ISCO and the hierarchy of characteristic frequencies, thereby shifting the nodal-precession scale and, in rapidly rotating regimes, allowing qualitative changes such as sign reversal (corresponding to a reversal in the direction of nodal precession). These trends provide a consistent pathway to connect strong-field dynamics with QPO phenomenology.

    \item [(vii)] Within the RP framework, the nodal precession frequency $\nu_{\rm nod}$ exhibits characteristic trends with spin and charge, including regimes where it can change sign, corresponding to a reversal of the nodal-precession orientation. Including charge introduces nontrivial modifications relative to the Kerr limit: at moderate spin, the change in $\nu_{\rm nod}(r_{\rm ISCO})$ remains comparatively small, preserving its natural association with the LF QPO band. However, as the spin approaches the rapidly rotating regime, the combined inward shift of $r_{\rm ISCO}$ and the enhanced vertical epicyclic response can drive $\nu_{\rm nod}$ into the hundreds-of-Hz range (towards HF QPO band) and even render it negative for sufficiently large charge. For $Q\neq 0$, the nodal-precession profile can also develop a finite maximum at a radius $r_p=\mathcal{O}(M)$, whose amplitude decreases monotonically with increasing $Q$. The black hole to naked singularity transition is accompanied by characteristic changes in these quantities, reinforcing the diagnostic value of combined precession observables.

    \item [(viii)] The above results collectively show that gyroscopic spin-precession observables encode robust qualitative signatures of horizons versus exposed singularities. Consequently, spin-precession effects offer, in principle, a viable theoretical framework for probing the nature of compact objects and for assessing strong-field departures associated with black hole/naked singularity transitions, thereby providing potential handles to test the cosmic censorship conjecture in Kerr-Newman spacetimes.

\end{itemize}

\section*{Acknowledgments}

AKC acknowledge the support of ICARD, Gurukula Kangri (Deemed to be University), Haridwar, India, for providing the necessary facilities. 
	

\end{document}